\documentclass{article}
\usepackage{arxiv}
\usepackage{graphicx}
\usepackage{mathptmx}
\usepackage[subrefformat=parens,labelformat=parens]{subcaption}
\usepackage{authblk}
\usepackage{comment}
\usepackage[titletoc]{appendix}
\usepackage{natbib}
\usepackage{longtable}
\usepackage{booktabs}
\usepackage{array}
\usepackage{amsfonts}
\usepackage{algorithm}
\usepackage{amsbsy}
\usepackage{amsthm}
\usepackage{amsmath}

\usepackage{algpseudocode}
\usepackage{bbm}
\usepackage{wrapfig}
\usepackage{booktabs}
\usepackage{multirow}
\usepackage{wrapfig,lipsum,booktabs}
\usepackage{hyperref}
\allowdisplaybreaks
\usepackage{colortbl}
\usepackage{seqsplit}
\usepackage{xcolor}
\definecolor{KWgreen}{RGB}{112,173,71} 
\definecolor{KWblue}{RGB}{0,112,192} 
\definecolor{KWred}{RGB}{192,0,0} 
\definecolor{KWpurple}{RGB}{112,48,160} 
\usepackage{float}

\title{GlycoMAC: A Multiscale Metabolic–Glycosylation Framework for Predicting Glycosylation Across Conditions in Mammalian Cell Cultures
}

\begin{document}

\author[1]{Yuming Zeng}
\author[ ~,2]{Sarah W. Harcum \thanks{Corresponding author: harcum@clemson.edu}}
\author[1]{Jinxiang Pei}
\author[ ~,1]{Wei Xie  \thanks{Corresponding author: w.xie@northeastern.edu}}

\affil[1]{Department of Mechanical and Industrial Engineering, Northeastern University, Boston, MA 02115, USA}
\affil[2]{Department of Bioengineering, Clemson University, Clemson, SC, USA}

\maketitle

\begin{abstract}
Antibody productivity and glycosylation quality in CHO cell cultures emerge from a dynamically evolving metabolic environment, yet existing models often work in isolation or at a single scale. Here, we present a multiscale mechanistic framework linking molecular, cellular, and process scales to predict how inputs shape bioprocess trajectories. 
The framework combines a single-cell kinetic model of metabolism and glycosylation with a stochastic population model that captures environment-dependent transitions among growth, production, and decline states. To characterize metabolic adaptation, we introduce the cumulative variation in oxygen uptake rate, a trajectory-based biomarker that quantifies the total metabolic adjustment experienced during culture.
Unlike population-averaged approaches, the model propagates cell-resolved metabolic states—including ammonia-regulated Golgi pH, nucleotide sugar availability, manganese cofactors, and synthesis rates—into glycan processing.
The framework was evaluated using CHO-K1 fed-batch cultures producing VRC01 IgG1 under targeted ammonia stress, matched control conditions, and 
a pyramid-feeding strategy with tighter control.
It accurately reproduced trajectories of cell growth, metabolites, productivity, and harvest glycosylation, including increased G0F abundance and reduced galactosylation under ammonia stress. By mechanistically linking process conditions to cell-state dynamics and glycosylation outcomes, the framework provides a unified foundation for digital bioprocessing, predictive biomanufacturing, and advanced process control.

\end{abstract}


\section{Introduction}

Chinese hamster ovary (CHO) cells are the dominant mammalian platform for recombinant therapeutic protein production, particularly monoclonal antibodies (mAbs), because of its scalability and ability to support proper protein folding, assembly, and complex post-translational modifications \citep{wurm2004production,durocher2009expression,butler2014choice,lalonde2017therapeutic}. In fed-batch CHO biomanufacturing, antibody productivity and glycosylation are coupled outcomes of a dynamically evolving metabolic environment rather than independent process endpoints \citep{xu2023progress}. Feeding strategies, substrate depletion, byproduct accumulation, and process controls such as pH and dissolved oxygen (DO) continuously reshape extracellular conditions and intracellular physiological states, thereby regulating cell growth, substrate uptake, central carbon metabolism, ammonia and lactate formation, antibody synthesis, and Golgi-resident glycosylation pathways \citep{gagnon2011high,mulukutla2012metabolic,hossler2009optimal,jiang2018ph,harcum2022pid,klaubert2026dynamic}.
These couplings create a central modeling challenge: predicting how process inputs drive metabolic-state evolution across molecular, cellular, and macroscopic scales and ultimately determine both culture productivity and glycosylation critical quality attributes (CQAs). Such metabolic shifts are reflected in changes in cellular energy metabolism and oxygen utilization  \citep{gagnon2011high,mulukutla2012metabolic}. Ammonia accumulation can perturb intracellular and Golgi pH, altering glycosylation kinetics \citep{gawlitzek2000ammonium,synoground2021transient}, while nutrient and amino acid availability regulate nucleotide-sugar precursor pools required for terminal glycan processing \citep{jimenez2013quantitative,aghamohseni2014effects}. As a result, feeding, ammonia stress, and pH perturbations can induce coordinated changes in growth, productivity, and glycosylation profiles, including agalactosylated, galactosylated, and sialylated species \citep{pacis2011effects,ivarsson2014evaluating,jiang2018ph}.

Substantial progress has been made in modeling CHO fed-batch culture dynamics, including viable cell density (VCD), nutrient consumption, byproduct accumulation, metabolic shifts, and recombinant protein production. Existing approaches include unstructured kinetic models \citep{robitaille2015single,kyriakopoulos2018kinetic}, structured intracellular metabolic models \citep{ahn2011metabolic,ahn2012towards,sha2018mechanistic}, stoichiometric and flux-balance frameworks \citep{selvarasu2012combined,huang2017quantitative}, and hybrid mechanistic--data-driven models \citep{pinto2023hybrid}. These models have advanced process understanding and prediction of culture trajectories, but most do not explicitly propagate metabolic-state evolution to detailed Golgi glycosylation processing. Recent single-cell-based multiscale culture models provide a basis for representing heterogeneous population dynamics and condition-dependent metabolic-state evolution \citep{wang2024multi}, but glycosylation CQAs remain less explicitly integrated.
In parallel, mechanistic glycosylation models have described Golgi glycan maturation through enzyme localization, nucleotide-sugar donor transport, Golgi residence time, and large-scale N-linked reaction networks \citep{umana1997mathematical,krambeck2005mathematical,hossler2007systems,jimenez2011dynamic,jimenez2013quantitative,krambeck2017model}. More recent studies have linked nutrient availability, nucleotide-sugar metabolism, process conditions, and CHO metabolic networks to mAb glycoform profiles \citep{villiger2016controlling,karst2017modulation,kotidis2019model,erklavec2021dynamic,kotidis2023choglyconet}. However, many culture-coupled glycosylation frameworks still rely on deterministic, averaged, or reduced metabolic states. This limits the ability to capture how heterogeneous single-cell metabolism, asynchronous phase transitions, and cell-to-cell variability in antibody synthesis and Golgi microenvironments collectively generate population-level glycoform distributions.

This limitation is critical because glycosylation is inherently heterogeneous. Mammalian cells produce distributions of glycoforms with varying levels of galactosylation, fucosylation, and sialylation, a phenomenon known as glycosylation microheterogeneity \citep{reusch2015fc,fisher2019n}. 
This heterogeneity arises from multiple factors, including Golgi enzyme localization, substrate competition, nucleotide-sugar availability, intracellular transport, and residence-time effects \citep{krambeck2005mathematical,jimenez2013quantitative,fisher2019n}.
At the same time, recombinant CHO cultures exhibit substantial heterogeneity in productivity, gene expression, and physiological state, even within clonally derived populations \citep{pilbrough2009intraclonal,chrysinas2024cell}. Consequently, identical population-averaged metabolite levels can correspond to distinct underlying cellular-state distributions and, therefore, different glycosylation outcomes, depending on process inputs, culture conditions, and cell fate trajectories.

To address this gap, we develop a stochastic, multiscale metabolism–glycosylation framework that links single-cell heterogeneity to population-level culture performance and product quality. At the single-cell level, each cell is described by a stochastic metabolic state and a discrete metabolic phase, with phase-dependent fluxes governing central carbon and amino acid metabolism, ammonia production, biomass growth, and antibody synthesis. 
A probabilistic metabolic phase-transition model captures cumulative metabolic adjustments over time while accounting for asynchronous state progression and phenotypic heterogeneity across the population. 
A mechanistic Golgi N-linked glycosylation model is then coupled to cell-specific metabolic trajectories through ammonia-mediated Golgi pH, nucleotide-sugar precursor availability, manganese availability, and antibody synthesis rate. Population-level VCD, extracellular metabolite concentrations, antibody titer, and glycoform distributions are obtained by aggregating heterogeneous single-cell outputs.

The framework is evaluated using recombinant CHO-K1 fed-batch cultures producing VRC01 IgG1 under three distinct conditions: a baseline process, an ammonia-stress condition, and a higher feeding strategy with tighter process control \citep{elliott2020spent,harcum2022pid,chitwood2023microevolutionary,synoground2021transient,klaubert2026dynamic}. These provide complementary conditions to the process–quality relationship: ammonia stress introduces an early byproduct burden, while the altered feeding and control strategies can modulate nutrient availability, growth dynamics, oxygen utilization, late-stage ammonia accumulation, and antibody productivity. Model performance is assessed by its ability to robustly predict culture dynamics and to mechanistically propagate process inputs to yield and glycosylation outcomes, including G0F abundance, terminal and total galactosylation, and sialylation.

The main contributions of this work are summarized as follows. First, we develop a bottom-up single-cell-to-population metabolism--glycosylation framework that couples stochastic single-cell metabolism with cell-specific Golgi N-linked glycosylation and aggregates heterogeneous outputs to predict culture dynamics and glycoform distributions. Second, we introduce a process-sensitive metabolic phase-transition mechanism driven by oxygen-uptake-rate dynamics, enabling feeding, ammonia stress, and pH behavior to shift metabolic-state timing and downstream glycosylation outcomes. Third, we demonstrate mechanistic propagation of process variations to glycosylation CQAs across a targeted ammonia-stress, its control, and a pyramid feeding protocol with tight process control. Finally, the multiscale model enables uncertainty-aware prediction of culture dynamics, antibody production, and glycosylation trajectories under dynamic process perturbations.

\section{Materials and Methods}
\label{sec:MaterialsMethods}



\subsection{Cell Culture Conditions}
\label{subsec:CultureCondition_expriments}

A recombinant CHO-K1 cell line (clone A11) expressing the anti-HIV monoclonal antibody VRC01 (IgG1) was used in all experiments. Cultures were performed in ambr250 bioreactors (Sartorius Stedim) at an initial seeding density of $0.4\times10^6$ cells mL$^{-1}$ in ActiPro medium supplemented with glutamine and maintained at 37$^\circ$C.

Three experimental conditions were evaluated:
\begin{itemize}
\item[Case A:] Control fed-batch cultures without ammonia stress (Elliott et al., 2020~\citep{elliott2020spent}).

\item[Case B:] Ammonia-stressed fed-batch cultures (10 mM ammonia added 12 h post-inoculation via NH$_4$Cl addition)~\citep{elliott2020spent}.

\item[Case C:] Pyramid fed-batch cultures with refined PID controls (Harcum et al., 2022~\citep{harcum2022pid}).
\end{itemize}
Cases A and B had an identical fed-batch strategy and differed only by the ammonia perturbation. The feeding strategy was constant daily feeding initiated on Day 3 at 3\% (v/v) Cell Boost 7a and 0.3\% (v/v) Cell Boost 7b. For Case~B, NH$_4$Cl was added 12 h post-inoculation to increase the ammonia concentration by 10 mM. To maintain consistent osmolarity and working volume, an equivalent volume of NaCl solution was added to the control cultures (Case~A).
Case~C employed a pyramid feeding strategy:  3\%/0.3\% (v/v) Days 3--5, 4\%/0.4\% Days 6--7, 5\%/0.5\% Days 8--9, 4\%/0.4\% Days 10--11, and 3\%/0.3\% Days 12--13.

All cultures were maintained at 50\% dissolved oxygen. {Antifoam was added as needed during fed-batch operation. For Cases~A and B,
a 10\% antifoaming solution (GE Healthcare) was used to control foaming
\cite{elliott2020spent}. For Case~C, a premixed 10\% antifoam solution
(Cytiva, Q7-2587) was added in $10~\mu\mathrm{L}$ increments,
typically beginning between Days~2 and~4\cite{harcum2022pid}. Reactor-specific addition times and dosing histories were
not available for Cases~A and B.} For Cases A and B, pH was maintained at 
pH 7.0 ($\pm0.2$ deadband). Case C was maintained at 7.0 throughout the culture. 
Additional experimental details and analytical methods are provided in Elliott et al. (2020) and Harcum et al. (2022). \cite{elliott2020spent, harcum2022pid}.

\textbf{Biomass and Target Protein.}
The biomass composition and dry cell weight of the CHO-K1 cell line were obtained from the characterization study reported by Sz{\'e}liov{\'a} et al. (2020)~\cite{szeliova2020cho}, which estimated a cellular dry weight of 252.3 pg/cell together with detailed biomass constituent information. The amino acid sequence of the VRC01 monoclonal antibody was reconstructed by translating the corresponding DNA sequence reported in Synoground et al.(2021)~\cite{synoground2021transient}. 


\subsection{Cell Culture Measurements}

\textbf{Offline Cell Culture Measurements.}
Daily samples were collected prior to feeding for offline analysis of critical process parameters (CPPs).
Viable cell density (VCD) and cell viability were measured using trypan blue exclusion with a Vi-Cell XR analyzer (Beckman Coulter). Extracellular concentrations of glucose, lactate, glutamine, glutamate, ammonia, and antibody titer were quantified using a Cedex Bioanalyzer (Roche Diagnostics). Additional amino acid concentrations were measured using capillary electrophoresis–high-pressure mass spectrometry (CE-HPMS; REBEL, 908 Devices).

\textbf{Online Cell Culture Measurements.} The ambr250 bioreactor system continuously monitored multiple process variables online throughout the fed-batch culture process, including working volume, feed and sampling volumes, pH, agitation speed, temperature, dissolved oxygen (DO), and the inlet and off-gas flow rates and gas compositions for air, O$_2$, and CO$_2$.

\textbf{Glycosylation Measurements.}
Samples for glycosylation
analysis were collected on Day 14 for all the fed-batch cultures.  Detailed procedures for
IgG purification, glycan labeling, and UPLC-based glycosylation
analysis are described in Synoground et al. (2021)~\cite{synoground2021transient}. {Because IgG was purified from the culture supernatant, the measured Day~14
glycan distribution represents the cumulative extracellular antibody pool
present at harvest rather than the instantaneous glycoform distribution
secreted specifically on Day~14. Accordingly, the experimental measurements
were compared with the model-predicted cumulative product-pool glycoform
distribution at Day~14.}

\subsection{Cell-Specific Oxygen Uptake Rate}

The cell-specific oxygen uptake rate ($q_{\mathrm{O}_2}$) is used to characterize cellular metabolic activity during fed-batch cultivation, particularly transitions between glycolytic and oxidative metabolism. Following Wang et al.~\cite{wang2024multi}, it is defined as
$
q_{\mathrm{O}_2}(t)
=
\frac{\mathrm{OUR}(t)}
{\mathrm{VCD}(t)},
$
where $\mathrm{VCD}(t)$ represents the VCD at time $t$. The oxygen uptake rate  $\mathrm{OUR}(t)$ is approximated by the oxygen transfer rate (OTR) derived from online measurements, assuming negligible transient accumulation of dissolved oxygen: $
\mathrm{OUR}(t)\approx \mathrm{OTR}(t).$
The OTR is estimated as $
\mathrm{OTR}
=
K\,n^a
\left(
\frac{M_f}{V}
\right)^b
C_{\mathrm{cal}}^*
\frac{y_{O_2}}{y_{O_2,\mathrm{cal}}}
\left(
1-\frac{\mathrm{DO}}{100}
\right),
$
where $n$ is the impeller speed, $M_f$ is the inlet gas flow rate, $V$ is the liquid volume, $y_{O_2}$ denotes the inlet oxygen concentration, and DO is the dissolved oxygen level expressed as percentage air saturation. The parameters $K$, $a$, and $b$ are empirical transfer coefficients.
OTR/OUR estimates are derived from online measurements of DO, gas flow rate, inlet oxygen composition, and agitation conditions. 

{Prior to calculating $q_{\mathrm{O}_2}$, the OTR time series was sorted by time and smoothed using a third-order Savitzky--Golay filter.
The resulting trajectory, 
$\widetilde{\mathrm{OTR}}(t)$, was subsequently used to compute both the cell-specific oxygen uptake rate and its cumulative variation, thereby reducing the influence of high-frequency measurement noise and short-term OTR fluctuations. Because OTR was inferred from online measurements through an empirical oxygen-transfer correlation rather than measured directly, it is treated as a soft-sensor estimate. Independent OTR measurements or run-specific cell-free $k_La$ validation data were not available in the source studies. Therefore, the uncertainty associated with the calculated OTR could not be quantified directly. Additional details regarding OTR estimation are provided by Wang et al.~\cite{wang2024multi}.}

\section{Model Development}
\label{sec:modelDevelopment}

\begin{figure}[t]
    \centering
    \includegraphics[width=0.9\linewidth]{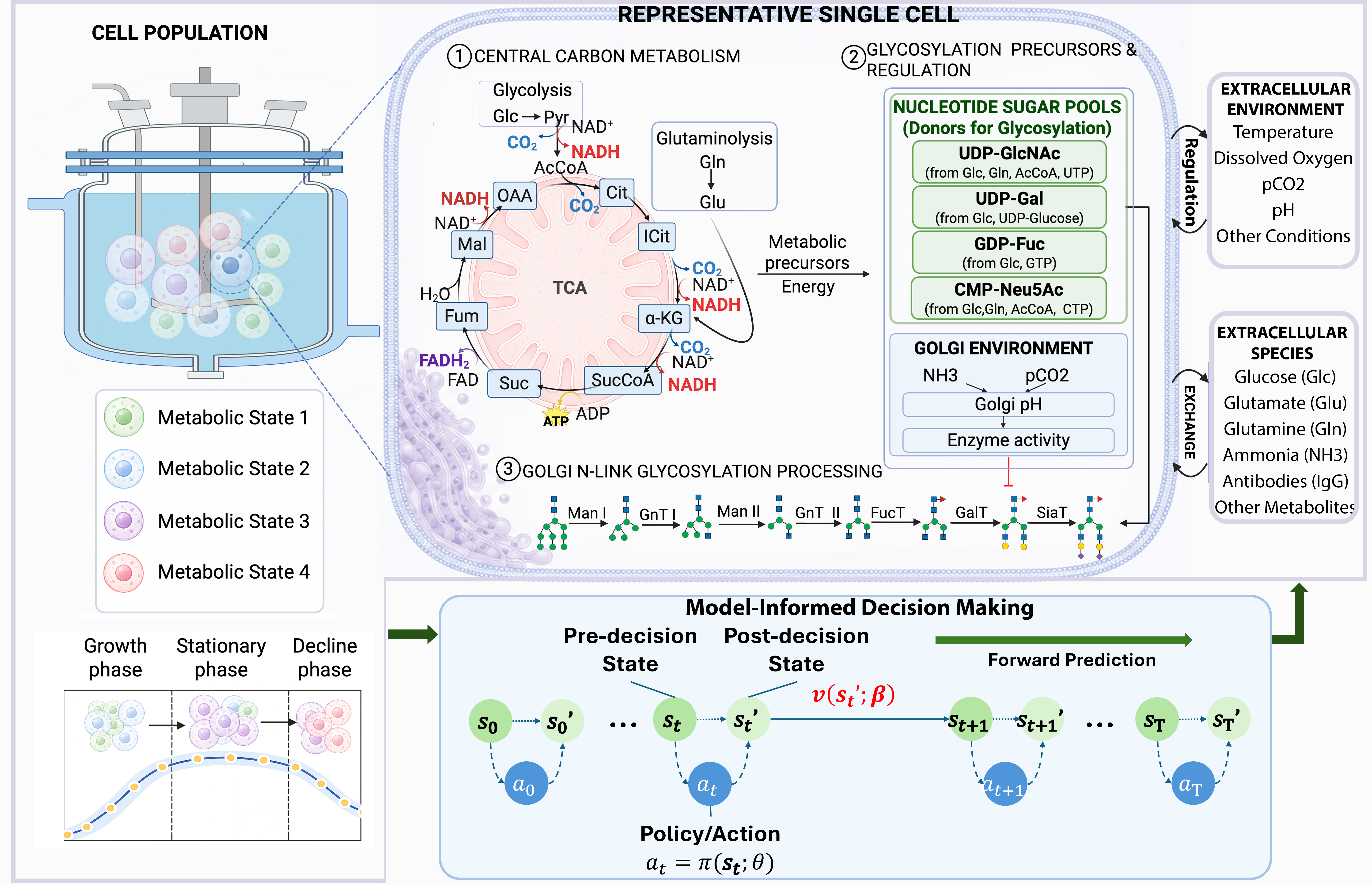}
    \caption{
    Overview of the multiscale metabolism--glycosylation modeling and decision-making framework.
    (a) Heterogeneous population dynamics constructed from stochastic single-cell trajectory models capturing metabolic-state transitions and cell-population evolution (top left).
    (b) Metabolism--glycosylation coupling linking intracellular metabolism and extracellular culture conditions to nucleotide-sugar synthesis, Golgi regulation, and glycoform formation (top right).
    (c) Model-informed decision-making framework based on pre-decision and post-decision states, where control actions and metabolic stochasticity jointly determine future state evolution and policy evaluation (bottom).
    }
    \label{fig:framework}
\end{figure}

We developed a modular multiscale framework with a hierarchical bottom-up/top-down architecture to capture causal relationships across molecular, cellular, and macroscopic bioreactor scales. The framework integrates stochastic single-cell metabolism, metabolic state transitions, Golgi N-linked glycosylation, and population-level culture dynamics (Fig.~\ref{fig:framework}).

Single-cell metabolism responds dynamically to extracellular conditions, including nutrient availability, metabolic byproducts, oxygen, pH, and antibody secretion. In turn, metabolic state influences glycosylation through ammonia-mediated Golgi pH shifts, nucleotide-sugar and manganese availability, and antibody synthesis rate. Population-level culture dynamics and glycosylation profiles are obtained by aggregating heterogeneous single-cell trajectories, thereby preserving intracellular variability and population-level product-quality attributes.

At the single-cell level, the model comprises three coupled modules: (i) a phase-dependent intracellular metabolic network, (ii) a metabolic-shift model describing transitions between metabolic states, and (iii) a mechanistic Golgi glycosylation network.
{Intracellular reaction dynamics are represented using a Chemical Langevin Equation (CLE) formulation, in which reaction events are modeled as state-dependent Poisson processes under the diffusion approximation. Population heterogeneity is captured through a hierarchical stochastic differential equation (SDE) framework that assigns cells distinct metabolic states and parameter realizations, while metabolic-state transitions are governed by probabilistic dynamics that depend on factors such as culture age and cellular stress. Consequently, cells exposed to the same extracellular environment may follow different intracellular trajectories, capturing both intrinsic molecular fluctuations and extrinsic population heterogeneity \citep{thomas2018sources,persson2022scalable,stamatakis2010mathematical}.
}

Cell-culture dynamics are formulated using modular SDEs,
$d\pmb{s}(t)
=\sum_\ell \pmb{\mu}_\ell(\pmb{s}_t;\pmb{\beta}_\ell)dt+
    \pmb{D}_\ell(\pmb{s}_t;\pmb{\beta}_\ell)d\mathbf{W}(t)$, 
    where drift and diffusion terms $(\pmb{\mu}_\ell,\pmb{D}_\ell)$ for each $\ell$-th module depend on reaction rates $\mathbf{v}_\ell(\pmb{s}_t;\pmb{\beta}_\ell)$ and $d\mathbf{W}_t$ represents stochastic fluctuations. Shared state variables $\pmb{s}_t$ enable seamless integration of the metabolic, phase-transition, and glycosylation modules, supporting multiscale simulations across diverse operating conditions.

Process decisions \(\pmb{a}_t\), including feeding, pH control, and dissolved-oxygen regulation, act on the extracellular environment and thereby influence single-cell metabolic trajectories. These effects propagate through the coupled metabolism--glycosylation network, ultimately affecting cell growth, metabolite dynamics, antibody production, and glycosylation critical quality attributes (CQAs).
Accordingly, the framework is formulated as a control-oriented state-transition system for model-informed decision making. At each decision epoch \(\tau_h\), the system is characterized by a pre-decision state \(\pmb{s}_h\) that summarizes the current culture, metabolic, and product-quality conditions. A control action is selected according to a policy
\(
\pmb{a}_h=\pi(\pmb{s}_h; \pmb{\theta}),
\)
where \(\pmb{a}_h\) denotes process interventions such as feeding, pH adjustment, or DO control. After accounting for known process effects (e.g., feed-induced dilution), the action maps the system to a post-decision state \(\pmb{s}_h^{\prime}\), enabling predictive evaluation and optimization of process-control strategies.

\subsection{Single-Cell Metabolic Model}

Built on the multiscale hybrid modeling framework of Wang et al.~\citep{wang2024multi}, the single-cell metabolic model integrates intracellular metabolic kinetics with discrete metabolic phase transitions. 
For each cell \(i\), let
\(Z_i(t)\in\{0,1,2,3\}\)
denote its metabolic phase, corresponding to early exponential growth, late
exponential growth, stationary phase, and decline phase, respectively. The intracellular metabolic state is represented by
\(\mathbf{u}_i(t)\in\mathbb{R}^m\), which contains the concentrations of $m$ metabolites, including glucose, lactate, glutamine, glutamate, ammonia, amino acids, and other species in the reaction network. Shared extracellular metabolite concentrations are denoted by
\(\mathbf{u}_{\mathrm{ex}}(t)\), while controlled bioreactor conditions are represented by
\(
\mathbf{c}(t)=\big[\mathrm{pH}(t),T(t),\mbox{DO}(t),\ldots\big]^\top
\), including pH, temperature, DO, and other environmental variables.

Single-cell metabolic dynamics are described by a phase-dependent chemical Langevin equation (CLE) 
\citep{gillespie2000chemical,schnoerr2017approximation}:
\begin{eqnarray}
\label{eq:single_cell_cle}
{d\mathbf{u}_i(t)
=N_{\mathrm{in}}
\mathbf{v}^{(Z_i(t))}
\left(
\mathbf{u}_i(t),
\mathbf{u}_{\mathrm{ex}}(t),
\mathbf{c}(t)
\right)dt }
+
\left[
N_{\mathrm{in}}\,
\Sigma^{(Z_i(t))}
\!\left(
\mathbf{u}_i(t),
\mathbf{u}_{\mathrm{ex}}(t),
\mathbf{c}(t)
\right)
\right]^{1/2}
d\mathbf{W}_i(t),
\end{eqnarray}
where \(N_{\mathrm{in}}\in\mathbb{R}^{m\times R}\) is the intracellular stoichiometric matrix and 
\(\mathbf{v}^{(z)}\) denotes the phase-dependent reaction-flux vector.
The CLE represents reaction events as Poisson processes and approximates their aggregate effect by a Wiener diffusion process. As a result, the diffusion covariance is determined by the reaction propensities, causing stochastic fluctuations to scale with the instantaneous reaction fluxes \citep{gillespie2000chemical,schnoerr2017approximation}.
 For a standard reaction-network CLE,
\(
\Sigma^{(z)}
\!\left(
\mathbf{u}_i(t),
\mathbf{u}_{\mathrm{ex}}(t),
\mathbf{c}(t)
\right)
=
\operatorname{diag}
\left(
\mathbf{v}^{(z)}
\!\left(
\mathbf{u}_i(t),
\mathbf{u}_{\mathrm{ex}}(t),
\mathbf{c}(t)
\right)
\right)
\).

Reaction fluxes depend on both intracellular states and the shared extracellular environment, thereby coupling single-cell metabolism to population-level culture conditions. The metabolic network includes glycolysis,
lactate metabolism, glutaminolysis, TCA-cycle-associated reactions,
amino-acid metabolism, antibody synthesis, and biomass synthesis. Fluxes are modeled using Michaelis–Menten-type kinetics with phase-dependent regulation, substrate limitation, metabolite inhibition, ammonia-associated effects, and pH-dependent enzyme activity.

The metabolic reaction network is summarized in
Table~\ref{tab:metabolic_network}, with corresponding kinetic expressions provided in Table~\ref{tab:flux_kinetics} and paramters in Appendix~\ref{app:model_parameters}. To account for experimentally observed pH sensitivity of CHO cell metabolism, pH effects are incorporated as a multiplicative modulation of reaction rates \citep{gagnon2011high,mulukutla2012metabolic}. For reaction $r$ in metabolic phase $z$, the pH-dependent activity factor is defined as
\begin{equation}
\label{eq:pH_modulation}
\phi_{r,\mathrm{pH}}^{(z)}
\!\big(\mathrm{pH}(t)\big)
=
\frac{1}{
1+
\left(\dfrac{10^{-\mathrm{pH}(t)}}{K_{1,r}^{(z)}}\right)^{n_r^{(z)}}
+
\left(\dfrac{K_{2,r}^{(z)}}{10^{-\mathrm{pH}(t)}}\right)^{n_r^{(z)}}
},
\end{equation}
where \(K_{1,r}^{(z)}\) and \(K_{2,r}^{(z)}\) define the lower and upper pH tolerance bounds, and \(n_r^{(z)}\) controls sensitivity to pH deviations. This dimensionless factor scales the maximum reaction rate, capturing reductions in effective enzyme activity under non-optimal pH conditions.
For each single-cell trajectory, the model computes a phase-dependent antibody production rate,
\(
q_{\mathrm{Ab},i}(t)
=
q_{\mathrm{Ab}}^{(Z_i(t))}
\!\left(
\mathbf{u}_i(t),
\mathbf{u}_{\mathrm{ex}}(t),
\mathbf{c}(t)
\right),
\)
which serves as the interface between intracellular metabolism and the downstream glycosylation model. Population-level quantities are obtained by aggregating single-cell outputs across the ensemble.

\subsection{Single-Cell Golgi Glycosylation Model}
\label{sec:glycosylation}

The glycosylation component is adapted from the mechanistic Golgi N-linked glycosylation model of Villiger et al. (2016)~\cite{villiger2016controlling}, including its transport–reaction structure, reaction network topology, enzyme localization, nucleotide-sugar transporter distribution, and quasi-steady-state representation of Golgi processing. The key contribution of the present work is the integration of this mechanistic Golgi model into a multiscale metabolism–glycosylation framework.

Specifically, cell-resolved metabolic states generated by the single-cell metabolism model directly regulate Golgi glycosylation through ammonia-associated pH perturbations, nucleotide-sugar precursor availability, manganese availability, and antibody synthesis rate. The resulting cell-specific glycosylation profiles are subsequently aggregated across the heterogeneous population to predict culture-level glycoform distributions. This formulation establishes a mechanistic link between intracellular metabolic heterogeneity, subcellular Golgi processing, and experimentally observed population-level product quality attributes.

For each cell \(i\), the Golgi state is defined as
\(
\mathbf{y}_i(\xi,t)
=
\big[
\mathbf{OS}_i(\xi,t)^\top,\,
\mathbf{NS}^{\mathrm{Golgi}}_i(\xi,t)^\top,\,
\mathbf{Nuc}^{\mathrm{Golgi}}_i(\xi,t)^\top
\big]^\top,
\)
where
\(\xi\in[0,1]\) represents the normalized position along the Golgi, from entry \((\xi=0)\) to exit \((\xi=1)\). Here,
\(\mathbf{OS}_i\) denotes the oligosaccharide distribution,
\(\mathbf{NS}^{\mathrm{Golgi}}_i\) the nucleotide-sugar concentrations, and \(\mathbf{Nuc}^{\mathrm{Golgi}}_i\) the corresponding nucleotide species.
Assuming a quasi-steady mapping of temporal progression onto spatial position, the transport–reaction dynamics are
\begin{equation}
\label{eq:single_cell_golgi}
\frac{\partial \mathbf{y}_i(\xi,t)}{\partial \xi}
=
\frac{\pi d^2}{4q_i(t)}
\begin{bmatrix}
\mathbf{V}_{OS}\mathbf{r}_i(\xi,t)\\
\mathbf{F}_{T,i}(\xi,t)+\mathbf{V}_{NS}\mathbf{r}_i(\xi,t)\\
-\mathbf{B}_{T,i}(\xi,t)+\mathbf{V}_{N}\mathbf{r}_i(\xi,t)
\end{bmatrix},
\end{equation}
where \(d\) is the Golgi diameter, \(q_i(t)\) the effective Golgi flow rate,
\(\mathbf{r}_i\) the glycosylation reaction rates,
\(\mathbf{V}_{OS}\), \(\mathbf{V}_{NS}\), \(\mathbf{V}_{N}\) stoichiometric matrices, and \(\mathbf{F}_{T,i}\), \(\mathbf{B}_{T,i}\)
the nucleotide-sugar and nucleotide transport fluxes.

Multiscale metabolic-glycosylation coupling is introduced through four metabolism-dependent inputs: intracellular ammonia
\(\mathrm{NH}_{3,i}(t)\), cytosolic nucleotide-sugar precursors
\(\mathbf{N}^{\mathrm{cyt}}_i(t)\), manganese concentration
\(\mathrm{Mn}_i(t)\), and antibody synthesis rate
$q_{\mathrm{Ab},i}(t)$.
These quantities are supplied by the single-cell metabolic model and regulate Golgi pH, nucleotide-sugar transport, cofactor availability, and glycoprotein influx, respectively.
The spatial organization of Golgi enzymes and nucleotide-sugar transporters is represented by Gaussian localization profiles:
\(
E_j(\xi)
=
E_j^{\max}
\exp
\left[
-\frac{1}{2}
\left(
\frac{\xi-\xi_j^{\max}}{\omega_j}
\right)^2
\right]\) and \(
TP_k(\xi)
=
TP_k^{\max}
\exp
\left[
-\frac{1}{2}
\left(
\frac{\xi-\xi_k^{\max}}{\omega_k}
\right)^2
\right],
\)
where \(E_j(\xi)\) and \(TP_k(\xi)\) denote the spatial distributions of enzyme \(j\) and transporter \(k\), respectively. These profiles, adopted from Villiger et al. (2016)~\citep{villiger2016controlling}, are treated as phase-independent structural properties of the Golgi apparatus.

\vspace{0.05in}
\textbf{(1) Ammonia-dependent Golgi pH and enzyme activity.}
Metabolically generated ammonia perturbs the intracellular and Golgi environment. As uncharged ammonia readily diffuses across membranes, Golgi pH is approximated as a monotonic function of the cell-specific ammonia level
\citep{villiger2016controlling,gawlitzek2000ammonium,synoground2021transient}:
\begin{equation}
\label{eq:single_cell_ph}
\mathrm{pH}_{G,i}(t)
=
pK_a
+
\log_{10}
\left(
\frac{\mathrm{NH}_{3,i}(t)}
{N_A-\mathrm{NH}_{3,i}(t)}
\right).
\end{equation}
The effective catalytic activity of enzyme \(j\) in cell \(i\) is modeled as a pH-dependent modulation of its maximal activity,
\begin{equation}
\label{eq:single_cell_kf_ph}
k_{f,j,i}(t)
=
k_{f,j}^{\max}
\exp
\left[
-\frac{1}{2}
\left(
\frac{
\mathrm{pH}_{G,i}(t)-\mathrm{pH}_{\mathrm{opt},j}
}
{\omega_{f,j}}
\right)^2
\right],
\end{equation}
where \(k_{f,j}^{\max}\) is the maximal catalytic activity,
\(\mathrm{pH}_{\mathrm{opt},j}\) is the optimal pH for enzyme \(j\), and
\(\omega_{f,j}\) controls pH sensitivity.

\vspace{0.05in}
\textbf{(2) Metabolism-dependent nucleotide-sugar transport.}
The transport flux for nucleotide-sugar \(k\) in cell \(i\) is
\begin{equation}
\label{eq:single_cell_transport}
\begin{aligned}
F_{T,k,i}(\xi,t)
=
k_{T,k}TP_k(\xi)
\left(
\frac{
NS^{\mathrm{cyt}}_{k,i}(t)
}{
K^{\mathrm{cyt}}_{NS,k}
+
NS^{\mathrm{cyt}}_{k,i}(t)
}
\right)
\times
\left(
\frac{
Nuc^{\mathrm{Golgi}}_{k,i}(\xi,t)
}{
K^{\mathrm{Golgi}}_{Nuc,k}
+
Nuc^{\mathrm{Golgi}}_{k,i}(\xi,t)
}
\right),
\end{aligned}
\end{equation}
where \(NS^{\mathrm{cyt}}_{k,i}(t)\) is the cytosolic nucleotide-sugar precursor concentration and
\(Nuc^{\mathrm{Golgi}}_{k,i}(\xi,t)\) is the corresponding Golgi nucleotide concentration. This formulation directly couples transport capacity to the cell’s metabolic state, linking intracellular metabolism to Golgi glycosylation dynamics.

\textbf{(3) Metabolism-dependent glycosylation reaction rates.}
The vector of glycosylation reaction rates for cell $i$ is written abstractly as
\(
\mathbf{r}_i(\xi,t)
=
\mathbf{r}
\left(
\mathbf{OS}_i(\xi,t),
\mathbf{NS}^{\mathrm{Golgi}}_i(\xi,t),
\mathbf{Nuc}^{\mathrm{Golgi}}_i(\xi,t),
\mathbf{k}_{f,i}(t),
\mathbf{E}(\xi),
\mathrm{Mn}_i(t)
\right),
\)
where
\(
\mathbf{k}_{f,i}(t)
=
\big[
k_{f,1,i}(t),
k_{f,2,i}(t),
\ldots,
k_{f,J,i}(t)
\big]^\top
\)
denotes the effective enzyme activities. Thus, local oligosaccharide availability, Golgi nucleotide-sugar and nucleotide pools, enzyme localization, ammonia-driven pH effects, and manganese availability jointly determine Golgi reaction fluxes. Detailed kinetic expressions follow Villiger et al. (2016)~\citep{villiger2016controlling} and are summarized in Appendix Table~\ref{tab:glyco_kinetics}. The model incorporates competitive Michaelis–Menten kinetics for mannosidases, sequential-order Bi–Bi kinetics for manganese-dependent glycosyltransferases, and random-order Bi–Bi kinetics for fucosylation and sialylation reactions.

\vspace{0.05in}

\textbf{(4) Metabolism-dependent glycoprotein influx and glycoform output.}
Glycosylation is initialized at the Golgi entry as 
\(
\mathbf{y}_i(0,t)
=
\mathbf{y}_{i,\mathrm{in}}(t),
\)
where the inflow depends on the cell-specific antibody synthesis rate \(q_{\mathrm{Ab},i}(t)\) and precursor availability. The secreted glycoform distribution is obtained at Golgi exit as
\(
\mathbf{g}_i(t)
=
\mathbf{g}
\left(
\mathbf{y}_i(1,t)
\right),
\)
where \(\mathbf{g}(\cdot)\) mapping terminal oligosaccharides to measured glycoform categories.

Although culture phase and time are known to influence CHO glycosylation \citep{sumit2019dissecting,radhakrishnan2017controlling}, this framework attributes phase-dependent behavior to the evolving intracellular and Golgi environment arising from the coupled stochastic culture model, rather than to phase-specific glycosyltransferase kinetic parameters. Accordingly, catalytic constants, dissociation constants, transporter parameters, and Golgi localization profiles are treated as phase-independent, while Golgi pH, nucleotide-sugar transport capacity, manganese availability, and antibody synthesis rate vary with the cell-specific metabolic state.


\subsection{Single-Cell Metabolic Shift Model}
\label{subsec:metabolic_shift_model}

For each cell \(i\), the discrete metabolic phase \(Z_i(t)\in\{0,1,2,3\}\)—representing early exponential, late exponential, stationary, and decline phases—is modeled as a continuous-time stochastic jump process. Extending Wang et al.~\citep{wang2024multi}, phase progression is predicted by the cumulative variation in the oxygen uptake rate (OUR):
\(
Q_{q_{\mathrm{O}_2}}(t_n)
=
\sum_{k=1}^{n}
\left|
q_{\mathrm{O}_2}(t_k)-q_{\mathrm{O}_2}(t_{k-1})
\right|,
\)
with
$
Q_{q_{\mathrm{O}_2}}(t)=Q_{q_{\mathrm{O}_2}}(t_n),$ for $
t_n\le t<t_{n+1}$
capturing total metabolic adjustment over time. Here, \(n\) denotes the discrete time-step index with \(t_n=n\Delta t\), and $\Delta t$ is the simulation timestep.
This formulation reflects the progressive, history-dependent metabolic remodeling in fed-batch CHO cultures, where cells adapt to evolving nutrient, oxygen, and stress conditions \citep{ahn2012towards,coulet2022metabolic}. As an online-accessible indicator of cellular activity, OUR has been widely linked to metabolic state and process dynamics \citep{martinez2019new,pappenreiter2019oxygen}. Using cumulative variation, rather than absolute OUR, emphasizes total respiratory adaptation while reducing sensitivity to baseline differences, yielding a compact history variable.

Let 
\(
\mathcal E=\{(0,1),(1,2),(2,3)\}
\)
denote the set of admissible transitions, enforcing ordered, forward-only progression. For transition \((a,b)\in\mathcal E\), the cell-specific transition intensity is
$
\lambda_{ab,i}(t)
=
\exp\bigl(\eta_{ab,i}(t)\bigr)$
with
\(
\eta_{ab,i}(t)
=
\alpha_{ab}
+
\beta_{ab}t
+
\gamma_{ab}
\left(
Q_{q_{\mathrm{O}_2}}(t)
\right)^2
\), where \(\beta_{ab}\) captures explicit time dependence and \(\gamma_{ab}\) quantifies sensitivity to accumulated oxygen-utilization variation. The log-linear form ensures nonnegative transition intensities.
The instantaneous transition hazard is
\begin{equation}
\label{eq:single_cell_phase_intensity}
\mathbb{P}
\left(
Z_i(t+dt)=b
\,\middle|\,
Z_i(t)=a
\right)
=
\lambda_{ab,i}(t)\,dt
+
o(dt).
\end{equation}
with zero intensity assigned to transitions not in \(\mathcal E\).

Over an interval \((\tau_{h-1},\tau_h)\), the transition probability is given by the survival function of an inhomogeneous Poisson process,
\begin{equation}
\label{eq:single_cell_interval_probability}
P_{ab,i}(\tau_h)
=
1-
\exp
\left[
-
\int_{\tau_{h-1}}^{\tau_h}
\lambda_{ab,i}(s)\,ds
\right].
\end{equation}
When \(\lambda_{ab,i}(t)\) varies slowly over the interval, this simplifies to
\(
P_{ab,i}(\tau_h)
\approx
1-
\exp
\left[
-
\exp
\left(
\eta_{ab,i}(\tau_h)
\right)
\Delta \tau_h
\right],
\)
where \(\Delta \tau_h=\tau_h-\tau_{h-1}\).

\subsection{Population-Level Cell-Culture Dynamics}

The population-level model is formulated as an ensemble diffusion approximation of the underlying stochastic single-cell system.  Individual cells retain intracellular metabolic states and phases, while interacting through a shared, well-mixed extracellular environment. The resulting culture dynamics form a hybrid jump–diffusion system: continuous stochastic evolution between feeding events, coupled with instantaneous jumps in extracellular variables induced by feeding.

Let \(\mathbf{u}_{\mathrm{ex}}(t)\) denote extracellular metabolite concentrations, and let \(X^{(z)}(t)\) represent the viable-cell density in metabolic phase \(z\in\{0,1,2,3\}\) with total viable-cell density
\(
X_{\mathrm{tot}}(t)
=
\sum_{z=0}^{3}X^{(z)}(t)
\).
Between feeding events
\(t\in(\tau_h,\tau_{h+1})\), extracellular dynamics are driven by aggregate cellular exchange:
\begin{equation}
\label{eq:extracellular_population_sde_revised}
\begin{aligned}
d\mathbf{u}_{\mathrm{ex}}(t)
=
\left[
\sum_{z=0}^{3}
X^{(z)}(t)
\bar{\mathbf{r}}_{\mathrm{ex}}^{(z)}(t)
\right]dt+
\left[
\frac{1}{\Omega_h}
\sum_{z=0}^{3}
X^{(z)}(t)
\bar{\Gamma}_{\mathrm{ex}}^{(z)}(t)
\right]^{1/2}
d\mathbf{W}_{\mathrm{ex}}(t),
\;
\end{aligned}
\end{equation}
where \(\Omega_h\) is an effective system-size parameter. The drift term aggregates phase-conditioned mean exchange fluxes,
$
\bar{\mathbf{r}}_{\mathrm{ex}}^{(z)}(t)
=
\mathbb{E}
\left[
N_{\mathrm{ex}}\mathbf{v}^{(z)}
(\cdot)
\,\middle|\,
Z_i(t)=z
\right],
$
while the diffusion term captures the corresponding covariance contributions,
$
\bar{\Gamma}_{\mathrm{ex}}^{(z)}(t)
=
\mathbb{E}
\left[
N_{\mathrm{ex}}
\Sigma^{(z)}
(\cdot)
N_{\mathrm{ex}}^\top
\,\middle|\,
Z_i(t)=z
\right],
$
with
\(N_{\mathrm{ex}}\)
denoting the extracellular stoichiometric matrix associated with the metabolic reaction network.

Phase-resolved viable-cell densities evolve via growth and interphase transitions:
\begin{eqnarray}
\label{eq:population_phase_sde_revised}
{dX^{(z)}(t)
=
\left[
\bar{\mu}^{(z)}(t)X^{(z)}(t)
+
\sum_{a\neq z}
\bar{\lambda}_{az}(t)X^{(a)}(t) \right. }
\left. \sum_{b\neq z}
\bar{\lambda}_{zb}(t)X^{(z)}(t)
\right]dt
+
\sigma_X^{(z)}
X^{(z)}(t)\,
dW_X^{(z)}(t),
\end{eqnarray}
where $
\bar{\mu}^{(z)}(t)
=
\mathbb{E}
\left[
\mu^{(z)}
(\cdot)
\mid
Z_i=z
\right]$ is the mean growth rate in phase $z$, 
$\bar{\lambda}_{ab}(t)
=
\mathbb{E}
\left[
\lambda_{ab,i}(t)
\,\middle|\,
Z_i(t)=a
\right]$ is the effective metabolic phase transition intensity, and \(\sigma_X^{(z)}\) captures stochastic growth variability within phase \(z\).

The population glycoform distribution is defined as a productivity-weighted average of cell-specific outputs:
\begin{equation}
\label{eq:population_glycoform_revised}
\bar{\mathbf{g}}(t)
=
\frac{
\sum_{i=1}^{N_c(t)}
q_{\mathrm{Ab},i}(t)\mathbf{g}_i(t)
}{
\sum_{i=1}^{N_c(t)}
q_{\mathrm{Ab},i}(t)
},
\end{equation}
where \(N_c(t)\) is the number of viable cells and \(\mathbf{g}_i(t)\) denotes the glycoform vector produced by cell \(i\) (Section~\ref{sec:glycosylation}). This weighting reflects the disproportionate contribution of highly productive cells to the experimentally observed glycoform profile.

Together, Eqs.~\eqref{eq:extracellular_population_sde_revised}–\eqref{eq:population_glycoform_revised} define a population-level jump–diffusion model that propagates single-cell metabolism, growth, and stochasticity to culture-scale dynamics. 
Drift terms represent aggregated effects of cellular metabolism, proliferation, and phase transitions, while diffusion terms capture the impact of single-cell variability on extracellular exchange and phase-resolved population dynamics.


\textbf{Numerical integration and uncertainty quantification.}
The coupled system is simulated over successive feeding intervals
\([\tau_h,\tau_{h+1})\). Extracellular states are updated instantaneously at feeding times, while intracellular stochastic dynamics are integrated between events using the Euler–Maruyama method. Phase-resolved cell populations evolve continuously through stochastic growth and probabilistic phase transitions, generating time-varying population heterogeneity throughout the culture.

{Using the Euler--Maruyama scheme, the single-cell metabolic state evolves as
\[
\mathbf{u}_{i,n+1}
={}
\mathbf{u}_{i,n}
+
N_{\mathrm{in}}\mathbf{v}_{i,n}\Delta t
+
N_{\mathrm{in}}
\left[
\operatorname{diag}(\mathbf{v}_{i,n})
\right]^{1/2}
\sqrt{\Delta t}\,
\boldsymbol{\xi}_{i,n},
\]
where 
\(
\boldsymbol{\xi}_{i,n}\sim\mathcal{N}(\mathbf{0},I)
\)
is independently sampled across cells and time steps. The resulting diffusion covariance is
$N_{\mathrm{in}}\operatorname{diag}(\mathbf{v}_{i,n})
N_{\mathrm{in}}^{\top}$. Phase transitions are sampled independently according to Eq.~(\ref{eq:single_cell_interval_probability}).
Uncertainty is quantified through independent ensemble realizations with identical model parameters, initial conditions, feeding strategy, and process inputs, differing only in Wiener increments and phase-transition outcomes. At each time point, predictions are summarized by the median (50th percentile), with uncertainty characterized by the 95\% prediction interval (2.5th--97.5th percentiles).
}

\subsection{{Parameter Estimation and Model Calibration}}
\label{sec:parameter_estimation}

{
Model calibration is performed hierarchically. The cell-culture sub-model is
first calibrated to the baseline condition (Case~A) to obtain a reference parameter vector, $\boldsymbol{\theta}^{0}$. Sensitivity and collinearity analyses identify 49 influential parameters, which are re-estimated using bounded, regularized nonlinear least squares with the Trust Region Reflective (TRF) algorithm, while all remaining parameters are fixed. The resulting parameterization is then applied to Cases B and C without further fitting.
The glycosylation sub-model is calibrated sequentially using the predicted cell-culture trajectories as inputs, with structural and transport parameters fixed from the literature and selected enzyme-activity, substrate-interaction, and Golgi-environment parameters estimated from glycoform measurements. Parameter uncertainty and practical
identifiability are assessed from the residual Jacobian using standard
errors, relative standard errors, and linearized confidence intervals.
Additional details are provided in Appendix~\ref{app:model_parameters}.}

{All simulations, parameter estimation, sensitivity analyses, and
uncertainty calculations are implemented in Python. The stochastic cell-culture and glycosylation models are solved using the Euler--Maruyama and LSODA methods, respectively. Parameter calibration is performed using
\texttt{lmfit} and the Trust Region Reflective algorithm in
\texttt{scipy.optimize.least\_squares}. Simulations were run on a standard desktop computer (Apple M4, 16~GB RAM), with a complete model run requiring approximately 2 minutes.}

\section{Results}

The datasets were selected to provide a controlled yet biologically informative testbed for evaluating the proposed metabolism–glycosylation modeling framework. The experiments included a targeted ammonia-stress perturbation and its parallel control cultures using a constant fed-rate, and pyramid feed fed-batch cultures with more refined control algorithms; enabling assessment of model predictions across multiple culture environments. The available offline measurements and online bioreactor signals further support model calibration and validation.

\begin{figure}[t]
    \centering
    \includegraphics[width=0.8\textwidth]{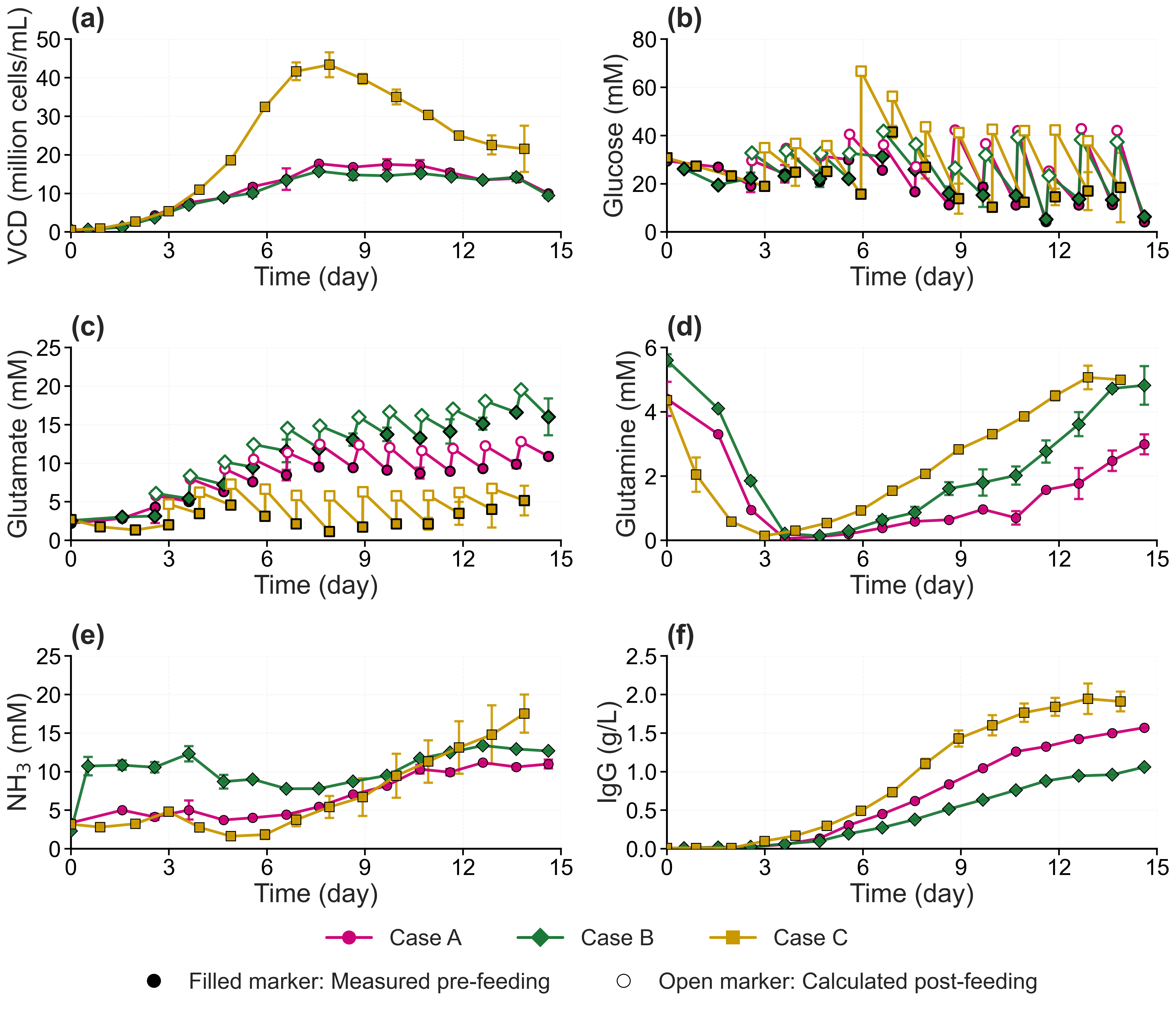}
    \caption{
Experimental data culture performance profiles.
(a)-VCD, (b)-glucose, (c)-glutamate, (d)-glutamine, (e)-ammonia, and (f)-IgG. Case A (magenta, circles) is the baseline fed-batch condition ($N = 2$). Case B (dark green, diamonds) is an ammonia-stressed fed-batch condition ($N = 2$). Case C (mustard, squares) is a pyramid fed-batch condition ($N = 3$). Filled markers represent measured values, and open markers represent calculated concentrations based on the feed mass balance.  Error bars represent the standard deviations. 
}
    \label{fig:raw_culture_profiles}
\end{figure}

\subsection{Cell Culture State Dynamics under Perturbations}
\label{subsec:culture_state_trajectories}

We first examine fed-batch culture profiles to identify key multivariate features motivating the multiscale metabolism–glycosylation framework. The three cases represent distinct processes: Case A (baseline), Case B (ammonia stress), and Case C (pyramid feeding).

Viable-cell-density and ammonia profiles reveal distinct culture trajectories (Fig.~\ref{fig:raw_culture_profiles}(a), (e)).
Case C reaches the highest peak VCD ($>$
\(40\times10^6\) cells/mL), while Cases A and B remain in a similar lower range ($\sim$\(15\)--\(18\times10^6\) cells/mL). The similarity of Cases A and B suggests ammonia stress does not primarily alter overall growth but instead perturbs metabolism and productivity. In contrast, Case C reflects enhanced growth followed by late-stage metabolic burden. The
ammonia profiles further distinguish externally imposed ammonia stress in Case B
from metabolically generated ammonia accumulation in Case C, a distinction that
is important because ammonia can influence CHO metabolism, recombinant protein
production, and glycosylation-related product quality
\citep{yang2000effects,gawlitzek2000ammonium,synoground2021transient,chitwood2023microevolutionary}.

Nutrient and amino-acid profiles support condition-dependent metabolic regulation (Fig.~\ref{fig:raw_culture_profiles}(b)--(d)). Glucose and
glutamate profiles show feeding-associated depletion and replenishment patterns. However, the
comparison between Cases A and B shows that nutrient availability alone cannot
explain the culture response: despite broadly similar feeding behavior, ammonia
stress in Case B leads to distinct glutamate, ammonia, and IgG profiles. In
particular, Case~B accumulates higher late-stage glutamate, whereas Case C
maintains lower glutamate despite achieving the highest VCD,
suggesting perturbation-specific changes in nitrogen and amino-acid metabolism
\citep{gagnon2011high,mulukutla2012metabolic,harcum2022pid,synoground2021transient}.

IgG profiles further show that productivity depends on both cell growth and metabolic state (Fig.~\ref{fig:raw_culture_profiles}(f)). Case C achieves the
highest final IgG concentration, consistent with its stronger culture expansion,
whereas Case B produces the lowest IgG level despite having a viable-cell-density
range comparable to Case A. This contrast indicates that product formation cannot
be represented as a function of VCD alone; it requires coupling
to metabolic state, phase-dependent productivity, and perturbation-driven
intracellular regulation \citep{kyriakopoulos2018kinetic,sha2018mechanistic,pinto2023hybrid}.

\begin{figure}[t]
    \centering
    \includegraphics[width=0.95\textwidth]{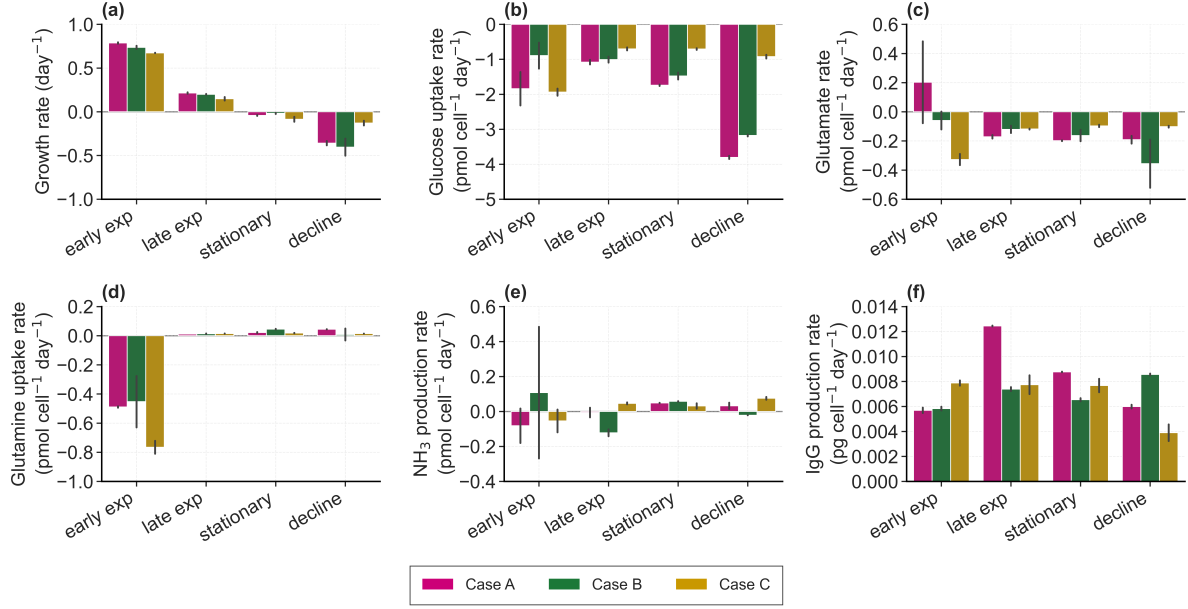}
    \caption{Phase-specific net rates for the three fed-batch culture conditions: {(a) Growth rate, (b) glucose rate, (c) glutamate rate, (d) glutamine rate, (e) ammonia rate, and (f) IgG productivity.} Positive values indicate production, whereas negative values indicate consumption. Case A (magenta) represents the baseline fed-batch condition ($N = 2$), Case B (dark green) represents the ammonia-stressed fed-batch condition ($N = 2$), and Case C (mustard) represents the pyramid fed-batch condition ($N = 3$). 
    Bars represent mean values, and error bars indicate the standard deviation (SD).
    }
    \label{fig:phase_rate_summary}
\end{figure}

Based on the raw VCD profiles in Fig.~\ref{fig:raw_culture_profiles}(a), the cultures do not share a single growth-transition time across all cases. We therefore examine the VCD trajectories on a logarithmic scale and fit piecewise linear segments over candidate exponential-growth windows (Fig.~\ref{fig:growth_qO2}(a)). Cases A and B show similar early-growth behavior, with apparent growth rates decreasing after approximately Day~4, whereas Case C maintains near-exponential growth until approximately Day~6 before slowing. Accordingly, growth-phase windows are assigned in a case-specific manner rather than using a single fixed time partition across all process conditions. This segmentation is consistent with condition-dependent metabolic-state progression in CHO fed-batch cultures \citep{mulukutla2012metabolic,wang2024multi}.

For Cases A and B, early and late exponential-growth regimes are defined as \(0\)--\(4\) and \(5\)--\(8\) days, followed by stationary-like and decline-like windows of \(9\)--\(13\) and \(>13\) days. For Case C, the corresponding windows are \(0\)--\(6\), \(7\)--\(8\), \(9\)--\(10\), and \(>10\) days. Cell-specific production or consumption rates are estimated by finite differences and normalized by the average VCD over adjacent sampling intervals:
\(
\hat q_{C,i}=\frac{(C_i-C_{i-1})/\Delta t_i}
{\left(X_i+X_{i-1}\right)/2},
\)
where \(C_i\) and \(X_i\) denote the measured concentration and VCD at time \(t_i\), respectively, and \(\Delta t_i=t_i-t_{i-1}\). Feeding points are excluded when applicable.

The phase-resolved rate summaries in Fig.~\ref{fig:phase_rate_summary} show clear regime-dependent behavior. Growth rates are highest in the early exponential window and decrease in later windows, with negative rates appearing in the decline-like regime. When case-specific time windows are used, the apparent growth rates in Cases A/B and Case C become comparable within corresponding regimes, suggesting that the main difference among conditions lies in the timing of metabolic-state progression rather than fundamentally different growth kinetics within each regime. Nutrient-consumption rates are strongest during early exponential growth and weaken or change direction later, reflecting feeding-induced replenishment, reduced growth demand, and metabolic adaptation. IgG, ammonia, and glutamate rates also show strong case dependence, indicating that productivity and byproduct formation are governed by both culture regime and perturbation condition rather than VCD alone \citep{robitaille2015single,wang2024multi}.

Extracellular pH trajectories further reflect the distinct process conditions (Fig.~\ref{fig:ph_trajectories}). Case C, operated under active pH control, maintains a comparatively stable pH near neutrality, whereas Cases A and B exhibit larger temporal variation with less tight pH control. Together, the VCD, rate, ammonia, nutrient, IgG, and pH profiles show three routes of culture-state evolution: the baseline trajectory in Case A, externally imposed ammonia stress in Case B, and feeding-enhanced growth with late-stage ammonia accumulation in Case C. These observations motivate the proposed stochastic multiscale metabolism--glycosylation framework, in which feeding acts on the shared extracellular environment, ammonia and pH perturbations modulate metabolic state, and heterogeneous phase-dependent single-cell dynamics govern growth, metabolite exchange, antibody production, and glycosylation outcomes.

\subsection{Stochastic Cell-Culture Model Captures Perturbation-Dependent Dynamics}

\begin{figure}[ht]
    \centering
    \includegraphics[width=0.95\textwidth]{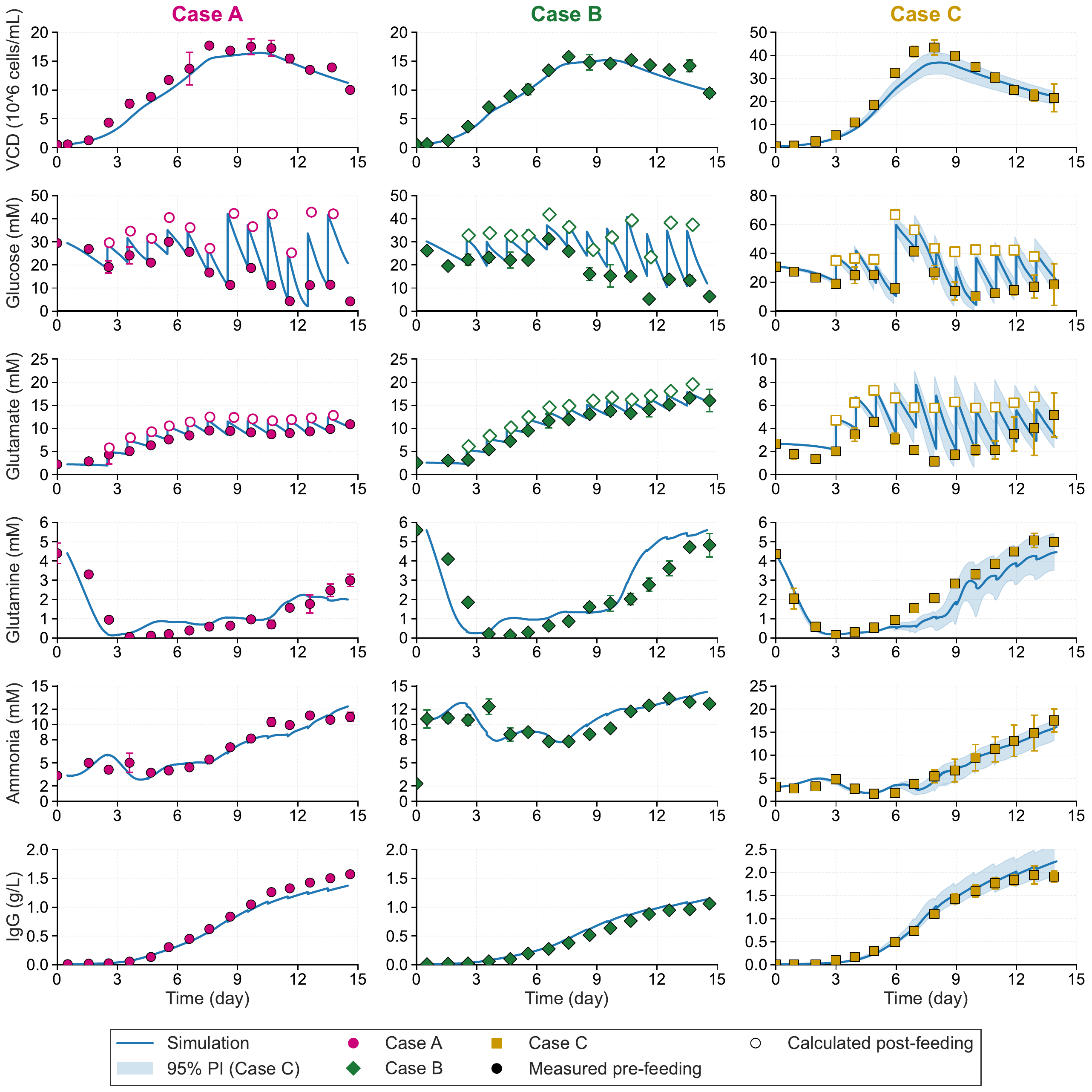}



    \caption{
Model predictions for fed-batch cell culture performance. Show by column: Case A (magenta, circles) is the baseline fed-batch condition ($N = 2$). Case B (dark green, diamonds) is an ammonia-stressed fed-batch condition ($N = 2$). Case C (mustard, squares) is a pyramid fed-batch condition ($N = 3$). Row 1 -  VCD, Row 2 - glucose, Row 3 - glutamate, Row 4 - glutamine, Row 5 - ammonia, and Row 6 - IgG. Filled markers represent measured values, and open markers represent calculated concentrations based on the feed mass balance. {Solid blue lines represent model predictions, and the shaded region
for Case~C represents the 95\% prediction interval.} Error bars represent the standard deviations.  
    }
    \label{fig:culture_model_fit_casesABC}
\end{figure}

After characterizing the perturbation-dependent culture-state trajectories in
Section~\ref{subsec:culture_state_trajectories}, we next evaluate whether the
proposed stochastic cell-culture model can reproduce these trajectories under
baseline, ammonia-stress, and altered-feeding conditions. The model is simulated
using case-specific initial conditions and feeding schedules while maintaining a
shared mechanistic structure for growth, metabolic exchange, ammonia generation,
and antibody production. Model parameters are estimated using only the baseline
condition (Case A, $N=2$), whereas Cases B and C are treated as perturbation tests
without re-estimating kinetic parameters, except for the experimentally specified
initial ammonia stress and feeding schedules.

As shown in Fig.~\ref{fig:culture_model_fit_casesABC}, the model captures the
dominant culture-state dynamics across all three cases while using the same
mechanistic structure. For the baseline condition (Case A), the model reproduces
the main temporal patterns in VCD, including early expansion,
transition to a plateau-like regime, and subsequent decline. The predicted
extracellular metabolite profiles also follow the measured trends, including
feeding-associated glucose fluctuations, early glutamine depletion followed by
later replenishment, progressive glutamic acid accumulation, late-stage ammonia
increase, and mAb titer accumulation.

More importantly, for the ammonia-stress condition (Case~B), the model captures the elevated
initial ammonia level and its subsequent evolution. Importantly, the simulated
VCD trajectory remains broadly comparable to that of Case~A,
consistent with the experimental observation. At the same time, the model
reproduces the distinct ammonia, glutamate, and mAb profiles under ammonia
stress, indicating that the perturbation affects metabolic state and effective
productivity without necessarily causing a large separation in VCD.

For the pyramid feeding condition (Case C), the model captures the substantially
stronger cell expansion, prolonged high-growth behavior, and increased antibody
accumulation due to the higher feeding rate. The metabolite predictions reproduce the dominant
feeding-associated patterns, including repeated glucose replenishment, early
glutamine depletion followed by stronger late-stage recovery, and progressive
ammonia accumulation during the high-density phase. Although the peak VCD is slightly underestimated, the model captures the major dynamic
differences induced by the pyramid feeding strategy.

Across all cases, the model reproduces both smooth biological dynamics and
discontinuous changes induced by feeding events. Sawtooth-like patterns in
glucose and glutamate reflect discrete feeding inputs, whereas smoother
viable-cell-density, ammonia, and mAb trajectories reflect the integrated effects
of growth, metabolic exchange, and product formation. These results support the
model structure in which feeding acts on the shared extracellular environment,
while phase-dependent metabolic kinetics and stochastic culture-state evolution
determine the resulting culture-scale observables. The qualitative agreement observed in Fig.~\ref{fig:culture_model_fit_casesABC}
is further supported by the quantitative error metrics summarized in
Table~\ref{tab:nrmse_summary}.

The model fit is also consistent with the metabolic-shift design introduced in
Section~\ref{subsec:metabolic_shift_model}. The piecewise VCD analysis in
Fig.~\ref{fig:growth_qO2}(a) and the experimentally derived cell-specific oxygen
uptake rate \(q_{\mathrm{O}_2}(t)\) in Fig.~\ref{fig:growth_qO2}(b) show that
the three cases differ not only in culture-state magnitude but also in the timing
of metabolic-state progression. By incorporating the cumulative
oxygen-utilization variation \(Q_{q_{\mathrm{O}_2}}(t)\), the phase-transition
model links feeding strategy, ammonia perturbation, and oxygen-utilization
dynamics to condition-dependent metabolic progression. This interpretation is
supported by the apparent rate analysis in Fig.~\ref{fig:phase_rate_summary},
where case-specific growth rates become comparable within corresponding regimes,
suggesting that differences among conditions are driven largely by transition
timing rather than fundamentally different growth kinetics within each regime.

\subsection{Coupled Metabolism--Golgi Model for Predicting Glycosylation Profiles}

After validating the stochastic cell-culture module, we evaluated whether predicted culture states could be propagated through the mechanistic glycosylation model to reproduce product-quality outcomes. The model uses ammonia-dependent Golgi pH, nucleotide-sugar availability, manganese availability, and antibody synthesis rates predicted by the cell-culture module as inputs. Predicted glycoform distributions were compared with Day~14 measurements for Cases~A--C.

{The Day~14 comparison assesses the cumulative glycoform distribution of the extracellular antibody pool at harvest, thereby evaluating whether the integrated histories of culture state, metabolism, productivity, and Golgi processing reproduce the observed final product-quality outcome. The glycosylation submodel was calibrated using only the harvest-time glycan measurements from Case~A and then applied to Cases~B and~C without re-estimation of any glycosylation parameters. Thus, the predictions for Cases~B and~C provide an independent cross-condition assessment of the model's ability to capture glycosylation changes arising from distinct metabolic environments.
This harvest-level evaluation is directly relevant to biomanufacturing, where Fc glycosylation is a critical quality attribute and glycan profiles are routinely used for product characterization and quality control \citep{reusch2015fc,luo2026glycan}. Consistent with previous CHO process studies, model performance is assessed using the final glycoform distribution in the culture supernatant \citep{kotidis2019model}.}

\begin{figure}[t]
    \centering
    \includegraphics[width=0.95\textwidth]{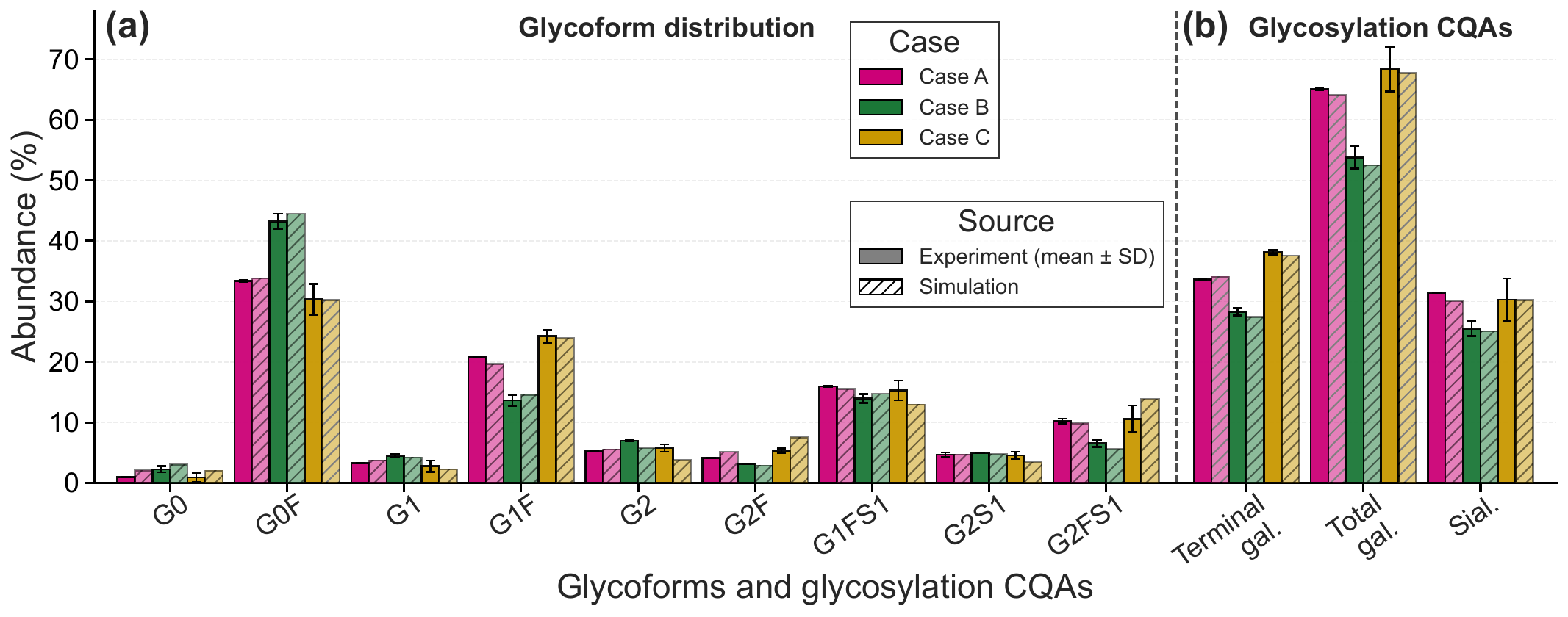}
    \caption{
    Comparison of experimental and simulated glycoform distributions and
    glycosylation CQAs across the three fed-batch culture conditions.
    (a) Glycoform distribution. (b) Derived glycosylation CQAs, including
    terminal galactosylation, total galactosylation, and sialylation.
    Solid bars represent experimental measurements, and hatched bars represent
    simulation results. Case A (magenta) represents the baseline fed-batch
    condition, Case B (dark green) represents the ammonia-stressed fed-batch
    condition, and Case C (mustard) represents the pyramid fed-batch condition.
    Experimental data are presented as mean $\pm$ SD
    ($N=2$ for Case A, $N=2$ for Case B, and $N=3$ for Case C).
    }
    \label{fig:glycosylation_bar_summary}
\end{figure}

As shown in Fig.~\ref{fig:glycosylation_bar_summary}, the coupled
metabolism--glycosylation framework reproduces the dominant glycosylation
differences across baseline, ammonia-stress, and pyramid feeding conditions.
In addition to fitting individual glycoform categories independently, the model
captures coordinated shifts among agalactosylation, terminal galactosylation,
total galactosylation, and sialylation. This is expected because
these CQAs are mechanistically coupled through sequential Golgi processing:
increased agalactosylation reflects reduced galactose extension, while terminal
galactosylation and sialylation depend on nucleotide-sugar donor availability,
Golgi pH, and residence-time-dependent processing capacity
\citep{hossler2009optimal,kochanowski2008influence,gramer2011modulation,aghamohseni2014effects}.

Under the baseline condition (Case A), the model reproduces the overall balance
among agalactosylated, galactosylated, and sialylated species. Predicted
terminal and total galactosylation levels agree well with experimental
measurements, indicating that the baseline metabolic trajectory provides a
consistent Golgi input environment. Minor deviations in sialylation suggest
sensitivity to downstream donor transport and terminal processing, but the
dominant CQA pattern is preserved.

Under ammonia-stress conditions (Case B), the experiment shows the highest
agalactosylated fraction and the lowest total galactosylation among the three
conditions. The model reproduces this shift, including elevated G0F and reduced
terminal and total galactosylation. These results support the proposed
ammonia--Golgi pH coupling mechanism, in which elevated ammonia perturbs the
Golgi microenvironment and reduces the efficiency of terminal galactose
extension. This interpretation is consistent with prior studies showing that
ammonia accumulation and pH variation can alter mammalian-cell glycosylation,
including reductions in galactosylation and sialylation
\citep{yang2000effects,gawlitzek2000ammonium,muthing2003effects,crowell2007amino,aghamohseni2014effects,synoground2021transient}.
Mechanistically, increased Golgi luminal pH can disrupt glycosyltransferase
localization and impair terminal glycan processing
\citep{axelsson2001neutralization,rivinoja2006elevated,rivinoja2009elevated,kopp2024golgi}.
The reduced sialylation in Case B is also consistent with incomplete
galactosylation, since terminal galactose residues are required acceptor sites
for sialylation. Thus, the observed CQA shift reflects altered Golgi processing
conditions rather than viable-cell-density differences alone.

Case C exhibits a distinct glycosylation response. Relative to the
ammonia-stress condition, Case C maintains higher total galactosylation and
achieves the highest terminal galactosylation among the three cases. The model
captures both the increase in terminal galactosylation and the overall upward
shift in galactosylation levels. This behavior is consistent with the increased feeding strategy, which changes nutrient availability, metabolic activity, and nucleotide-sugar precursor supply. Previous studies have shown that feeding
strategies involving galactose, uridine, and manganese can enhance galactosylation by increasing precursor availability and supporting
glycosyltransferase activity \citep{gramer2011modulation,kotidis2019model}.
At the same time, late-stage ammonia accumulation in Case C may limit the extent
of favorable terminal processing
\citep{yang2000effects,gawlitzek2000ammonium,muthing2003effects,crowell2007amino,aghamohseni2014effects,synoground2021transient,sumit2019dissecting}.
Therefore, Case~C illustrates competing glycosylation mechanisms: improved
nutrient availability and productivity can support glycan processing, whereas
high-density metabolic burden and ammonia accumulation can counteract favorable
terminal processing.

Overall, the comparison across Cases A--C demonstrates that the proposed
framework captures perturbation-dependent glycosylation behavior through
upstream metabolic-state dynamics rather than case-specific tuning of Golgi
kinetic parameters. A single set of glycosylation parameters is maintained across
conditions, and differences in CQA profiles emerge from variations in ammonia
exposure, nutrient availability, metabolic phase progression, oxygen
utilization, and antibody production predicted by the cell-culture model. These
results support the central premise of the multiscale framework: process-level
perturbations first reshape the extracellular and intracellular metabolic
environment, and these changes are then propagated through mechanistic Golgi
processing to determine glycosylation CQAs
\citep{pacis2011effects,kochanowski2008influence,aghamohseni2014effects,kotidis2019model}.

\subsection{Independent Trajectory Comparison and Uncertainty Propagation Analysis}





\begin{figure}[t]
    \centering
    \includegraphics[width=0.86\linewidth]{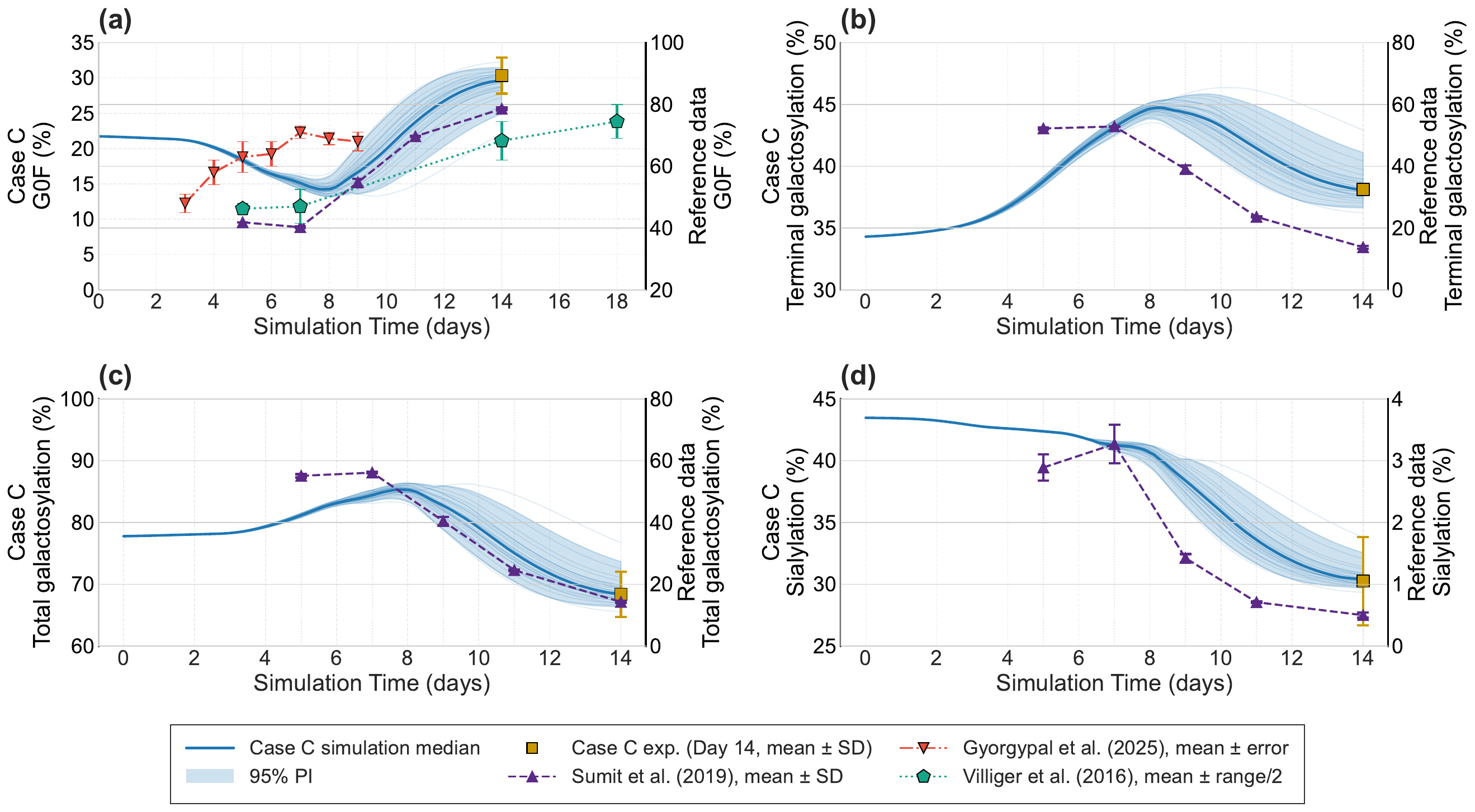}
    \caption{{
 Predicted Case~C glycosylation trajectories with uncertainty propagation from stochastic metabolic variability, compared with independent literature data \citep{sumit2019dissecting,gyorgypal2025temporal,villiger2016controlling}. Panels show (a) G0F,
    (b) terminal galactosylation, (c) total galactosylation, and
    (d) sialylation. Solid blue lines denote median model predictions and shaded regions indicate 95\% prediction intervals. Case~C harvest measurements (Day~14, $N=3$, mean
    $\pm$ SD) are shown as mustard squares. Independent measurements are shown as purple upward triangles (Sumit et al., 2019), red downward triangles (Gyorgypal et al., 2025), and teal pentagons (Villiger et al., 2016). Left and right y-axes correspond to the Case~C and literature datasets, respectively.
    }}
    \label{fig:uncertainty_glycosylation}
\end{figure}

To further assess model prediction robustness, we considered two complementary analyses. First, 
{predicted glycosylation trajectories are compared qualitatively with independent time-course measurements reported in the literature~\cite{villiger2016controllingI,sumit2019dissecting,gyorgypal2025temporal}.}
Second, stochastic variability in cellular metabolism is propagated through the integrated metabolism--glycosylation framework to quantify uncertainty in culture-state and glycosylation predictions.

{
The primary validation endpoint is the Day~14 harvest glycosylation profile (Fig.~\ref{fig:uncertainty_glycosylation}), which represents the cumulative extracellular antibody pool and the principal product-quality outcome in upstream biomanufacturing \citep{reusch2015fc,luo2026glycan,kotidis2019model}. Because harvest measurements do not directly assess intermediate glycosylation dynamics, published time-course studies provide complementary evidence for assessing trajectory plausibility.}


{
Figure~\ref{fig:uncertainty_culture} compares the Case~C culture trajectories with three literature datasets, among which the HiPDOG condition of Sumit et al.~(2019) \cite{sumit2019dissecting} shows the closest culture-state behavior to Case~C. Figure~\ref{fig:uncertainty_glycosylation} then compares the predicted Case~C glycosylation trajectories with literature data. The HiPDOG condition serves as the primary reference because of its similarity to Case~C and the availability of time-course measurements of G0F, terminal galactosylation, total galactosylation, and sialylation. Additional G0F trajectories from Villiger et al.~(2016) and Gyorgypal et al.~(2025) provide further temporal benchmarks \cite{villiger2016controllingI,gyorgypal2025temporal}. 
Given differences in cell lines, products, media, and operating conditions, as well as incomplete experimental information, the comparison emphasizes temporal trends rather than absolute glycoform abundances.
}

{As shown in Fig.~\ref{fig:uncertainty_glycosylation}, the predicted trajectories are broadly consistent with published observations. All reference studies report increasing G0F during culture, particularly at later stages. The model also reproduces the rise-and-fall patterns of terminal and total galactosylation, as well as the progressive decline in sialylation reported by Sumit et al.~(2019). 
These comparisons provide a qualitative assessment of trajectory plausibility rather than quantitative external validation, as the literature studies do not report key inputs, such as feeding strategies and media formulations, or the complete set of process measurements needed to infer cell line-specific mechanistic parameters.
}

{
Together, these comparisons indicate that the multiscale framework captures temporal glycosylation dynamics consistent with observations from independent fed-batch CHO studies, complementing the cross-condition extrapolation validation based on Day~14 harvest glycosylation profiles.
}

To quantify stochastic variability, we performed ensemble simulations of the coupled framework. Stochasticity arises
from metabolic flux fluctuations and probabilistic phase-transition dynamics,
and propagates from intracellular metabolism to culture-scale observables and
downstream glycosylation CQAs. Figure~\ref{fig:uncertainty_glycosylation} shows median trajectories and 95\% prediction intervals 
for culture-state variables and major glycosylation CQAs under Case~C. Ensemble-based approaches are widely used to characterize variability in stochastic biochemical systems and quantify uncertainty propagation \citep{schnoerr2017approximation,warne2019simulation,kurdyaeva2021uncertainty}.

As shown in Fig.~\ref{fig:culture_model_fit_casesABC} Case C, uncertainty varies across
culture variables. Viable-cell density exhibits relatively narrow prediction
intervals, indicating that population growth is comparatively robust to moderate
intracellular stochasticity. In contrast, extracellular metabolites show broader
uncertainty, especially for glucose and glutamate, reflecting feeding-induced
discontinuities and stochastic variation in metabolic exchange rates. Ammonia
uncertainty increases during late culture, consistent with cumulative variability
in nitrogen metabolism and byproduct accumulation. Most experimental
measurements fall within the predicted intervals, suggesting that the model
captures both mean trends and realistic process variability.

Uncertainty propagation to glycosylation CQAs
(Fig.~\ref{fig:uncertainty_glycosylation}) is strongly time dependent. Prediction
intervals remain relatively narrow early in culture but expand during the
mid-to-late stage, especially around days 8--12, when G0F increases sharply and
terminal galactosylation, total galactosylation, and sialylation decrease most
rapidly. This pattern indicates that variability is amplified during periods of
rapid glycosylation-state evolution rather than accumulating uniformly over time.

{Together, the trajectory comparison and uncertainty analysis indicate that the framework produces biologically plausible glycosylation dynamics while quantifying the impact of stochastic metabolic variability on predicted CQAs.}



\section{Discussion}

This study presents a multiscale framework for CHO culture dynamics that integrates nonlinear, state-dependent metabolism and glycosylation through coupled single-cell states, phase transitions, and Golgi processing. The modular structure enables mechanistic modeling across process conditions by linking bottom-up intracellular dynamics with top-down process controls, such as feeding and pH regulation. This integration provides a mechanistic representation linking metabolism,
population heterogeneity, glycan processing, and macroscopic culture behavior.

\textbf{Multiscale metabolism--glycosylation coupling.}
This framework models metabolism–glycosylation coupling at the single-cell level, where each cell’s dynamic metabolic state and phase jointly determine flux activity, ammonia production, productivity, and glycosylation precursor availability. Population-level outputs—cell density, extracellular metabolites, titer, and glycoform distributions—emerge from aggregating heterogeneous single-cell trajectories. This bottom-up formulation provides a mechanistic pathway for linking intracellular heterogeneity to population-level CQAs. Prior studies have reported substantial intraclonal variability and dynamically evolving heterogeneity in CHO cell cultures \citep{pilbrough2009intraclonal,moller2020quantification}. While population-averaged models can reproduce bulk culture behavior, they inherently mask cell-to-cell differences in metabolic activity and productivity that critically shape the final glycoform distribution. In this context, explicitly resolving stochastic single-cell dynamics offers a principled approach for propagating variability across scales.

Glycosylation results should therefore be interpreted as coordinated CQA shifts rather than independent glycoform fits. Because G0F, galactosylation, and sialylation arise from sequential Golgi processing, they are jointly governed by nucleotide-sugar availability, Golgi conditions, enzyme activity, and residence time. The ability of the framework to reproduce glycosylation responses across conditions supports an integrated metabolism–glycosylation mechanism and underscores the importance of propagating stochastic single-cell dynamics to population-level glycoform distributions.

\textbf{Oxygen-utilization-driven phase progression.}
A key insight is that culture time alone does not capture metabolic-state evolution in fed-batch processes. Phase-resolved analysis shows that conditions differ mainly in the timing of metabolic state transitions—e.g., Case C sustains a high-growth state longer than Cases A and B—rather than in within-phase kinetics.
An oxygen-uptake-rate–driven transition provides a compact, measurable proxy for this progression. Since \(q_{\mathrm{O}_2}(t)\) can be inferred from oxygen-transfer, off-gas, and viable-cell-density data, it links routine process measurements to latent metabolic states. Within this framework, feeding, ammonia stress, and pH control reshape oxygen-utilization trajectories, thereby modulating phase-transition probabilities and downstream glycosylation outcomes.

\textbf{Distinct ammonia contexts explain condition-dependent glycosylation responses.}
Comparison of Cases A--C reveals that the impact of ammonia on glycosylation depends strongly on the underlying biological context.
In Case B, externally imposed ammonia stress increases G0F abundance and reduces galactosylation relative to the baseline condition, consistent with previous reports that elevated ammonia can impair terminal glycan processing through modulation of Golgi pH and enzymatic activity \citep{yang2000effects,gawlitzek2000ammonium,synoground2021transient,axelsson2001neutralization,rivinoja2006elevated}. 
In contrast, ammonia accumulation in Case~C arises progressively during the late high-density phase as a consequence of intensified nutrient feeding and enhanced cellular activity.
Although the more aggressive feeding strategy supports increased cell growth and antibody production, it also elevates metabolic burden and byproduct accumulation. As a result, glycosylation outcomes reflect a dynamic balance between enhanced precursor availability and ammonia-associated constraints on terminal glycan processing. 
These findings demonstrate that the effects of feeding and ammonia on glycosylation cannot be considered independently; rather, they emerge from coupled influence on metabolic-state evolution and Golgi processing.

\textbf{Implications for uncertainty-aware prediction, monitoring, and control.}
Simulations reveal that uncertainty propagates unevenly across scales: VCD remains relatively well constrained, whereas extracellular metabolites and glycosylation CQAs show greater uncertainty, especially during periods when glycosylation is most sensitive to metabolic-state changes \citep{schnoerr2017approximation,warne2019simulation,kurdyaeva2021uncertainty}. This uncertainty-aware prediction provides information beyond a deterministic trajectory by identifying sensitive periods in which small metabolic perturbations may produce larger CQA shifts. Because the phase-transition mechanism is linked to oxygen-utilization dynamics and online process measurements, the framework provides a foundation for uncertainty-aware monitoring, design of experiments, and future quality-aware model predictive control of CHO fed-batch processes \citep{ma2025quasi,ma2026adaptive}.

\textbf{Limitations and future work.}
Several limitations should be noted. First, model evaluation was limited to a small number of process conditions and glycosylation datasets. Broader validation across feeding strategies, pH-control schemes, ammonia perturbations, cell lines, and products is needed to establish generalizability and further improve parameter identifiability.

{Second, intracellular and Golgi mechanisms are represented in simplified form. In particular, intracellular nucleotide-sugar pools were not measured directly and were instead inferred through literature-based mechanistic relationships. As a result, some estimated glycosylation parameters may compensate for uncertainty in unobserved intracellular states and should be interpreted as effective rather than uniquely identifiable biochemical parameters. This interpretation is consistent with the identifiability analysis, which indicates that several glycosylation parameters are only weakly constrained by the available bulk glycoform measurements.
Future studies incorporating intracellular nucleotide-sugar, metabolite, and multi-omics measurements could provide stronger mechanistic constraints and improve parameter identifiability. Nevertheless, a practical advantage of the framework is that it can be applied using standard process measurements without requiring routine intracellular sampling.
}

{
Third, glycan measurements were available only at harvest (Day~14). Although these data validate the final glycoform distribution, they do not capture intermediate glycosylation dynamics. While the predicted trajectories are qualitatively consistent with independent time-course observations reported in the literature, longitudinal glycan measurements from the same platform would enable a more rigorous evaluation. Future studies should therefore include glycosylation sampling throughout the exponential, stationary, and decline phases of culture.}

Overall, the results indicate that CHO culture performance and antibody glycosylation are jointly governed by process‑driven metabolic state evolution. By integrating stochastic single‑cell metabolism, population aggregation, oxygen‑uptake‑rate–driven phase transitions, and mechanistic Golgi glycosylation, the framework provides a mechanistic, uncertainty‑aware basis for predicting how process perturbations impact glycosylation CQAs.
Future work could couple this framework with model-based reinforcement learning to enable end-to-end, quality-aware control that simultaneously improves yield and CQA consistency.


\section*{AUTHOR CONTRIBUTIONS}
\textbf{Yuming Zeng:} Conceptualization, Methodology, Investigation, Coding, Software, Visualization, Interpretation, and Writing – original draft. 
\textbf{Sarah W. Harcum:} Acquisition of data, Visualization, Interpretation, Supervision, and Writing – review \& editing.
\textbf{Jinxiang Pei:} Methodology, and Writing – review \& editing.
\textbf{Wei Xie:} Conceptualization, Methodology, Resources, Supervision, and Writing – review \& editing.

\section*{ACKNOWLEDGMENTS}
The authors acknowledge support from the National Institute of Standards and Technology (Grants 70NANB24H293, 70NANB17H002, 70NANB21H086) and the National Science Foundation (CMMI-2442970) to Wei Xie, and from the National Science Foundation (OIA-1736123, EEC-2100442) to Sarah W. Harcum. Drs. Xie and Harcum are co-corresponding authors.

\bibliographystyle{plain}
\bibliography{metabolite}

\newpage
\onecolumn
\appendix

\renewcommand{\thefigure}{S\arabic{figure}}
\setcounter{figure}{0}

\renewcommand{\thetable}{S\arabic{table}}
\setcounter{table}{0}

\section{Metabolic Network and Kinetic Equations}

\begin{table}[htbp]
\centering
\normalsize
\renewcommand{\arraystretch}{1.25}
\caption{ ~Reactions for the metabolic network
}
\label{tab:metabolic_network}
\begin{tabular}{|l|p{14cm}|}
\hline
\rowcolor[HTML]{C0C0C0} 
\textbf{No.} & \multicolumn{1}{c|}{\textbf{Pathway}}                    \\ \hline 
\hline
\textbf{1}   & EGLC $\rightarrow$ G6P           \\ \hline
\textbf{2}   & G6P $\rightarrow$ 2 PYR          \\ \hline
\textbf{3}   & PYR  $\leftrightarrow$ LAC               \\ \hline
\textbf{4}   & LAC $\leftrightarrow$ ELAC               \\ \hline 
\textbf{5}  & GLU + PYR $\leftrightarrow$ AKG + ALA                     \\ \hline
\textbf{6}  & ALA $\rightarrow$ EALA                   \\ \hline
\textbf{7}  & SER $\rightarrow$ PYR + NH$_3$               \\ \hline
\textbf{8}  & ESER $\rightarrow$ SER               \\ \hline
\textbf{9}  & PYR $\rightarrow$ AcCoA + CO$_2$         \\ \hline
\textbf{10}  & AcCoA + OAA $\rightarrow$ AKG +CO$_2$ \\ \hline
\textbf{11}  & AKG $\rightarrow$ SUC +  CO$_2$   \\ \hline
\textbf{12}  & SUC $\rightarrow$ MAL +  CO$_2$   \\ \hline
\textbf{13}  & MAL$ \rightarrow$ OAA               \\ \hline
\textbf{14}  & MAL $\rightarrow$ PYR + CO$_2$          \\ \hline
\textbf{15}  & EGLN $\rightarrow$ GLN               \\ \hline
\textbf{16}  & GLN $\leftrightarrow$ GLU + NH$_3$           \\ \hline
\textbf{17}  & GLU $\leftrightarrow$ AKG + NH$_3$      \\ \hline
\textbf{18}  & EGLU $\rightarrow$  GLU              \\ \hline
\textbf{19}  & ASP + AKG $\leftrightarrow$ GLU + OAA + NH$_3$               \\ \hline
\textbf{20}  & EASP $\rightarrow$ ASP              \\ \hline
\textbf{21}  & LEU + AKG $\rightarrow$ GLU + 3 AcCoA \\ \hline
\textbf{22}  & ELEU $\rightarrow$ LEU              \\ \hline
\textbf{23}  & VAL + AKG $\rightarrow$ GLU + SUC + CO$_2$              \\ \hline
\textbf{24}  & EVAL $\rightarrow$ VAL              \\ \hline
\textbf{25}  & ILE + AKG $\rightarrow$ GLU + SUC + AcCoA \\ \hline
\textbf{26}  & EILE $\rightarrow$ ILE              \\ \hline
{\textbf{27}}  & {ENH$_3$ $\leftrightarrow$ NH$_3$}              \\ \hline
\textbf{28}  & 0.43 EALA + 0.36 EASP + 0.40 EGLN + 0.44 EGLU + 1.08 ESER + 0.70 ELEU + 0.25 EILE + 0.79 EVAL$\rightarrow$ ANTI \\ \hline 
\textbf{29}  & ANTI $\rightarrow$ EANTI \\ \hline 
\textbf{30}  & 0.39 EALA + 0.26 EASP + 0.32 EGLN + 0.32 EGLU + 0.34 ESER + 0.22 LEU + 0.14 ILE + 0.22 EVAL + 0.11 EGLC $\rightarrow$ BIOM \\ \hline
\end{tabular}
\end{table}

\clearpage

\begin{table}[t]
\centering
\caption{\;Independent metabolic flux expressions used in the stochastic multi-scale model. }
\label{tab:flux_kinetics}
\centering
\fontsize{9.5pt}{11.5pt}\selectfont
\renewcommand{\arraystretch}{1.9}
\begin{tabular}{|l|p{14cm}|}
\hline
\rowcolor[HTML]{C0C0C0} 
Flux & Kinetic expression \\
\hline

$v_2^{(z)}$
&
$\displaystyle
\phi_{2,\mathrm{pH}}^{(z)}\!\big(\mathrm{pH}(t)\big)\,
V_2^{(z)}
\frac{\mathrm{EGLC}}
{K_{m,\mathrm{EGLC}}^{(z)}+\mathrm{EGLC}}
\frac{K_{i,\mathrm{LAC}}^{(z,2)}}
{K_{i,\mathrm{LAC}}^{(z,2)}+\mathrm{ELAC}}
$
\\\hline

$v_3^{(z)}$
&
$\displaystyle
V_{3f}^{(z)}
\frac{\mathrm{EGLC}}
{K_{m,\mathrm{EGLC}}^{(z)}+\mathrm{EGLC}}
\frac{\mathrm{NH}_3}
{K_{m,\mathrm{NH}_3}^{(z)}+\mathrm{NH}_3}
-
\phi_{3,\mathrm{pH}}^{(z)}\!\big(\mathrm{pH}(t)\big)\,
V_{3r}^{(z)}
\frac{\mathrm{ELAC}}
{K_{m,\mathrm{ELAC}}^{(z)}+\mathrm{ELAC}}
$
\\\hline

$v_5^{(z)}$
&
$\displaystyle
\phi_{5,\mathrm{pH}}^{(z)}\!\big(\mathrm{pH}(t)\big)\,
V_{5f}^{(z)}
\frac{\mathrm{EGLC}}
{K_{m,\mathrm{EGLC}}^{(z)}+\mathrm{EGLC}}
-
V_{5r}^{(z)}
\frac{\mathrm{EALA}}
{K_{m,\mathrm{EALA}}^{(z)}+\mathrm{EALA}}
$
\\\hline

$v_7^{(z)}$
&
$\displaystyle
V_{7}^{(z)}
\frac{\mathrm{ESER}}
{K_{m,\mathrm{ESER}}^{(z)}+\mathrm{ESER}}
$
\\\hline

$v_{15}^{(z)}$
&
$\displaystyle
V_{15f}^{(z)}
\frac{\mathrm{EGLN}}
{K_{m,\mathrm{EGLN}}^{(z)}+\mathrm{EGLN}}
-
\phi_{15,\mathrm{pH}}^{(z)}\!\big(\mathrm{pH}(t)\big)\,
V_{15r}^{(z)}
\frac{\mathrm{GLN}}
{K_{m,\mathrm{GLN}}^{(z)}+\mathrm{GLN}}
$
\\\hline

$v_{16}^{(z)}$
&
$\displaystyle
V_{16f}^{(z)}
\frac{\mathrm{EGLN}}
{K_{m,\mathrm{EGLN}}^{(z)}+\mathrm{EGLN}}
\frac{K_{i,\mathrm{LAC}}^{(z,16)}}
{K_{i,\mathrm{LAC}}^{(z,16)}+\mathrm{ELAC}}
-
\phi_{16,\mathrm{pH}}^{(z)}\!\big(\mathrm{pH}(t)\big)\,
V_{16r}^{(z)}
\frac{\mathrm{EGLU}}
{K_{m,\mathrm{EGLU}}^{(z)}+\mathrm{EGLU}}
\frac{\mathrm{NH}_3}
{K_{m,\mathrm{NH}_3}^{(z)}+\mathrm{NH}_3}
$
\\\hline

$v_{17}^{(z)}$
&
$\displaystyle
\phi_{17,\mathrm{pH}}^{(z)}\!\big(\mathrm{pH}(t)\big)\,
V_{17f}^{(z)}
\frac{\mathrm{EGLN}}
{K_{m,\mathrm{EGLN}}^{(z)}+\mathrm{EGLN}}
-
V_{17r}^{(z)}
\frac{\mathrm{ENH}_3}
{K_{m,\mathrm{ENH}_3}^{(z)}+\mathrm{ENH}_3}
$
\\\hline

$v_{18}^{(z)}$
&
$\displaystyle
V_{18f}^{(z)}
\frac{\mathrm{EGLU}}
{K_{m,\mathrm{EGLU}}^{(z)}+\mathrm{EGLU}}
\frac{K_{i,\mathrm{NH}_3}^{(z)}}
{K_{i,\mathrm{NH}_3}^{(z)}+\mathrm{ENH}_3}
-
V_{18r}^{(z)}
\frac{\mathrm{EGLC}}
{K_{m,\mathrm{EGLC}}^{(z)}+\mathrm{EGLC}}
$
\\\hline

$v_{19}^{(z)}$
&
$\displaystyle
V_{19f}^{(z)}
\frac{\mathrm{EASP}}
{K_{m,\mathrm{EASP}}^{(z)}+\mathrm{EASP}}
-
V_{19r}^{(z)}
\frac{\mathrm{EGLU}}
{K_{m,\mathrm{EGLU}}^{(z)}+\mathrm{EGLU}}
\frac{\mathrm{NH}_3}
{K_{m,\mathrm{NH}_3}^{(z)}+\mathrm{NH}_3}
$
\\\hline

$v_{21}^{(z)}$
&
$\displaystyle
V_{21}^{(z)}
\frac{\mathrm{ELEU}}
{K_{m,\mathrm{ELEU}}^{(z)}+\mathrm{ELEU}}
$
\\\hline

$v_{23}^{(z)}$
&
$\displaystyle
V_{23}^{(z)}
\frac{\mathrm{EVAL}}
{K_{m,\mathrm{EVAL}}^{(z)}+\mathrm{EVAL}}
$
\\\hline

$v_{25}^{(z)}$
&
$\displaystyle
V_{25}^{(z)}
\frac{\mathrm{EILE}}
{K_{m,\mathrm{EILE}}^{(z)}+\mathrm{EILE}}
$
\\\hline

$v_{27}^{(z)}$
&
$\displaystyle
V_{27}^{(z)}
\big(\mathrm{ENH}_3-\mathrm{NH}_3\big)
$
\\\hline

$v_{28}^{(z)}$
&
$\displaystyle
V_{28}^{(z)}
\frac{\mathrm{EGLN}}
{K_{m,\mathrm{EGLN}}^{(z,28)}+\mathrm{EGLN}}
\frac{\mathrm{EGLU}}
{K_{m,\mathrm{EGLU}}^{(z,28)}+\mathrm{EGLU}}
\frac{\mathrm{EALA}}
{K_{m,\mathrm{EALA}}^{(z,28)}+\mathrm{EALA}}
\frac{\mathrm{EASP}}
{K_{m,\mathrm{EASP}}^{(z,28)}+\mathrm{EASP}}
\frac{\mathrm{ESER}}
{K_{m,\mathrm{ESER}}^{(z,28)}+\mathrm{ESER}}
$\\
&
$\displaystyle
\frac{\mathrm{ELEU}}
{K_{m,\mathrm{ELEU}}^{(z,28)}+\mathrm{ELEU}}
\frac{\mathrm{EILE}}
{K_{m,\mathrm{EILE}}^{(z,28)}+\mathrm{EILE}}
\frac{\mathrm{EVAL}}
{K_{m,\mathrm{EVAL}}^{(z,28)}+\mathrm{EVAL}}
\frac{K_{i,\mathrm{ENH}_3}^{(z,28)}}
{K_{i,\mathrm{ENH}_3}^{(z,28)}+\mathrm{ENH}_3}
$
\\\hline

$v_{30}^{(z)}$
&
$\displaystyle
V_{30}^{(z)}
\frac{\mathrm{EGLC}}
{K_{m,\mathrm{EGLC}}^{(z,30)}+\mathrm{EGLC}}
\frac{\mathrm{EGLN}}
{K_{m,\mathrm{EGLN}}^{(z,30)}+\mathrm{EGLN}}
\frac{\mathrm{EGLU}}
{K_{m,\mathrm{EGLU}}^{(z,30)}+\mathrm{EGLU}}
\frac{\mathrm{EALA}}
{K_{m,\mathrm{EALA}}^{(z,30)}+\mathrm{EALA}}
\frac{\mathrm{EASP}}
{K_{m,\mathrm{EASP}}^{(z,30)}+\mathrm{EASP}}
$\\
&
$\displaystyle
\frac{\mathrm{ESER}}
{K_{m,\mathrm{ESER}}^{(z,30)}+\mathrm{ESER}}
\frac{\mathrm{ELEU}}
{K_{m,\mathrm{ELEU}}^{(z,30)}+\mathrm{ELEU}}
\frac{\mathrm{EILE}}
{K_{m,\mathrm{EILE}}^{(z,30)}+\mathrm{EILE}}
\frac{\mathrm{EVAL}}
{K_{m,\mathrm{EVAL}}^{(z,30)}+\mathrm{EVAL}}
$
\\
\hline
\multicolumn{2}{|l|}{
\textbf{Dependent fluxes determined from quasi-steady-state constraints}
}
\\
\hline

\multicolumn{2}{|c|}{
$\displaystyle
\begin{array}{lllll}
v_1^{(z)}=v_2^{(z)}, &
v_4^{(z)}=v_3^{(z)}, &
v_6^{(z)}=v_5^{(z)}, &
v_8^{(z)}=v_7^{(z)}, &
v_{20}^{(z)}=v_{19}^{(z)},
\\
v_{22}^{(z)}=v_{21}^{(z)}, &
v_{24}^{(z)}=v_{23}^{(z)}, &
v_{26}^{(z)}=v_{25}^{(z)}, &
v_{29}^{(z)}=v_{28}^{(z)}
\end{array}
$
}
\\
\hline

\multicolumn{2}{|c|}{
$\displaystyle
\begin{array}{lll}
v_{14}^{(z)}
=
v_{18}^{(z)}
+
v_{19}^{(z)}
+
v_{23}^{(z)}
+
v_{25}^{(z)},
&
v_9^{(z)}
=
2v_2^{(z)}
+
v_7^{(z)}
+
v_{14}^{(z)}
-
v_3^{(z)}
-
v_5^{(z)},
&
v_{10}^{(z)}
=
v_9^{(z)}
+
3v_{21}^{(z)}
+
v_{25}^{(z)},
\\

v_{11}^{(z)}
=
v_{10}^{(z)}
+
v_{18}^{(z)},
&
v_{12}^{(z)}
=
v_{11}^{(z)}
+
v_{23}^{(z)}
+
v_{25}^{(z)},
&
v_{13}^{(z)}
=
v_{12}^{(z)}
-
v_{14}^{(z)}
\end{array}
$
}
\\
\hline

\end{tabular}
\end{table}

\clearpage

\clearpage
\begin{table}[p]
\centering
\caption{\;Representative Golgi glycosylation reaction-rate expressions adapted from Villiger et al. (2016)~\cite{villiger2016controlling}.}
\label{tab:glyco_kinetics}
\renewcommand{\arraystretch}{1.55}
\setlength{\tabcolsep}{3pt}
\begin{tabular}{|p{3.6cm}|p{11.9cm}|}
\hline
\rowcolor[HTML]{C0C0C0}
Reaction class & Representative rate expression \\
\hline

Michaelis--Menten kinetics
(ManI and ManII)
&
\[
r_{j,\ell,i}(\xi,t)
=
\frac{
k_{f,j,i}(t)\,
E_j(\xi)\,
[\mathrm{OS}_{\ell,i}(\xi,t)]
}{
K_{d,\ell,j}
\left(
1
+
\dfrac{[\mathrm{OS}_{\ell,i}(\xi,t)]}{K_{d,\ell,j}}
+
\sum_{m=1}^{N_{\mathrm{OS}}}
\dfrac{[\mathrm{OS}_{m,i}(\xi,t)]}{K_{d,m,j}}
+
\dfrac{[\mathrm{OS}_{\ell-1,i}(\xi,t)]}{K_{d,\ell-1,j}}
\right)
}.
\]
\\
\hline

Sequential-order bi--bi kinetics
with Mn cofactor
(GnTI, GnTII, and GalT)
&
\[
r_{j,\ell,i}(\xi,t)
=
\frac{
k_{f,j,i}(t)\,
E_j(\xi)\,
[\mathrm{Mn}_{i}(t)]\,
[\mathrm{NS}_{k,i}^{\mathrm{Golgi}}(\xi,t)]\,
[\mathrm{OS}_{\ell,i}(\xi,t)]
}{
K_{d,\mathrm{Mn},j}
K_{d,k,j}
K_{d,\ell,j}
\Theta_{j,\mathrm{seq},\ell,i}(\xi,t)
}.
\]
\[
\begin{aligned}
\Theta_{j,\mathrm{seq},\ell,i}(\xi,t)
=
&\,1
+
\frac{[\mathrm{Mn}_{i}(t)]}{K_{d,\mathrm{Mn},j}}
+
\frac{[\mathrm{Mn}_{i}(t)]}{K_{d,\mathrm{Mn},j}}
\frac{[\mathrm{NS}_{k,i}^{\mathrm{Golgi}}(\xi,t)]}{K_{d,k,j}}
\\
&+
\frac{[\mathrm{Mn}_{i}(t)]}{K_{d,\mathrm{Mn},j}}
\frac{[\mathrm{NS}_{k,i}^{\mathrm{Golgi}}(\xi,t)]}{K_{d,k,j}}
\frac{[\mathrm{OS}_{\ell,i}(\xi,t)]}{K_{d,\ell,j}}
\\
&+
\frac{[\mathrm{Mn}_{i}(t)]}{K_{d,\mathrm{Mn},j}}
\frac{[\mathrm{NS}_{k,i}^{\mathrm{Golgi}}(\xi,t)]}{K_{d,k,j}}
\sum_{m=1}^{N_{\mathrm{OS}}}
\frac{[\mathrm{OS}_{m,i}(\xi,t)]}{K_{d,m,j}}
\\
&+
\frac{[\mathrm{OS}_{\ell+1,i}(\xi,t)]}{K_{d,\ell+1,j}}
\frac{[\mathrm{Nuc}_{k,i}^{\mathrm{Golgi}}(\xi,t)]}{K_{d,N_k,j}}
+
\frac{[\mathrm{Nuc}_{k,i}^{\mathrm{Golgi}}(\xi,t)]}{K_{d,N_k,j}} .
\end{aligned}
\]
\\
\hline

Random-order bi--bi kinetics
(FucT and SiaT)
&
\[
r_{j,\ell,i}(\xi,t)
=
\frac{
k_{f,j,i}(t)\,
E_j(\xi)\,
[\mathrm{NS}_{k,i}^{\mathrm{Golgi}}(\xi,t)]\,
[\mathrm{OS}_{\ell,i}(\xi,t)]
}{
K_{d,k,j}
K_{d,\ell,j}
\Theta_{j,\mathrm{rand},\ell,i}(\xi,t)
}.
\]
\[
\begin{aligned}
\Theta_{j,\mathrm{rand},\ell,i}(\xi,t)
=
&\,1
+
\frac{[\mathrm{NS}_{k,i}^{\mathrm{Golgi}}(\xi,t)]}{K_{d,k,j}}
+
\frac{[\mathrm{OS}_{\ell,i}(\xi,t)]}{K_{d,\ell,j}}
+
\sum_{m=1}^{N_{\mathrm{OS}}}
\frac{[\mathrm{OS}_{m,i}(\xi,t)]}{K_{d,m,j}}
\\
&+
\frac{[\mathrm{NS}_{k,i}^{\mathrm{Golgi}}(\xi,t)]}{K_{d,k,j}}
\frac{[\mathrm{OS}_{\ell,i}(\xi,t)]}{K_{d,\ell,j}}
+
\frac{[\mathrm{NS}_{k,i}^{\mathrm{Golgi}}(\xi,t)]}{K_{d,k,j}}
\sum_{m=1}^{N_{\mathrm{OS}}}
\frac{[\mathrm{OS}_{m,i}(\xi,t)]}{K_{d,m,j}}
\\
&+
\frac{[\mathrm{Nuc}_{k,i}^{\mathrm{Golgi}}(\xi,t)]}{K_{d,N_k,j}}
\frac{[\mathrm{OS}_{\ell+1,i}(\xi,t)]}{K_{d,\ell+1,j}}
+
\frac{[\mathrm{Nuc}_{k,i}^{\mathrm{Golgi}}(\xi,t)]}{K_{d,N_k,j}}
+
\frac{[\mathrm{OS}_{\ell+1,i}(\xi,t)]}{K_{d,\ell+1,j}} .
\end{aligned}
\]
\\
\hline

\end{tabular}
\end{table}






\begin{table}[h!]
\centering
\renewcommand{\arraystretch}{1.25}
\caption{~Metabolite abbreviations and full names}
\begin{tabular}{|c|c|}
\hline
\rowcolor[HTML]{C0C0C0} 
\textbf{Abbreviation} & \textbf{Full Name} \\
\hline
AcCoA & Acetyl-Coenzyme A \\
\hline
AKG & $\alpha$-Ketoglutarate \\
\hline
ALA & Alanine \\
\hline
ANTI & Antibody \\
\hline
ASP & Aspartate \\
\hline
BIOM & Biomass \\
\hline
CO$_2$ & Carbon dioxide \\
\hline
EALA & Alanine, extracellular \\
\hline
EANTI & Antibody, extracellular \\
\hline
EASP & Aspartate, extracellular \\
\hline
EGLC & Glucose, extracellular \\
\hline
EGLN & Glutamine, extracellular \\
\hline
EGLU & Glutamate, extracellular \\
\hline
EILE & Isoleucine, extracellular \\
\hline
ELAC & Lactate, extracellular \\
\hline
ELEU & Leucine, extracellular \\
\hline
ESER & Serine, extracellular \\
\hline
EVAL & Valine, extracellular \\
\hline
ENH$_3$ & Ammonia, extracellular \\
\hline
G6P & Glucose-6-phosphate \\
\hline
GLN & Glutamine \\
\hline
GLU & Glutamate \\
\hline
ILE & Isoleucine \\
\hline
LAC & Lactate \\
\hline
LEU & Leucine \\
\hline
MAL & Malate \\
\hline
NH$_3$ & Ammonia \\
\hline
OAA & Oxaloacetate \\
\hline
PYR & Pyruvate \\
\hline
SER & Serine \\
\hline
SUC & Succinate \\
\hline
VAL & Valine \\
\hline
\end{tabular}
\label{tab:metabolite}
\end{table}

\clearpage

\section{Supplementary Statistical Tests}

\begin{table}[htbp]
\centering
\caption{\;Bootstrap-based normalized prediction errors for culture-state variables across Cases A--C. Values are reported as mean $\pm$ half-width of the 95\% bootstrap confidence interval. {For each case and observable, bootstrap confidence intervals are calculated
by resampling ensemble realizations (stochastic trajectories) with replacement. For each of $B=2000$
bootstrap realizations, NRMSE and NMAE are recalculated using
observed range (max - min) of the measured time series. The 2.5th and 97.5th percentiles of the
bootstrap distributions define the corresponding 95\% confidence intervals.
Table~S5 reports the bootstrap mean and confidence-interval half-width.
This bootstrap calculation quantifies finite-sample uncertainty in the
reported error metrics. It is distinct from the stochastic ensemble
prediction interval and is not used to define or modify the diffusion
variance in the mechanistic model.}}
\label{tab:nrmse_summary}

\begin{tabular}{lccc}
\toprule
Variable &
Case A &
Case B &
Case C \\
\midrule

\multicolumn{4}{l}{\textit{NRMSE}} \\
VCD            & 0.080 $\pm$ 0.006 & 0.080 $\pm$ 0.008 & 0.091 $\pm$ 0.009 \\
Glucose        & 0.201 $\pm$ 0.009 & 0.170 $\pm$ 0.011 & 0.174 $\pm$ 0.012 \\
Glutamine      & 0.134 $\pm$ 0.002 & 0.183 $\pm$ 0.001 & 0.118 $\pm$ 0.019 \\
Glutamic Acid  & 0.088 $\pm$ 0.021 & 0.053 $\pm$ 0.013 & 0.251 $\pm$ 0.047 \\
Ammonia        & 0.137 $\pm$ 0.007 & 0.234 $\pm$ 0.003 & 0.122 $\pm$ 0.012 \\
mAb Titer      & 0.050 $\pm$ 0.011 & 0.073 $\pm$ 0.016 & 0.100 $\pm$ 0.028 \\
\midrule

\multicolumn{4}{l}{\textit{NMAE}} \\
VCD            & 0.065 $\pm$ 0.005 & 0.062 $\pm$ 0.006 & 0.060 $\pm$ 0.009 \\
Glucose        & 0.115 $\pm$ 0.012 & 0.136 $\pm$ 0.008 & 0.135 $\pm$ 0.011 \\
Glutamine      & 0.107 $\pm$ 0.002 & 0.149 $\pm$ 0.001 & 0.099 $\pm$ 0.016 \\
Glutamic Acid  & 0.073 $\pm$ 0.019 & 0.044 $\pm$ 0.010 & 0.189 $\pm$ 0.044 \\
Ammonia        & 0.118 $\pm$ 0.007 & 0.169 $\pm$ 0.002 & 0.102 $\pm$ 0.009 \\
mAb Titer      & 0.037 $\pm$ 0.008 & 0.058 $\pm$ 0.013 & 0.076 $\pm$ 0.019 \\
\bottomrule
\end{tabular}
\end{table}

\clearpage

\section{Supplementary Process Dynamics}

\begin{figure}[htbp]
    \centering

    \begin{minipage}{0.7\textwidth}
        \centering
        \includegraphics[width=\textwidth]
        {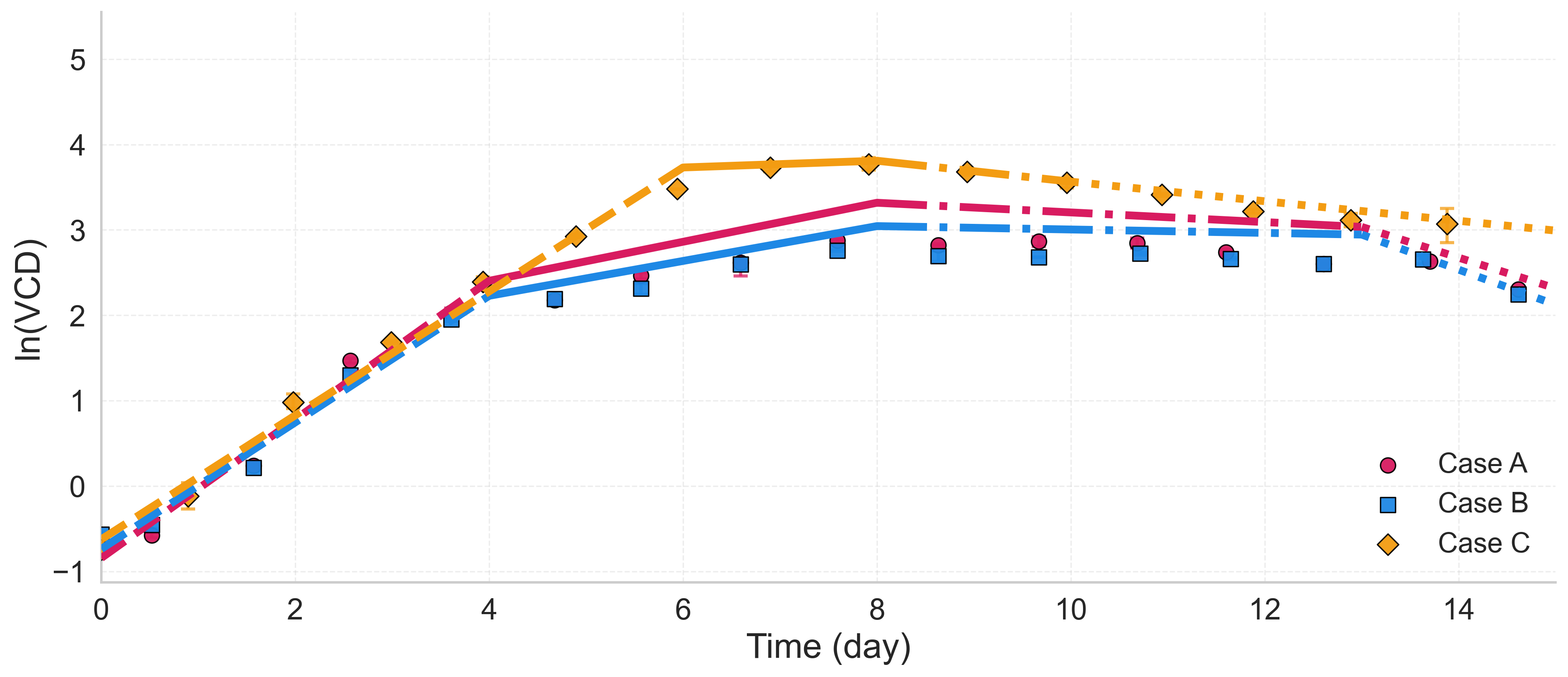}

        \vspace{0.2em}
        \textbf{(a)} Piecewise exponential growth analysis of VCD.
    \end{minipage}

    \vspace{0.8em}

    \begin{minipage}{0.7\textwidth}
        \centering
        \includegraphics[width=\textwidth]
        {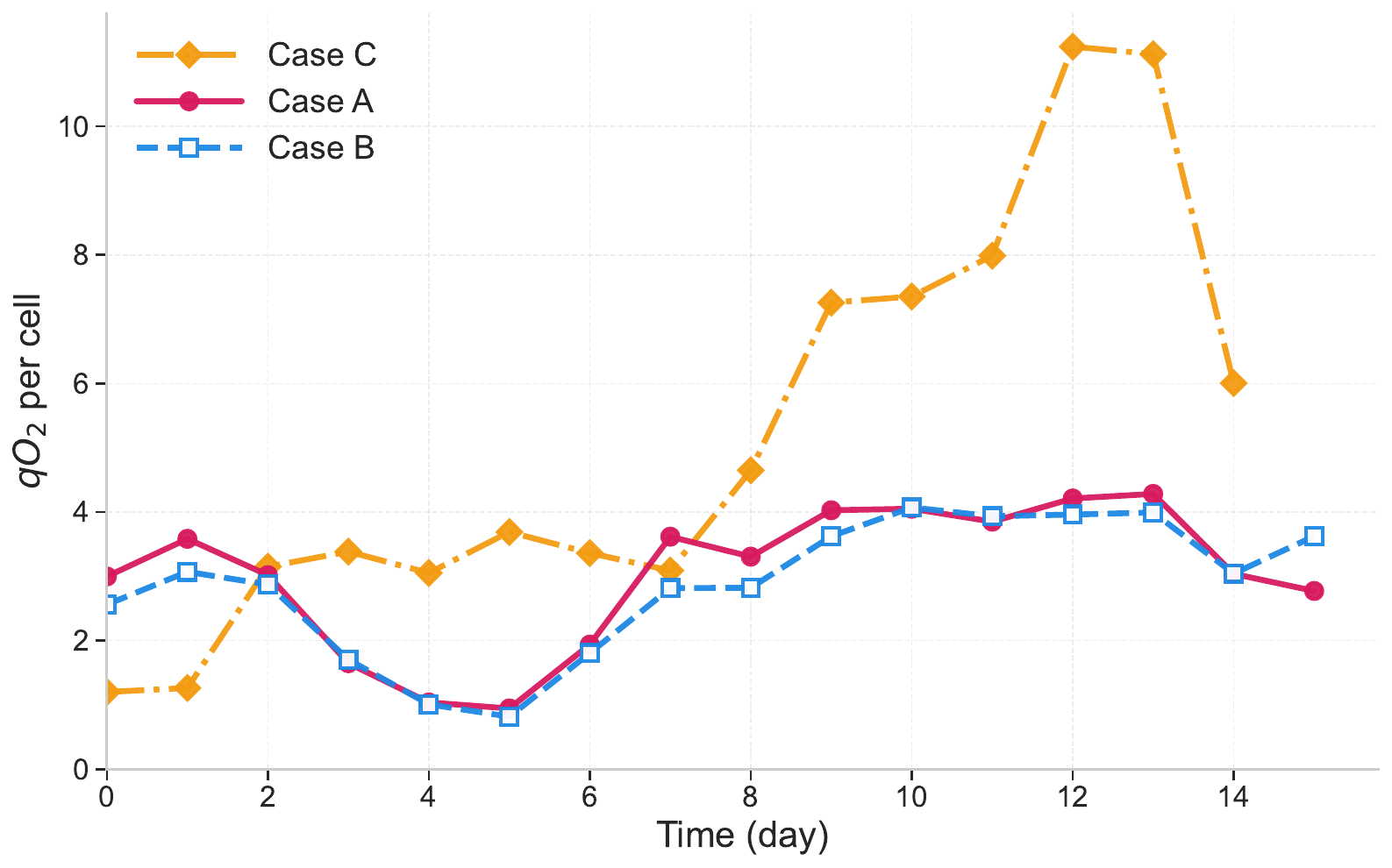}

        \vspace{0.2em}
        \textbf{(b)} Simulated cell-specific oxygen uptake rate
        \(q_{\mathrm{O}_2}(t)\) for Cases A--C.
    \end{minipage}

    \caption{\;
    Growth-phase characterization and oxygen-utilization dynamics across Cases A--C.
    Panel (a) shows piecewise exponential growth analysis of VCD. Experimental VCD measurements are transformed to
    $\ln(\mathrm{VCD})$
    and fitted using exponential growth models of the form
    $X_v(t)=X_{v,0}\exp(\mu t)$
    over selected culture intervals. The resulting growth-rate estimates quantify the transition from rapid exponential expansion to slower population growth under different cultivation conditions.
    Panel (b) shows simulated cell-specific oxygen uptake rate
    \(q_{\mathrm{O}_2}(t)\)
    for Cases A--C. The distinct temporal profiles indicate that feeding strategy and ammonia perturbation induce different oxygen-utilization dynamics. These oxygen-utilization changes are subsequently summarized through
    \(Q_{q_{\mathrm{O}_2}}(t)\)
    and used to characterize metabolic-state progression and phase-transition dynamics in the proposed framework.
    }
    \label{fig:growth_qO2}
\end{figure}

\clearpage
\begin{figure}[t]
    \centering
    \includegraphics[width=0.95\linewidth]{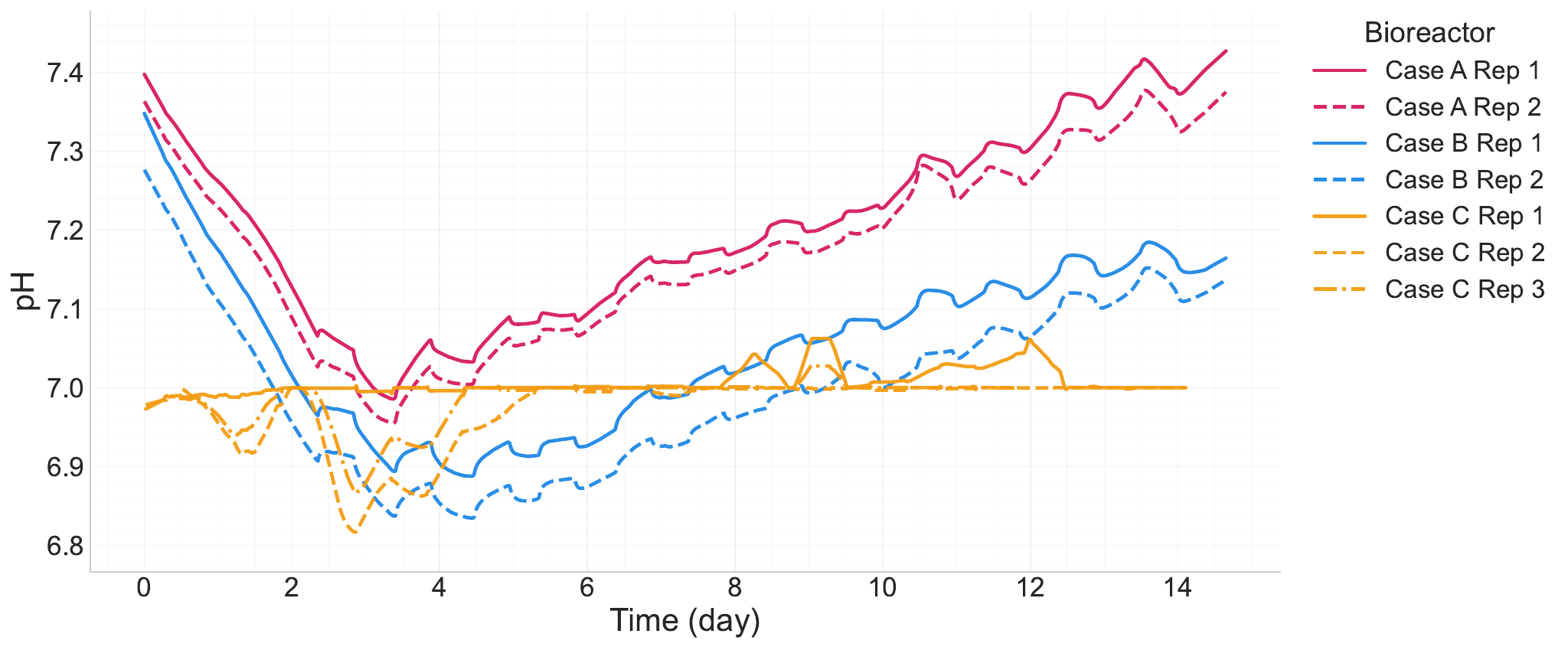}
    \caption{\;Time-course pH trajectories for Cases A--C across replicate bioreactors. The profiles show the smoothed pH evolution over the culture duration, with distinct replicate-to-replicate variability across cases.}
    \label{fig:ph_trajectories}
\end{figure}

\clearpage

\section{Cell-Resolved Glycosylation Heterogeneity under Identical Extracellular Conditions}

{
To isolate the contribution of intracellular heterogeneity, we performed a computational experiment with 100 virtual cells under the Case~C condition. All cells experienced identical extracellular concentrations and process inputs, while intracellular metabolic dynamics and phase transitions evolved independently.
Cell-specific metabolic states were then propagated through the Golgi glycosylation model to generate individual glycoform profiles. Consequently, all predicted glycosylation variability arises from stochastic intracellular heterogeneity rather than differences in extracellular exposure.
}

\begin{figure}[!htbp]
    \centering
    \includegraphics[
        width=0.85\textwidth
    ]{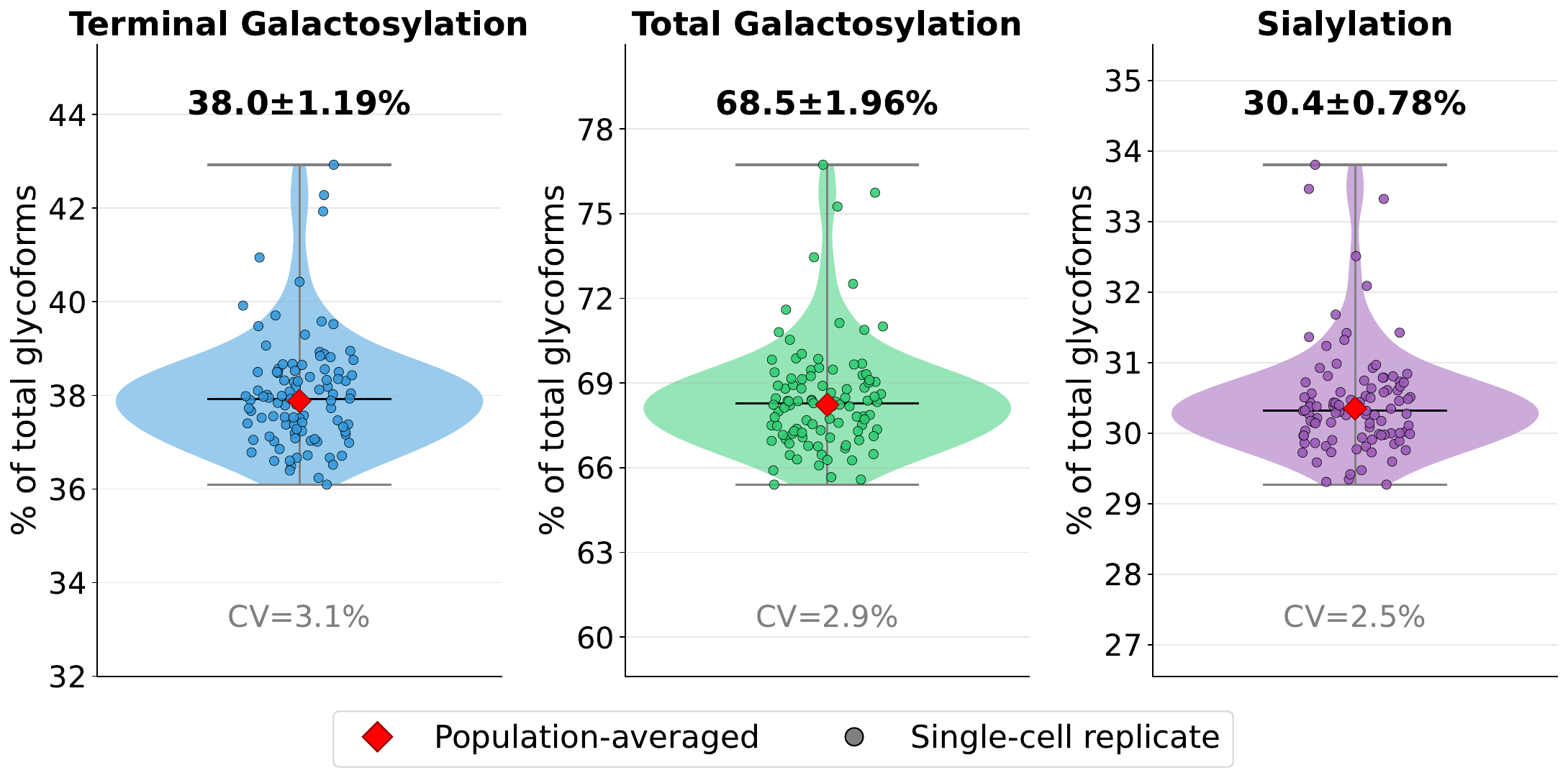}

    \caption{{\;
 Predicted cell-to-cell distributions of aggregated glycosylation CQAs at harvest for Case~C. All 100 virtual cells experience the same extracellular conditions and process inputs, while independent intracellular stochastic dynamics generate cell-specific metabolic states and glycosylation profiles. Distributions are shown for the non-galactosylated fraction, terminal galactosylation, total galactosylation, and sialylation.
    }}
    \label{fig:cell_resolved_glycosylation}
\end{figure}

{
Harvest-level variability is summarized in Table~\ref{tab:response_cell_resolved_glyco_statistics}. Across the nine glycoforms, coefficients of variation (CVs) ranged from 1.2\% to 9.8\% (mean: 5.2\%). G0 exhibited the greatest variability (CV = 9.8\%), whereas G1FS1 and G2S1 were the least variable (CVs of 1.2\% and 2.5\%, respectively), indicating that intracellular stochasticity propagates unevenly through different glycosylation pathways. 
Figure~\ref{fig:cell_resolved_glycosylation} shows harvest distributions for aggregated glycosylation CQAs. The non-galactosylated fraction exhibited the greatest variability (mean: 31.5\%, CV: 6.2\%), while total galactosylation, terminal galactosylation, and sialylation showed lower variability, with CVs of 2.9\%, 3.1\%, and 2.5\%, respectively. 
}

{
These results highlight the value of the cell-resolved framework. Even under identical extracellular conditions, the model predicts distributions of intracellular states and glycosylation outcomes, whereas a deterministic population-averaged model would yield only a single value for each glycoform or CQA.
The framework therefore provides a mechanistic link between intracellular heterogeneity and population-level variability in glycosylation quality attributes.
Because glycoform-resolved measurements at the single-cell level remain beyond current experimental capabilities, these distributions should be interpreted as model-predicted heterogeneity rather than experimentally validated single-cell glycosylation profiles.
}

\begin{table}[!htbp]
    \centering
    \caption{%
 {\;Predicted cell-to-cell variability in harvest glycoforms and aggregated glycosylation CQAs for Case~C (100 virtual cells). Mean, standard deviation (SD), interquartile range (IQR), minimum, and maximum are reported in percentage points; CV denotes the coefficient of variation.}
    }
    \label{tab:response_cell_resolved_glyco_statistics}
\begin{tabular}{lrrrrrr}
    \toprule
    Model output
    & Mean
    & SD
    & IQR
    & Minimum
    & Maximum
    & CV (\%) \\
    \midrule
    \multicolumn{7}{l}{\textit{Individual glycoforms}} \\
    G0
    & 1.92 & 0.188 & 0.188 & 1.14 & 2.17 & 9.8 \\
    G0F
    & 29.6 & 1.77 & 1.70 & 22.1 & 32.5 & 6.0 \\
    G1
    & 2.26 & 0.150 & 0.205 & 1.85 & 2.60 & 6.6 \\
    G1F
    & 24.5 & 0.953 & 1.05 & 22.8 & 28.3 & 3.9 \\
    G2
    & 3.74 & 0.143 & 0.192 & 3.46 & 4.08 & 3.8 \\
    G2F
    & 7.60 & 0.518 & 0.582 & 6.57 & 9.36 & 6.8 \\
    G1FS1
    & 13.1 & 0.157 & 0.219 & 12.7 & 13.4 & 1.2 \\
    G2S1
    & 3.41 & 0.0860 & 0.119 & 3.27 & 3.63 & 2.5 \\
    G2FS1
    & 14.0 & 0.903 & 0.990 & 12.2 & 17.2 & 6.5 \\
    \midrule
    \multicolumn{7}{l}{\textit{Aggregated glycosylation traits}} \\
    Non-galactosylated fraction
    & 31.5 & 1.96 & 1.88 & 23.3 & 34.6 & 6.2 \\
    Terminal galactosylation
    & 38.0 & 1.19 & 1.24 & 36.1 & 42.9 & 3.1 \\
    Total galactosylation
    & 68.5 & 1.96 & 1.88 & 65.4 & 76.7 & 2.9 \\
    Sialylation
    & 30.4 & 0.775 & 0.667 & 29.3 & 33.8 & 2.5 \\
    \bottomrule
\end{tabular}%
\end{table}

\clearpage

\section{Comparison of Case C Culture-State Predictions with Literature Data}

\begin{figure}[htbp]
    \centering
    \includegraphics[width=\linewidth]{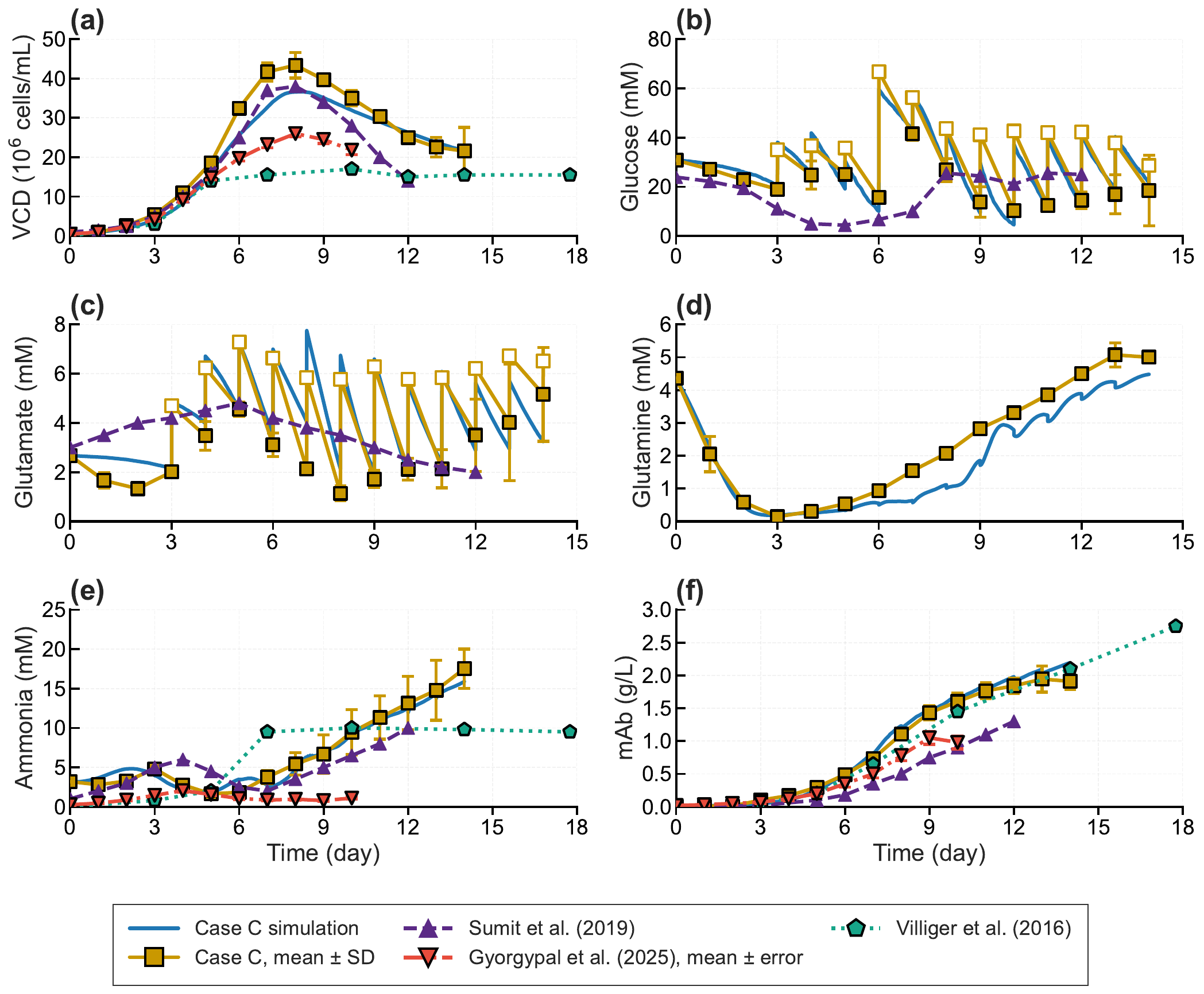}
    \caption{\; {
Comparison of predicted Case~C culture-state trajectories with experimental measurements and reference fed-batch datasets from Sumit et al.~(2019), Gyorgypal et al.~(2025), and Villiger et al.~(2016) \citep{sumit2019dissecting,gyorgypal2025temporal,villiger2016controlling}.
Panels show (a) viable cell density (VCD), (b) glucose, (c) glutamate, (d) glutamine, (e) ammonia, and (f) mAb concentration. Solid blue lines denote Case~C model predictions. Case~C measurements are shown as mustard squares (mean $\pm$ SD), while literature data are shown as purple upward triangles (Sumit et al., 2019), red downward triangles (Gyorgypal et al., 2025), and teal pentagons (Villiger et al., 2016). Error bars for the literature data are included where reported.
}}
    \label{fig:uncertainty_culture}
\end{figure}

\clearpage
\section{MODEL PARAMETERS, ESTIMATION, AND IDENTIFIABILITY}
\label{app:model_parameters}

\subsection{Parameter Classification and Calibration Strategy}

{
The model comprises two parameter groups: phase-dependent cell-culture kinetic parameters and Golgi glycosylation parameters. To improve transparency and avoid re-estimating established mechanistic quantities, parameters are classified as either estimated from the present dataset or fixed from the literature. An overview is provided in Table~\ref{tab:param_overview}, with complete definitions, values, and sources listed in Tables~\ref{tab:params}--\ref{tab:kd_overrides}.}

{The cell-culture model is calibrated to the baseline fed-batch condition (Case~A). The resulting parameter set is then applied unchanged to Cases~B and~C, which serve as perturbation tests under altered ammonia and feeding conditions. Estimated parameters include phase-dependent reaction rates, substrate-affinity constants, inhibition and activation coefficients, growth/death parameters, and phase-transition coefficients.
}

{The glycosylation sub-model is calibrated sequentially using the predicted cell-culture trajectories as inputs. Structural parameters describing Golgi organization, enzyme localization, nucleotide-sugar transport, and established kinetic constants are fixed from prior studies, whereas selected cell-line-specific parameters related to enzyme activity and Golgi environmental conditions are estimated from glycoform data.
This approach is consistent with the model's one-way coupling, whereby cell-culture states drive glycosylation, but glycosylation does not feed back into the cell-culture dynamics.
}

\subsection{Parameter Estimation Procedure}

{The cell-culture kinetic parameters are estimated by minimizing a
regularized normalized least-squares objective,
\begin{equation}
J(\boldsymbol{\theta})
=
\sum_{s}\sum_{t}
\frac{
\left[
y_s^{\mathrm{obs}}(t)
-
\hat{y}_s(t,\boldsymbol{\theta})
\right]^2
}{
\sigma_s^2
}
+
\lambda
\sum_i
\left(
\frac{\theta_i-\theta_i^0}{|\theta_i^0|}
\right)^2,
\label{eq:param_estimation_objective}
\end{equation}
where $y_s^{\mathrm{obs}}(t)$ denotes the measured value of species $s$
at time $t$, $\hat{y}_s(t,\boldsymbol{\theta})$ is the corresponding
model prediction, and $\sigma_s$ is the interquartile range of the
observations for species $s$. The first term provides a normalized
least-squares measure of model-data mismatch, while the second term is a
Tikhonov regularization term centered at the reference parameter values
$\boldsymbol{\theta}^0$. A regularization weight of $\lambda=5$ is used throughout.}

{Reference parameter values are selected from published kinetics, stoichiometric constraints, reported parameter ranges, and consistency with the experimental trajectories. To maintain physiological plausibility, estimated parameters are constrained within $\pm30\%$ of their reference values,
$
\theta_i
\in
\left[
\theta_i^0(1-\delta),
\theta_i^0(1+\delta)
\right],$ with $
\delta=0.30$.
\label{eq:param_bounds}
}

{An initial calibration is performed using \texttt{lmfit} to obtain the reference parameter set $\boldsymbol{\theta}^{0}$ for sensitivity analysis. After sensitivity-based parameter reduction, the retained active parameters are re-estimated using the Trust Region Reflective (TRF) algorithm implemented in \texttt{scipy.optimize.least\_squares}. The resulting solution is used for subsequent uncertainty and practical-identifiability analysis. Glycosylation parameters are estimated separately, following cell-culture calibration, using the same TRF optimization framework.
}

\subsection{Glycosylation Parameter Estimation}

{The glycosylation model is calibrated after estimation of the upstream
cell-culture model. The fitted parameter vector is
\begin{equation}
\boldsymbol{\theta}_{\mathrm{glyco}}
=
\left[
\mathbf{k}_{f\max},
\mathbf{e}_{\max},
\boldsymbol{\omega}_{f},
n_a,
K_{d,\mathrm{SiaT}},
K_{d,\mathrm{GalT}},
K_{d,\mathrm{GalT\_f}},
K_{dk,\mathrm{GalT}},
K_{dk,\mathrm{SiaT}}
\right]^{\top},
\label{eq:glyco_param_vector}
\end{equation}
which includes parameters governing enzyme activity, effective enzyme abundance, Golgi pH sensitivity, and selected glycosyltransferase substrate interactions. Structural and transport parameters are fixed at literature-derived values (Table~\ref{tab:param_overview}).}

{Calibration is performed sequentially,
\begin{equation}
\boldsymbol{\theta}_{\mathrm{CC}}
\longrightarrow
\left\{
\mathrm{VCD}(t),
[\mathrm{NH}_3](t),
c_{\mathrm{mAb}}(t),
\ldots
\right\}
\longrightarrow
\boldsymbol{\theta}_{\mathrm{glyco}},
\label{eq:sequential_calibration}
\end{equation}
where the calibrated cell-culture model provides the dynamic inputs required by the Golgi model. Glycosylation parameters are then estimated from the measured glycoform distributions while the cell-culture parameters remain fixed.}

\subsection{Parameter Sensitivity and Identifiability Analyses}
\label{app:identifiability}

{
Because the mechanistic model contains substantially more kinetic parameters
than can be uniquely constrained by the available measurements, parameter
uncertainty was assessed using three complementary analyses: output-sensitivity
screening, local first-order identifiability analysis, and
Jacobian-based practical-identifiability analysis. Sensitivity screening
was used to identify parameters that materially influence the measured
trajectories and to reduce the dimension of the cell-culture calibration
problem. Local identifiability was then assessed from the rank structure of the output-sensitivity Jacobian, which quantifies the number of independently distinguishable parameter directions. Finally, the residual Jacobian at the calibrated solution was used to estimate practical parameter uncertainty and confidence intervals.
}

\paragraph{Output-sensitivity analysis and active-set construction.}

{
For each candidate cell-culture parameter $\theta_j$, local output sensitivity
was evaluated using centered finite differences with a relative perturbation
magnitude of $\delta=0.01$. For observable $y_k(t)$, the normalized sensitivity
was defined as
}

\begin{equation}
S_{kj}(t)
=
\frac{
y_k\!\left(t;\theta_j^{*}(1+\delta)\right)
-
y_k\!\left(t;\theta_j^{*}(1-\delta)\right)
}{
2\delta\,y_k^{\mathrm{ref}}(t)
},
\label{eq:local_sensitivity}
\end{equation}
{where $\theta_j^{*}$ denotes the reference-calibrated parameter value and
$y_k^{\mathrm{ref}}(t)$ denotes the corresponding model output evaluated at the reference parameter vector. An overall sensitivity score for parameter $\theta_j$ was defined as the root-mean-square of the normalized sensitivities across all measured outputs and sampling times,
}

\begin{equation}
S_j
=
\left[
\frac{1}{N_{\mathrm{obs}}}
\sum_{k,t}
S_{kj}(t)^2
\right]^{1/2},
\label{eq:sensitivity_score}
\end{equation}
{where $N_{\mathrm{obs}}$ is the total number of output--time-point
combinations included in the analysis.
}

{
The change in the complete data-residual vector under the same perturbation was
also calculated as
}
\begin{equation}
\Delta R_j
=
\left\|
\mathbf{r}\!\left(\theta_j^{*}(1+\delta)\right)
-
\mathbf{r}\!\left(\theta_j^{*}(1-\delta)\right)
\right\|_2,
\label{eq:residual_sensitivity}
\end{equation}
{where $\mathbf{r}$ denotes the data-residual vector. Parameters satisfying $\Delta R_j<0.005$ were classified as locally
insensitive under the current observation design and were held at their
reference-calibrated values during the reduced calibration. The remaining parameters, which produced numerically stable and non-negligible residual changes, were retained as the sensitivity-active parameter set.
}

{
The phase-transition coefficients denoted by $K1_{*}$ and $K2_{*}$ were
treated separately from this screening procedure. Their reference-calibrated
magnitudes were approximately $10^{-8}$, yielding absolute finite-difference steps of
}

\begin{equation}
h_j
=
\delta\left|\theta_j^{*}\right|
\approx
10^{-10}.
\label{eq:small_parameter_step}
\end{equation}
{At this scale, the centered finite difference involved subtraction of nearly
identical simulated trajectories and was insufficiently resolved numerically.
The resulting sensitivity scores were approximately $10^{6}$--$10^{7}$ and
were strongly dependent on the finite-difference step size. These values were
therefore interpreted as numerical amplification associated with the extremely
small absolute perturbations, rather than as reliable evidence of exceptionally
large physical output sensitivity. Accordingly, the $K1_{*}$ and $K2_{*}$ coefficients were excluded from the active calibration set and retained at their
reference-calibrated values.
}

\begin{figure}[htbp]
    \centering
    \includegraphics[width=0.92\textwidth]
    {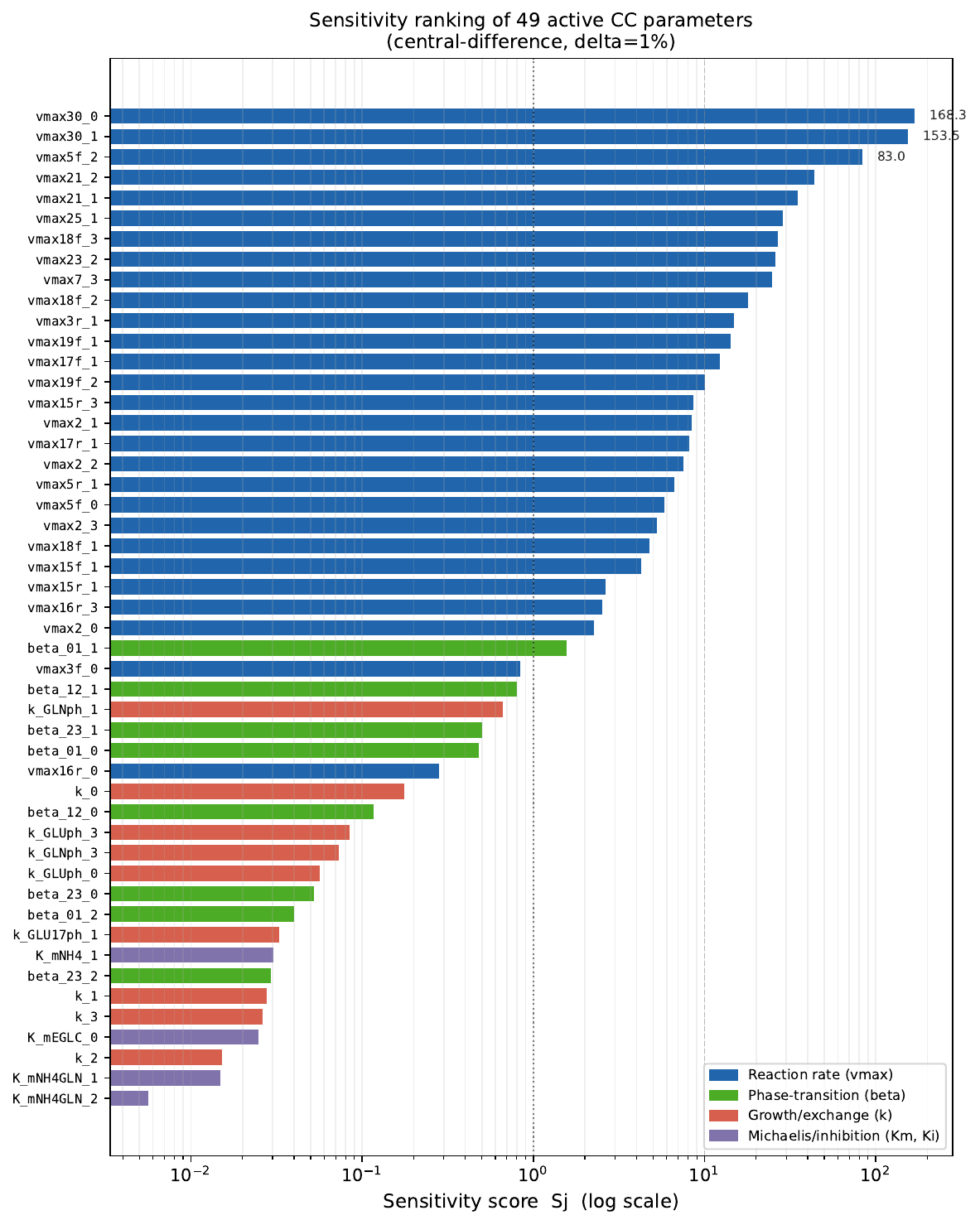}
    \caption{\;{
    Ranked output-sensitivity scores for the 49 sensitivity-active
    cell-culture parameters. Sensitivity scores were calculated using centered
    finite differences with a relative perturbation magnitude of
    $\varepsilon=0.01$ and are displayed on a logarithmic scale. Parameters are
    ordered by decreasing $S_j$ and colored according to parameter class:
    maximum reaction rates ($v_{\max}$), phase-transition coefficients
    ($\beta$), growth and exchange coefficients ($k$), and Michaelis or
    inhibition coefficients ($K_m$ and $K_i$). The dotted and dashed vertical
    lines indicate $S_j=1$ and $S_j=10$, respectively, and are included as
    visual reference levels rather than formal statistical thresholds.
    Reaction-rate parameters dominate the upper portion of the ranking,
    although the sensitivity scores span more than four orders of magnitude
    across the complete active set. The $K1_{*}$ and $K2_{*}$
    phase-transition coefficients are not included because their calibrated
    magnitudes of approximately $10^{-8}$ produce insufficiently resolved
    finite-difference steps and numerically amplified, step-size-dependent
    sensitivity estimates.
    }}
    \label{fig:cc_sensitivity_ranking}
\end{figure}

{Application of these criteria yielded a reduced cell-culture calibration set containing 49 parameters. Figure~\ref{fig:cc_sensitivity_ranking} shows their sensitivity ranking. The reliable sensitivity scores $S_j$ span more
than four orders of magnitude, indicating substantial variation in the responsiveness of the measured trajectories to individual parameters. The largest sensitivities were associated with $v_{\max,30}^{(0)}$, $v_{\max,30}^{(1)}$, and
$v_{\max,5f}^{(2)}$, with scores of 168.3, 153.5, and 83.0, respectively.
Maximum reaction-rate parameters dominate the upper portion of the ranking, whereas phase-transition, growth, exchange, Michaelis, and inhibition parameters are concentrated in the middle and lower portions.
}

{Although the ranking quantifies the magnitude of model-output responses to individual parameter perturbations, sensitivity alone does not establish parameter identifiability. Parameters with substantial individual sensitivities may still produce highly correlated output responses and therefore be difficult to distinguish. Potential parameter confounding was therefore examined separately through analysis of the rank and correlation structure of the full sensitivity Jacobian.
}

\paragraph{Local first-order theoretical identifiability.}

{
Local first-order identifiability was assessed separately for the
cell-culture and glycosylation submodels using their respective output-sensitivity Jacobians. For a submodel with $p$ fitted parameters, let
$\mathbf{J}\in\mathbb{R}^{N_{\mathrm{obs}}\times p}$ denote the sensitivity
Jacobian evaluated at the calibrated parameter vector. To remove differences in sensitivity magnitude and focus on parameter distinguishability, each column of $\mathbf{J}$ was normalized by its Euclidean norm,
\begin{equation}
\overline{\mathbf{J}}_{:j}
=
\frac{
\mathbf{J}_{:j}
}{
\left\|\mathbf{J}_{:j}\right\|_2
},
\qquad
j=1,\ldots,p,
\label{eq:normalized_sensitivity_jacobian}
\end{equation}
and the corresponding column-normalized Fisher information matrix (FIM) was defined as,
$
\mathbf{F}_{\mathrm{norm}}
=
\overline{\mathbf{J}}^{\top}
\overline{\mathbf{J}}.
\label{eq:normalized_fim}
$
Equivalently, if
$
\mathbf{F}
=
\mathbf{J}^{\top}\mathbf{J}$ and $
\mathbf{D}
=
\operatorname{diag}
\left(
\left\|\mathbf{J}_{:1}\right\|_2,
\ldots,
\left\|\mathbf{J}_{:p}\right\|_2
\right)$,
then $
\mathbf{F}_{\mathrm{norm}}
=
\mathbf{D}^{-1}
\mathbf{F}
\mathbf{D}^{-1}
\label{eq:normalized_fim_equivalence}
$.
This normalization separates linear dependence among parameter-sensitivity
directions from differences in their absolute sensitivity magnitudes. Provided
that no sensitivity column is identically zero, column normalization does not
change the rank of the sensitivity Jacobian and therefore does not change the
number of locally distinguishable first-order parameter directions.
}

{Let
$\lambda_1\geq\lambda_2\geq\cdots\geq\lambda_p$
denote the eigenvalues of $\mathbf{F}_{\mathrm{norm}}$, and let
$\sigma_1\geq\sigma_2\geq\cdots\geq\sigma_p$
denote the singular values of $\overline{\mathbf{J}}$. Since
$
\lambda_i
=
\sigma_i^2,
\label{eq:eigen_singular_relation}
$
the eigenvalue and singular-value spectra provide equivalent numerical-rank
diagnostics. Eigenvalues satisfying
$
\lambda_i
>
10^{-8}\lambda_{\max}
\label{eq:eigenvalue_rank_threshold}
$
were classified as numerically nonzero. The equivalent singular-value
criterion was
$
\sigma_i
>
10^{-4}\sigma_{\max}.
\label{eq:singular_value_rank_threshold}
$
To assess pairwise parameter confounding, the absolute entries of the column-normalized FIM were also examined:
\begin{equation}
\left|
F_{\mathrm{norm},ij}
\right|
=
\left|
\frac{
\mathbf{J}_{:i}^{\top}\mathbf{J}_{:j}
}{
\left\|\mathbf{J}_{:i}\right\|_2
\left\|\mathbf{J}_{:j}\right\|_2
}
\right|,
\label{eq:fim_cosine_similarity}
\end{equation}
which corresponds to the absolute cosine similarity between sensitivity vectors. Values near one indicate nearly parallel or antiparallel first-order
output responses and therefore potential pairwise parameter confounding.
However, these pairwise entries do not determine the dimension of the complete
near-null space; numerical rank was determined from the full sensitivity
matrix.
}

\vspace{0.05in}
\textit{{Cell-culture submodel.}}

\begin{figure}[htbp]
    \centering
    \includegraphics[width=\textwidth]
    {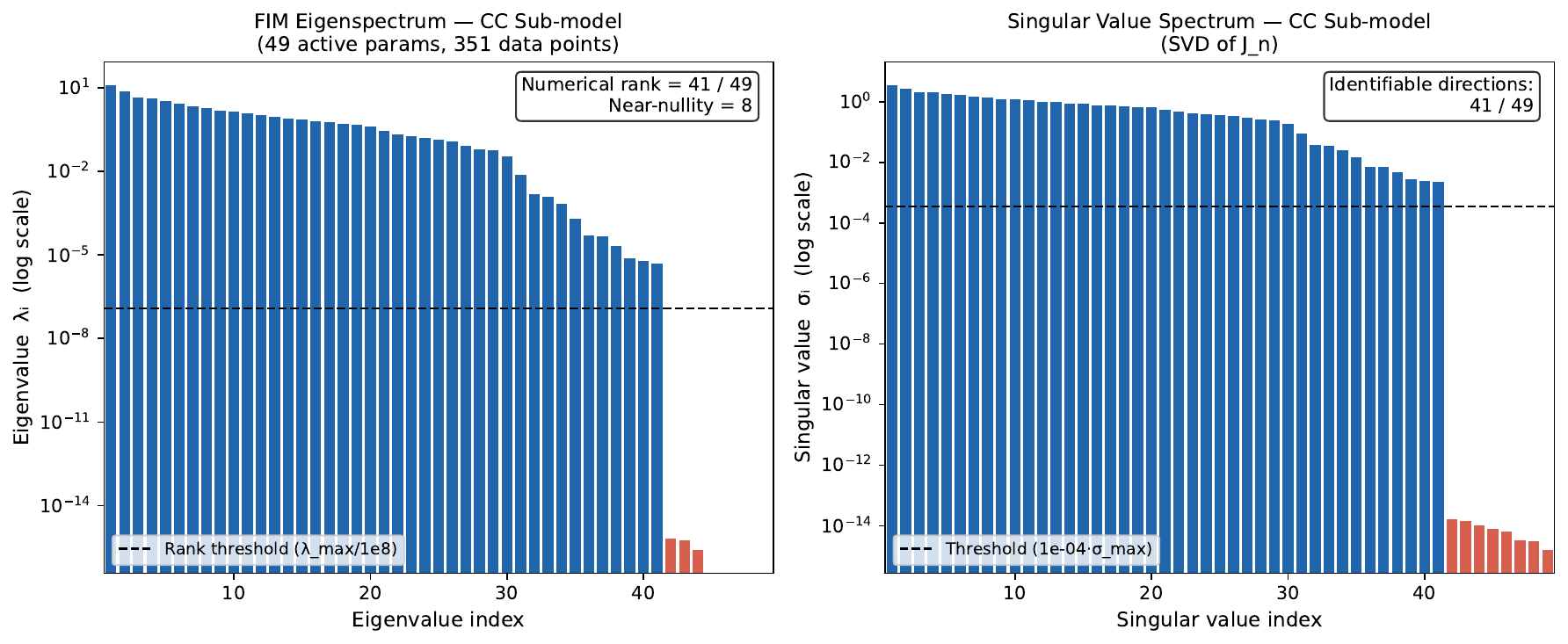}
    \caption{\;{
    Local first-order identifiability analysis for the 49
    sensitivity-active parameters. The left panel shows the eigenvalues of
    the normalized sensitivity Gram matrix $\mathbf{G}=\overline{\mathbf{J}}^{\top}\overline{\mathbf{J}}$,
    and the right panel shows the singular values of
    $\overline{\mathbf{J}}$. The numerical-rank thresholds are
    $\lambda_i>10^{-8}\lambda_{\max}$ and, equivalently,
    $\sigma_i>10^{-4}\sigma_{\max}$. Under these criteria, 41 independent
    first-order sensitivity directions are retained, while eight parameter
    combinations lie in the numerical near-null space.}
    }
    \label{fig:local_identifiability_spectrum}
\end{figure}

\begin{figure}[htbp]
    \centering
    \includegraphics[width=0.92\textwidth]
    {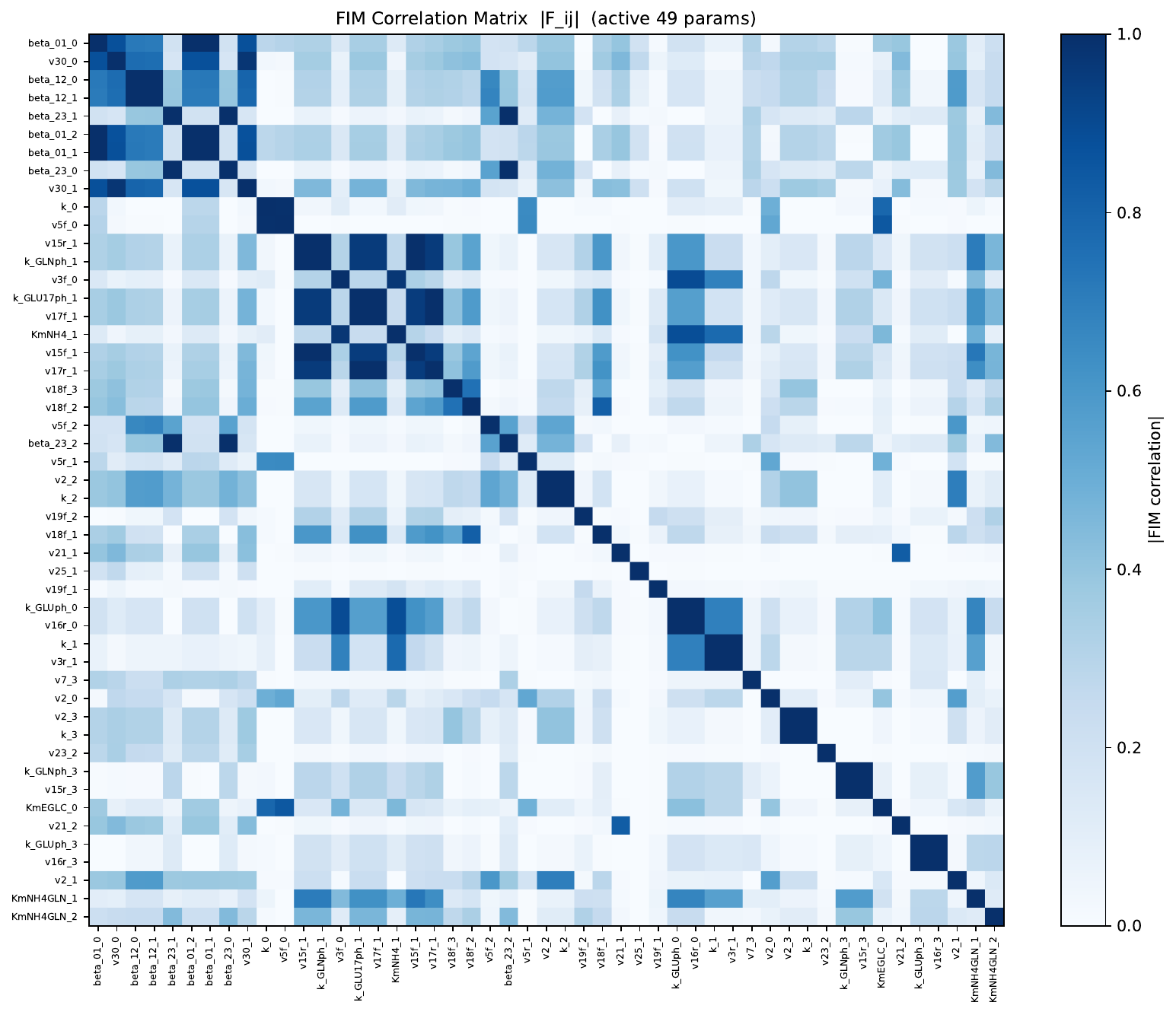}
    \caption{\;{
    Absolute normalized sensitivity Gram matrix for the 49
    sensitivity-active parameters. Each entry represents the absolute cosine
    similarity between two normalized parameter-sensitivity vectors. Values
    near one indicate nearly parallel or antiparallel first-order output
    responses and therefore potential pairwise parameter confounding under
    the current observation design.
    }}
    \label{fig:local_identifiability_gram}
\end{figure}

{For the cell-culture submodel, the sensitivity Jacobian associated with the
49 sensitivity-active parameters had numerical rank 41 under the criteria in
Eqs.~\eqref{eq:eigenvalue_rank_threshold} and
\eqref{eq:singular_value_rank_threshold}. Thus, the retained parameters spanned 41 locally distinguishable directions, with eight parameter combinations residing in the numerical near-null space. The corresponding eigenvalue and singular-value spectra are shown in Fig.~\ref{fig:local_identifiability_spectrum}.
Figure~\ref{fig:local_identifiability_gram} summarizes pairwise relationships among the normalized sensitivity vectors. Large
off-diagonal entries indicate parameter pairs that produce similar normalized
first-order changes in the measured cell-culture outputs. This heatmap is used
as a pairwise confounding diagnostic only; the eight-dimensional near-null
space is determined from the numerical rank of the complete cell-culture
sensitivity matrix.
}

\vspace{0.05in}

\textit{{Glycosylation submodel.}}

\begin{figure}[htbp]
    \centering
    \includegraphics[width=\textwidth]
    {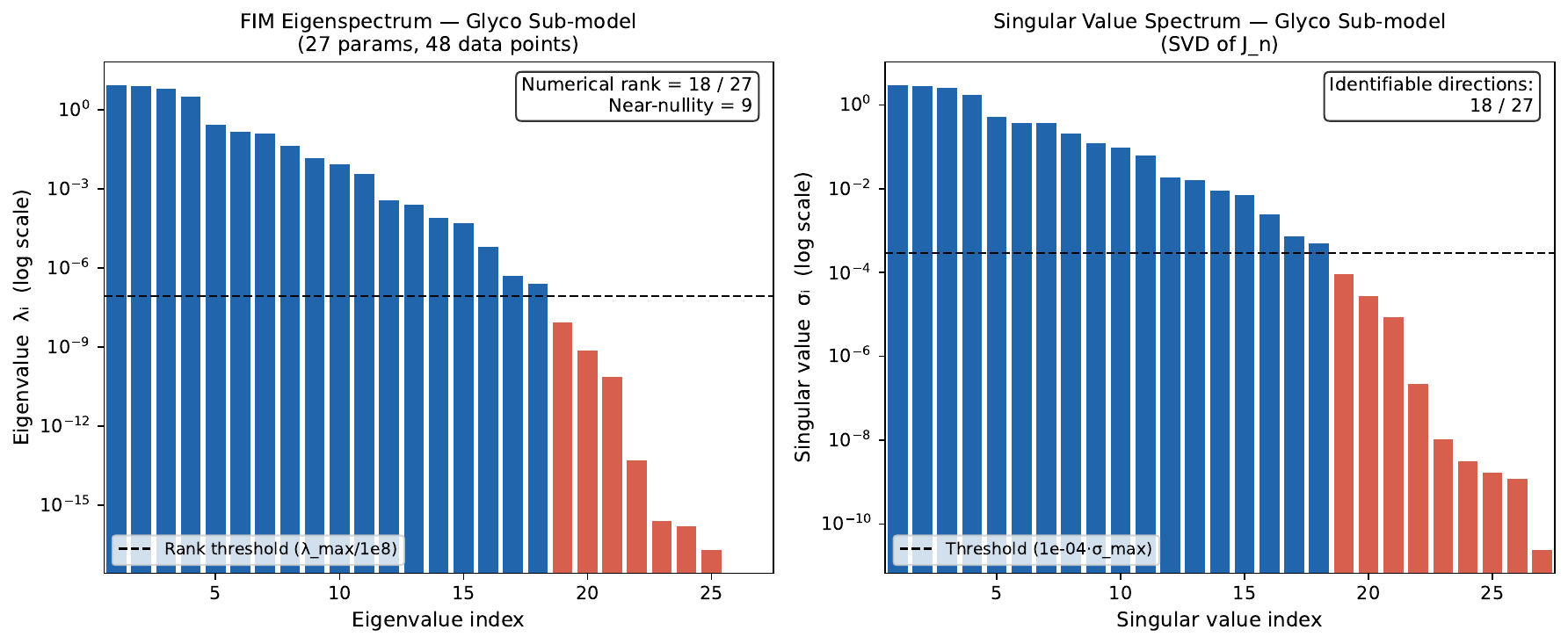}
    \caption{\;{
    Local first-order theoretical-identifiability analysis for the 27 fitted
    glycosylation parameters using 48 glycosylation observations. The left panel
    shows the eigenvalues of the column-normalized Fisher-information matrix
    $\mathbf{F}_{\mathrm{norm,gly}}
    =\overline{\mathbf{J}}_{\mathrm{gly}}^{\top}
    \overline{\mathbf{J}}_{\mathrm{gly}}$, and the right panel shows the singular
    values of $\overline{\mathbf{J}}_{\mathrm{gly}}$. The numerical-rank criteria
    are $\lambda_i>10^{-8}\lambda_{\max}$ and, equivalently,
    $\sigma_i>10^{-4}\sigma_{\max}$. Under these criteria, 18 independent
    first-order sensitivity directions are retained, while nine parameter
    combinations lie in the numerical near-null space. Because the matrix is
    numerically rank deficient, its full numerical condition number is treated
    as effectively infinite.
    }}
    \label{fig:glyco_identifiability_spectrum}
\end{figure}

\begin{figure}[htbp]
    \centering
    \includegraphics[width=0.92\textwidth]
    {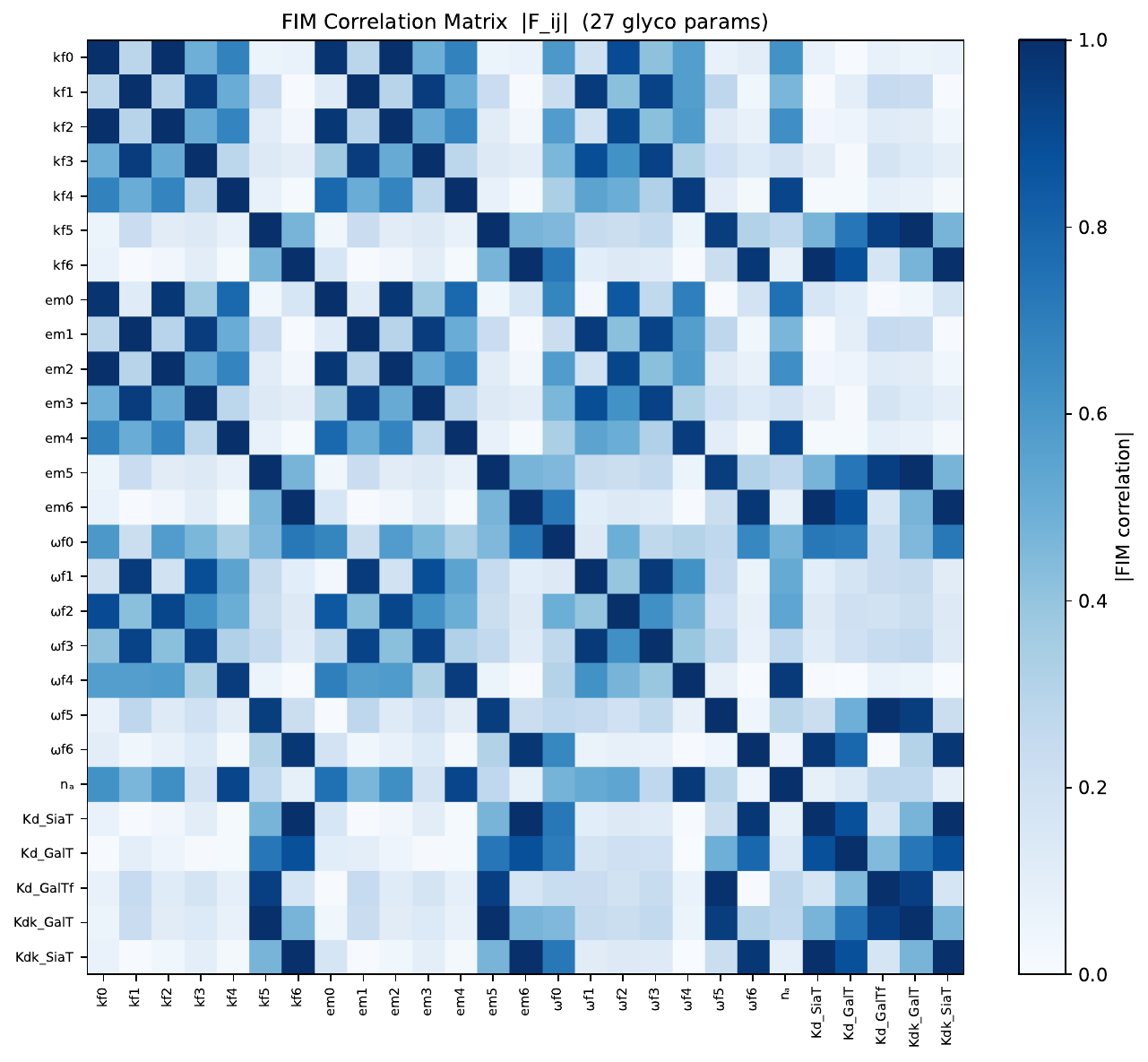}
    \caption{\;{
    Absolute column-normalized Fisher-information matrix for the 27 fitted
    glycosylation parameters. Each entry represents the absolute cosine
    similarity between two normalized parameter-sensitivity vectors. Values near
    one indicate that the corresponding parameters generate nearly parallel or
    antiparallel first-order changes in the measured glycosylation outputs and
    therefore indicate potential pairwise parameter confounding under the current
    observation design. This heatmap is a pairwise diagnostic only; the numerical
    rank of 18 and the nine-dimensional numerical near-null space are determined
    from the complete glycosylation sensitivity matrix, as shown in
    Fig.~\ref{fig:glyco_identifiability_spectrum}.
    }}
    \label{fig:glyco_identifiability_heatmap}
\end{figure}

{The same analysis was applied to all 27 fitted parameters of the
glycosylation submodel. The corresponding sensitivity Jacobian contained
48 glycosylation data points and 27 parameter columns and had numerical rank 18 under the same criteria. Consequently, the parameter set spanned 18 locally distinguishable directions, while nine parameter combinations formed a numerical near-null space. The corresponding spectra are shown in Fig.~\ref{fig:glyco_identifiability_spectrum}. 
Pairwise relationships among the glycosylation parameter-sensitivity
directions are shown in Fig.~\ref{fig:glyco_identifiability_heatmap}. As for
the cell-culture analysis, this heatmap is interpreted only as a diagnostic of
potential pairwise confounding. The nine-dimensional near-null space is determined from the rank of the complete sensitivity matrix rather than from individual correlation coefficients.
}

{Both column-normalized FIMs were numerically rank deficient under the adopted
criterion. Their full numerical condition numbers were therefore treated as
effectively infinite rather than reported as finite quantitative values. The resulting near-nullities should be interpreted as eight cell-culture and nine glycosylation parameter combinations that cannot be distinguished locally, rather than as eight or nine individually non-identifiable parameters.
}

{Because these analyses are evaluated at the calibrated parameter vectors, they provide only local information and depend on the measured outputs, sampling times, and experimental conditions included in the study. They therefore do not constitute global structural-identifiability proofs for the stochastic multiscale model. Instead, they quantify local parameter distinguishability and complement the practical-identifiability analysis described below.
}

\paragraph{Practical-identifiability analysis.}

{
After convergence of each parameter-estimation run, practical uncertainty was quantified from the Jacobian of the data residuals with respect to the fitted parameters.
The residual Jacobian used in this analysis contained only the data-misfit
residuals and excluded all Tikhonov regularization terms. 
Let
\(
\mathbf{r}_{\mathrm{data}}(\boldsymbol{\theta})
=
[r_1(\boldsymbol{\theta}),
\ldots,
r_{n_{\mathrm{data}}}(\boldsymbol{\theta})
]^{\top}
\)
denote the vector of data-misfit residuals. Its Jacobian evaluated at the
fitted parameter vector $\hat{\boldsymbol{\theta}}$ is
\begin{equation}
\mathbf{J}_r
:=
\left.
\frac{\partial\mathbf{r}_{\mathrm{data}}(\boldsymbol{\theta})}
{\partial\boldsymbol{\theta}^{\top}}
\right|_{\boldsymbol{\theta}=\hat{\boldsymbol{\theta}}}
=
\left[
\left.
\frac{\partial r_k(\boldsymbol{\theta})}
{\partial\theta_j}
\right|_{\boldsymbol{\theta}=\hat{\boldsymbol{\theta}}}
\right]_{\substack{
k=1,\ldots,n_{\mathrm{data}}\\
j=1,\ldots,n_{\mathrm{fit}}
}}
\in
\mathbb{R}^{n_{\mathrm{data}}\times n_{\mathrm{fit}}}.
\label{eq:residual_jacobian}
\end{equation}
The local covariance matrix was
approximated as
\begin{equation}
\mathbf{C}
\approx
s^2
\left(
\mathbf{J}_r^{\top}\mathbf{J}_r
\right)^{+},
\qquad
s^2
=
\frac{
J_{\mathrm{data}}
}{
n_{\mathrm{data}}-n_{\mathrm{fit}}
},
\label{eq:local_covariance}
\end{equation}
where $(\cdot)^{+}$ denotes the Moore--Penrose pseudoinverse,
$J_{\mathrm{data}}$ is the data-residual sum of squares,
$n_{\mathrm{data}}$ is the number of experimental observations, and
$n_{\mathrm{fit}}$ is the number of fitted parameters included in the
corresponding calibration. The standard error (SE) and relative standard error (RSE) of
parameter $\theta_i$ were calculated as
$
\mathrm{SE}_i
=
\sqrt{C_{ii}}$ and $
\mathrm{RSE}_i
=
100
\frac{
\mathrm{SE}_i
}{
\left|\hat{\theta}_i\right|
}.
\label{eq:rse}
$
Approximate local 95\% confidence intervals were estimated using the linearized Gaussian approximation:
$
\hat{\theta}_i
\pm
1.96\,\mathrm{SE}_i.
\label{eq:local_ci}
$
}

{
These intervals quantify local uncertainty around the fitted solution and
should not be interpreted as formal profile-likelihood confidence intervals.
For descriptive reporting, parameters with $\mathrm{RSE}<20\%$ were classified
as well constrained, parameters with
$20\%\leq\mathrm{RSE}\leq50\%$ as moderately constrained, and parameters with
$\mathrm{RSE}>50\%$ as weakly constrained. These thresholds were used as
descriptive uncertainty categories rather than as formal identifiability
criteria.
}

{
Parameters that converged at or near an imposed optimization bound were
additionally classified as boundary-limited. For such parameters, a small Jacobian-based standard error, particularly when obtained from a pseudoinverse-based covariance estimate, was not considered evidence of precise identification by the experimental data.
}

\vspace{0.05in}
\textit{{Cell-culture parameters.}}

{
Sensitivity screening of the cell-culture parameter set retained 49 parameters
for the final reduced calibration, with all remaining parameters held fixed at
their reference-calibrated values. The reduced Trust Region Reflective optimization converged successfully, yielding a normalized data misfit of
$
\frac{
J_{\mathrm{data}}
}{
n_{\mathrm{data}}
}
=
\frac{271.1}{351}
=
0.77.
\label{eq:cc_normalized_data_misfit}
$
The residual-Jacobian analysis identified 11 cell-culture parameters with
$\mathrm{RSE}<20\%$, 6 parameters with RSE values between 20\% and 50\%, and
32 parameters with $\mathrm{RSE}>50\%$. The latter group was therefore
classified as weakly constrained by the current experimental dataset. Their
inclusion in the sensitivity-active set indicates that they influence one or
more measured trajectories, but their large local uncertainty shows that they
cannot be independently estimated with high precision from the available
data. Parameter-specific fitted values, standard errors, RSE values, local
confidence intervals, and uncertainty classifications are reported in
Table~\ref{tab:active_params}.
}

\vspace{0.05in}
\textit{{Glycosylation parameters.}}

{
The same residual-Jacobian analysis was applied to the 27 fitted
glycosylation parameters. Among these, 6 were classified as well constrained ($\mathrm{RSE}<20\%$), 1 as moderately constrained ($20\%\leq\mathrm{RSE}\leq50\%$), and 20 as weakly constrained ($\mathrm{RSE}>50\%$). The well-constrained subset was dominated by enzyme
turnover-rate parameters, whereas peak enzyme concentrations, pH-activity
widths, and several Golgi kinetic parameters were more weakly constrained by
the available bulk glycoform measurements.
}

{
Accordingly, the weakly constrained glycosylation parameters should not be
interpreted as independently identified biochemical constants. Rather, they
represent effective calibrated parameters within the coupled glycosylation
model and may partially compensate for uncertainty in unmeasured intracellular
states, including intracellular nucleotide-sugar availability. Fitted values, standard errors, RSE values, confidence intervals, and constraint classifications are reported in Table~\ref{tab:glyco_params_ci}.
}

\clearpage

\begin{longtable}{@{}p{1.2cm}p{5.2cm}cp{5.0cm}@{}}
  \caption{\;{Summary of model parameters, indicating whether each parameter was fixed or estimated during calibration.
           Phase index $p\in\{0,1,2,3\}$ correspond to the Early Growth, Late Exponential, Stationary, and Decline phases, respectively.}}
  \label{tab:param_overview}\\
  \toprule
  \textbf{Sub-model} & \textbf{Parameter group} & \textbf{Count} &
  \textbf{Source / notes} \\
  \midrule
  \endfirsthead
  \multicolumn{4}{c}{\tablename~\thetable{} (continued)}\\
  \toprule
  \textbf{Sub-model} & \textbf{Parameter group} & \textbf{Count} &
  \textbf{Source / notes} \\
  \midrule
  \endhead
  \bottomrule
  \endfoot

  \multicolumn{4}{l}{\textit{Cell culture kinetic sub-model (all estimated from data)}}\\[2pt]
  CC & Max.\ reaction rates $v_{\max,p}$ & 111 &
       Michaelis--Menten $v_{\max}$ for 20+ metabolic reactions; phase-specific \\
  CC & Michaelis constants $K_{m,p}$ & 126 &
       Substrate affinities for metabolism, IgG synthesis, and biomass \\
  CC & Inhibition constants $K_{i,p}$ & 15 &
       Lactate and ammonium product inhibition \\
  CC & Activation constants $K_{a,p}$ & 4 &
       Glutamine activation of transport \\
  CC & Growth/death constants $k_p$ & 14 &
       Net growth and apoptosis rate per phase \\
  CC & Phase-transition logistic coefficients $\beta$ & 12 &
       OUR-driven logistic regression for phase transitions (3 transitions $\times$ 4 coefficients) \\
  CC & Miscellaneous ($K_1$, $K_2$, $\alpha$) & 9 & \\
  \cmidrule(l){3-3}
  CC & \textbf{Subtotal} & \textbf{291} & All parameters were included in the fitting process; however, only 49 were updated, while the remaining parameters retained their literature-reported values. \\[4pt]

  \multicolumn{4}{l}{\textit{Glycosylation sub-model -- estimated parameters}}\\[2pt]
  Glyco & Max.\ enzyme turnover rates $k_{f\max}$ & 7 &
          One per enzyme (Man1, Man2, GnT1, GnT2, FucT, GalT, SiaT);
          cell-line-specific $k_{\text{cat}}$ \\
  Glyco & Peak Golgi enzyme concentrations $e_{\max}$ & 7 &
          Cell-line-specific enzyme expression levels \\
  Glyco & Enzyme pH-activity widths $\omega_f$ & 7 &
          Golgi pH environment of this cell line \\
  Glyco & Sialyltransferase dissociation constants $K_{d,\text{SiaT}}$,
          $K_{dk,\text{SiaT}}$ & 2 &
          Adjusted from \cite{villiger2016controlling} defaults \\
  Glyco & Galacto\-syltrans\-ferase dissociation constants
          $K_{d,\text{GalT}}$, $K_{d,\text{GalT\_f}}$,
          $K_{dk,\text{GalT}}$ & 3 &
          Adjusted from \cite{villiger2016controlling} defaults \\
  Glyco & Golgi pH buffering constant $n_a$ & 1 &
          Adjusted from \cite{villiger2016controlling} default \\
  \cmidrule(l){3-3}
  Glyco & \textbf{Subtotal} & \textbf{27} & Estimated from glycoform data \\[4pt]

  \multicolumn{4}{l}{\textit{Glycosylation sub-model -- fixed from literature}}\\[2pt]
  Glyco & Golgi geometry ($a$, $v$, $d$, $q$) & 4 &
          \cite{villiger2016controlling}, SI Table~1 \\
  Glyco & Enzyme localisation ($z_{\max,j}$, $\omega_j$) & 14 &
          \cite{villiger2016controlling}, SI Tables~6--7 \\
  Glyco & Transport protein params ($t_{p\max}$, $z_{\max,k}$, $\omega_k$) & 12 &
          \cite{villiger2016controlling}, SI Table~7 \\
  Glyco & Transport kinetics ($k_{tk}$, $k_{cyt,ns}$, $k_{Golgi,n}$) & 11 &
          \cite{jimenez2011dynamic}, Table~3 \\
  Glyco & Dissociation constants (Man1, Man2, GnT1, GnT2, FucT) & 10 &
          \cite{villiger2016controlling}, SI Table~9 \\
  Glyco & UDP-Gal balance ($m_{UDP}$, $k_{UDP}$, $k_{Gal,UDP}$) & 3 &
          \cite{jimenez2011dynamic}, Table~3 \\
  Glyco & Optimal Golgi pH $\text{pH}_{\text{opt}}$, apparent p$K_A$ & 2 &
          \cite{villiger2016controlling}, SI Tables~4--5 \\
  Glyco & Mn dissociation constants $K_{d,Mn,\text{GnT}}$, $K_{d,Mn,\text{GalT}}$ & 2 &
          \cite{jimenez2011dynamic}, Table~3 \\
  Glyco & Cytosolic nucleotide sugar concentrations $\mathbf{ns}_{\text{cyt}}$,
          nucleotide concentrations $\mathbf{n}_{\text{cyt}}$ & 7 & \cite{jimenez2011dynamic} \\
  \cmidrule(l){3-3}
  Glyco & \textbf{Subtotal} & \textbf{65} & All fixed at literature values \\
\end{longtable}

\clearpage

\begin{longtable}{@{}lrrrrrr@{}}
  \caption{\;{Cell-culture sub-model parameters retained after sensitivity screening ($n=49$).  $\theta^0$ denotes the reference value and 
           $\hat{\theta}$ the TRF estimate (Case A, $\chi^2/n=0.77$).
          SE and RSE are the Jacobian-based standard and relative standard errors, respectively;
           95\%~CI are calculated as $\hat{\theta}\pm1.96\,\mathrm{SE}$.
           Parameters with RSE$\approx0$ lie at or near a parameter bound, resulting in near-zero pseudoinverse-based SE estimates.}}
  \label{tab:active_params}\\
  \toprule
  \textbf{Parameter} & $\theta^0$ & $\hat{\theta}$ & SE & RSE\,\% &
  CI$_{95}$ lo & CI$_{95}$ hi \\
  \midrule
  \endfirsthead
  \multicolumn{7}{c}{\tablename~\thetable{} (continued)}\\
  \toprule
  \textbf{Parameter} & $\theta^0$ & $\hat{\theta}$ & SE & RSE\,\% &
  CI$_{95}$ lo & CI$_{95}$ hi \\
  \midrule
  \endhead
  \bottomrule
  \endfoot
  \multicolumn{7}{l}{\textit{Well constrained (RSE $<$ 20\%)}}\\
  \texttt{k\_3} & 10 & 7.87 & $\ll$0.001 & $\approx$0 & 7.87 & 7.87 \\
  \texttt{k\_1} & 10 & 10.75 & 5.62e-05 & $\approx$0 & 10.75 & 10.75 \\
  \texttt{k\_2} & 20.01 & 16.91 & 0.00217 & $\approx$0 & 16.9 & 16.91 \\
  \texttt{k\_GLNph\_3} & 3.5 & 2.45 & 3.43e-04 & $\approx$0 & 2.45 & 2.45 \\
  \texttt{k\_GLUph\_3} & 3 & 2.58 & 7.01e-04 & $\approx$0 & 2.58 & 2.58 \\
  \texttt{k\_GLUph\_0} & 5 & 5.56 & 0.182 & 3.3 & 5.21 & 5.92 \\
  \texttt{k\_GLU17ph\_1} & 15 & 19.5 & 1.15 & 5.9 & 17.25 & 21.75 \\
  \texttt{vmax30\_0} & 0.0310 & 0.0258 & 0.00185 & 7.2 & 0.0221 & 0.0294 \\
  \texttt{vmax18f\_3} & 0.0150 & 0.0159 & 0.00219 & 13.8 & 0.0116 & 0.0202 \\
  \texttt{vmax2\_3} & 0.0500 & 0.0396 & 0.00655 & 16.5 & 0.0268 & 0.0525 \\
  \texttt{vmax7\_3} & 0.0112 & 0.0145 & 0.00264 & 18.2 & 0.0093 & 0.0197 \\
  \midrule[0.3pt]
  \multicolumn{7}{l}{\textit{Moderately constrained (RSE 20--50\%)}}\\
  \texttt{vmax16r\_3} & 0.1000 & 0.0861 & 0.0188 & 21.9 & 0.0492 & 0.123 \\
  \texttt{vmax25\_1} & 0.0100 & 0.0070 & 0.00226 & 32.3 & 0.0026 & 0.0114 \\
  \texttt{vmax15r\_3} & 0.0300 & 0.0210 & 0.00732 & 34.9 & 0.0066 & 0.0354 \\
  \texttt{beta\_23\_2} & -11.99 & -11.05 & 4.17 & 37.7 & -19.21 & -2.88 \\
  \texttt{vmax30\_1} & 0.0100 & 0.0088 & 0.00406 & 46.2 & 8.26e-04 & 0.0167 \\
  \texttt{vmax21\_1} & 0.0082 & 0.0103 & 0.00485 & 46.9 & 8.31e-04 & 0.0199 \\
  \midrule[0.3pt]
  \multicolumn{7}{l}{\textit{Weakly constrained (RSE $>$ 50\%)}}\\
  \texttt{k\_GLNph\_1} & 0.8 & 0.733 & 0.383 & 52.3 & -0.0178 & 1.48 \\
  \texttt{vmax2\_1} & 0.0300 & 0.0210 & 0.0167 & 79.5 & -0.0117 & 0.0538 \\
  \texttt{vmax16r\_0} & 1 & 1.13 & 0.898 & 79.7 & -0.633 & 2.89 \\
  \texttt{vmax2\_0} & 0.12 & 0.0840 & 0.0673 & 80.1 & -0.0478 & 0.216 \\
  \texttt{vmax18f\_1} & 0.0620 & 0.0434 & 0.0386 & 89.0 & -0.0323 & 0.119 \\
  \texttt{vmax3r\_1} & 0.0187 & 0.0199 & 0.0306 & 153.3 & -0.0400 & 0.0799 \\
  \texttt{beta\_23\_0} & -34.99 & -26.73 & 48 & 179.5 & -120.8 & 67.3 \\
  \texttt{vmax19f\_1} & 0.0200 & 0.0140 & 0.0276 & 197.1 & -0.0401 & 0.0681 \\
  \texttt{vmax5r\_1} & 0.0500 & 0.0650 & 0.15 & 230.3 & -0.228 & 0.358 \\
  \texttt{K\_mEGLC\_0} & 10.37 & 13.47 & 32.7 & 242.8 & -50.65 & 77.6 \\
  \texttt{beta\_01\_0} & -11 & -12.06 & 31.7 & 262.9 & -74.18 & 50.06 \\
  \texttt{vmax3f\_0} & 0.621 & 0.435 & 1.15 & 263.7 & -1.81 & 2.68 \\
  \texttt{beta\_12\_0} & -30 & -30.77 & 83.1 & 270.1 & -193.6 & 132.1 \\
  \texttt{vmax5f\_0} & 0.15 & 0.105 & 0.307 & 292.8 & -0.498 & 0.708 \\
  \texttt{k\_0} & 5 & 3.5 & 10.5 & 300.6 & -17.12 & 24.12 \\
  \texttt{beta\_12\_1} & 3.7 & 3.38 & 11 & 326.6 & -18.27 & 25.03 \\
  \texttt{K\_mNH4GLN\_1} & 7.2 & 9.36 & 34.7 & 370.6 & -58.6 & 77.32 \\
  \texttt{beta\_01\_2} & 49.99 & 35.93 & 134 & 374.2 & -227.6 & 299.5 \\
  \texttt{K\_mNH4\_1} & 15 & 10.55 & 41.5 & 393.7 & -70.84 & 91.94 \\
  \texttt{vmax17f\_1} & 0.0400 & 0.0520 & 0.289 & 555.4 & -0.514 & 0.618 \\
  \texttt{vmax15r\_1} & 0.2 & 0.184 & 1.53 & 832.7 & -2.81 & 3.18 \\
  \texttt{vmax17r\_1} & 0.0500 & 0.0355 & 0.324 & 913.0 & -0.6 & 0.672 \\
  \texttt{vmax15f\_1} & 0.1000 & 0.129 & 1.27 & 985.8 & -2.36 & 2.62 \\
  \texttt{beta\_01\_1} & 1.2 & 0.874 & 10.4 & >999 & -19.45 & 21.2 \\
  \texttt{beta\_23\_1} & 4 & 4.57 & 65.1 & >999 & -123 & 132.1 \\
  \texttt{vmax23\_2} & 0.0100 & 0.0070 & 0.13 & >999 & -0.248 & 0.262 \\
  \texttt{vmax19f\_2} & 0.0300 & 0.0210 & 0.442 & >999 & -0.846 & 0.888 \\
  \texttt{vmax18f\_2} & 0.0200 & 0.0203 & 0.598 & >999 & -1.15 & 1.19 \\
  \texttt{vmax2\_2} & 0.0400 & 0.0343 & 1.07 & >999 & -2.06 & 2.13 \\
  \texttt{vmax21\_2} & 0.0059 & 0.0061 & 0.207 & >999 & -0.399 & 0.411 \\
  \texttt{vmax5f\_2} & 0.0043 & 0.0044 & 0.299 & >999 & -0.582 & 0.59 \\
  \texttt{K\_mNH4GLN\_2} & 7.39 & 7.52 & 2.45e+03 & >999 & -4794 & 4809 \\
\end{longtable}

\clearpage

\begin{longtable}{@{}lrrrrrr@{}}
  \caption{\;{Glycosylation sub-model parameter estimates ($n=27$) obtained from 48 glycoform observations (8 glycoforms across 6 time points; Case A).
  Parameters were estimated using bounded TRF optimization ($\pm50\%$ bounds) with LSODA simulation and a 10-core parallel Jacobian.
  The final fit yielded         $\chi^2_{\text{data}}/n = 60.7$ and $s^2 = 62.1$.
           SEs and 95\% CIs were computed from the Jacobian pseudoinverse. Parameters at or near a bound have RSE$\approx0$.}}
  \label{tab:glyco_params_ci}\\
  \toprule
  \textbf{Parameter} & $\theta^0$ & $\hat{\theta}$ & SE & RSE\,\% &
  CI$_{95}$ lo & CI$_{95}$ hi \\
  \midrule
  \endfirsthead
  \multicolumn{7}{c}{\tablename~\thetable{} (continued)}\\
  \toprule
  \textbf{Parameter} & $\theta^0$ & $\hat{\theta}$ & SE & RSE\,\% &
  CI$_{95}$ lo & CI$_{95}$ hi \\
  \midrule
  \endhead
  \bottomrule
  \endfoot
  \multicolumn{7}{l}{\textit{Well constrained (RSE $<$ 20\%)}}\\
  $K_{d,\text{SiaT}}$ & 4810 & 5717 & 59.7 & 1.0 & 5600 & 5834 \\
  $k_{f\max,1}$ & 1324 & 1807 & 21.2 & 1.2 & 1766 & 1849 \\
  $k_{f\max,6}$ & 2300 & 1598 & 25.1 & 1.6 & 1548 & 1647 \\
  $k_{f\max,2}$ & 460 & 684.8 & 11.7 & 1.7 & 661.8 & 707.8 \\
  $k_{f\max,3}$ & 600 & 762.5 & 43 & 5.6 & 678.2 & 846.7 \\
  $k_{f\max,5}$ & 1200 & 639.9 & 39.7 & 6.2 & 562.1 & 717.7 \\
  \midrule[0.3pt]
  \multicolumn{7}{l}{\textit{Moderately constrained (RSE 20--50\%)}}\\
  $k_{f\max,4}$ & 291 & 310.9 & 146 & 47.0 & 24.69 & 597.1 \\
  \midrule[0.3pt]
  \multicolumn{7}{l}{\textit{Weakly constrained (RSE $>$ 50\%)}}\\
  $K_{d,\text{GalT}}$ & 4160 & 4957 & 1.27e+05 & >999 & -2.44e+05 & 2.54e+05 \\
  $K_{dk,\text{GalT}}$ & 70 & 70.77 & 4.44e+03 & >999 & -8629 & 8770 \\
  $K_{dk,\text{SiaT}}$ & 10 & 10.01 & 704 & >999 & -1370 & 1390 \\
  $e_{\max,4}$ & 0.183 & 0.199 & 17.8 & >999 & -34.74 & 35.14 \\
  $n_a$ & 14 & 13.46 & 1.88e+03 & >999 & -3677 & 3704 \\
  $\omega_{f,5}$ & 0.5 & 0.25 & 38.4 & >999 & -75.1 & 75.6 \\
  $K_{d,\text{GalT\_f}}$ & 1600 & 1909 & 3.31e+05 & >999 & -6.46e+05 & 6.50e+05 \\
  $e_{\max,5}$ & 0.855 & 0.428 & 127 & >999 & -249.4 & 250.3 \\
  $e_{\max,6}$ & 0.626 & 0.51 & 159 & >999 & -311.8 & 312.8 \\
  $\omega_{f,0}$ & 1.72 & 1.99 & 1.17e+03 & >999 & -2294 & 2298 \\
  $\omega_{f,4}$ & 1.5 & 1.53 & 949 & >999 & -1858 & 1861 \\
  $k_{f\max,0}$ & 400 & 593.6 & 4.95e+05 & >999 & -9.70e+05 & 9.71e+05 \\
  $e_{\max,2}$ & 0.114 & 0.171 & 151 & >999 & -296.5 & 296.8 \\
  $e_{\max,0}$ & 0.232 & 0.341 & 331 & >999 & -647.8 & 648.4 \\
  $e_{\max,3}$ & 0.102 & 0.138 & 235 & >999 & -461.3 & 461.5 \\
  $e_{\max,1}$ & 0.141 & 0.197 & 562 & >999 & -1101 & 1101 \\
  $\omega_{f,6}$ & 1 & 0.988 & 3.15e+03 & >999 & -6172 & 6174 \\
  $\omega_{f,3}$ & 0.96 & 1.19 & 5.88e+03 & >999 & -1.152e+04 & 1.152e+04 \\
  $\omega_{f,2}$ & 1.08 & 1.36 & 6.79e+03 & >999 & -1.331e+04 & 1.331e+04 \\
  $\omega_{f,1}$ & 1.39 & 1.59 & 3.04e+04 & >999 & -5.965e+04 & 5.965e+04 \\
\end{longtable}

\clearpage

\clearpage

\clearpage

\clearpage

\clearpage

\clearpage

\begin{longtable}{l r r r r}
\caption{\; Kinetic parameter values by growth phase $z$. Dashes denote parameters not active in that phase.}\label{tab:params}\\
\toprule
Parameter & $z=0$ & $z=1$ & $z=2$ & $z=3$ \\
\midrule
\endfirsthead
\multicolumn{5}{c}{\tablename\ \thetable{} -- continued}\\
\toprule
Parameter & $z=0$ & $z=1$ & $z=2$ & $z=3$ \\
\midrule
\endhead
\midrule \multicolumn{5}{r}{\textit{continued on next page}}\\
\endfoot
\bottomrule
\endlastfoot
$V_{2}^{(z)}$ & 0.12 & 0.03 & 0.04 & 0.05 \\
$V_{3f}^{(z)}$ & 0.6207 & 0.1073 & 0.1073 & 0.03073 \\
$V_{3r}^{(z)}$ & 0.08219 & 0.01866 & 0.1187 & 0.05187 \\
$V_{5f}^{(z)}$ & 0.15 & 0.02 & 0.0043 & 0 \\
$V_{5r}^{(z)}$ & 0.05 & 0.05 & 0 & 0.002 \\
$V_{7}^{(z)}$ & 0.0316 & $2.487\times10^{-6}$ & $6.002\times10^{-4}$ & 0.01116 \\
$V_{16f}^{(z)}$ & 1 & 0.05 & 0.3 & 0.08 \\
$V_{16r}^{(z)}$ & 1 & 0.8 & 0.02 & 0.1 \\
$V_{18f}^{(z)}$ & 0.02 & 0.062 & 0.02 & 0.015 \\
$V_{18r}^{(z)}$ & 0.012 & 0.01 & 0.004 & 0.008 \\
$V_{19f}^{(z)}$ & 0.01 & 0.02 & 0.03 & 0.04 \\
$V_{19r}^{(z)}$ & 0.09 & 0.07 & 0.05 & 0.04 \\
$V_{21}^{(z)}$ & 0.01 & 0 & 0 & 0.0056 \\
$V_{23}^{(z)}$ & 0.001 & 0.01 & 0.01 & 0.01 \\
$V_{25}^{(z)}$ & 0.01 & 0.01 & 0.01 & 0.01 \\
$V_{28}^{(z)}$ & 0.001 & 0.0018 & 0.0035 & $3.5\times10^{-4}$ \\
$V_{30}^{(z)}$ & 0.031 & 0.01 & 0.001 & -0.004 \\
$K_{m,\mathrm{EGLC}}^{(z)}$ & 10.37 & 25 & 25 & 10.37 \\
$K_{i,\mathrm{LAC}}^{(z,2)}$ & 600 & 600 & 600 & 600 \\
$K_{i,\mathrm{LAC}}^{(z,16)}$ & 20 & 5 & 500 & 500 \\
$K_{m,\mathrm{ELAC}}^{(z)}$ & 10.37 & 5.365 & 5.365 & 4.957 \\
$K_{m,\mathrm{EALA}}^{(z)}$ & 10 & 1.3 & 1.3 & 1.3 \\
$K_{m,\mathrm{ESER}}^{(z)}$ & 7.316 & 1.849 & 14.89 & 14.07 \\
$K_{m,\mathrm{EGLN}}^{(z)}$ & 10 & 5 & 100 & 20 \\
$K_{m,\mathrm{NH}_3}^{(z)}$ & 100 & 15 & 30 & 50 \\
$K_{m,\mathrm{EGLU}}^{(z)}$ & 10 & 20 & 2 & 2 \\
$K_{m,\mathrm{EASP}}^{(z)}$ & 1.852 & 1.749 & 2.249 & 2.065 \\
$K_{m,\mathrm{ELEU}}^{(z)}$ & 6 & 6 & 6 & 4 \\
$K_{m,\mathrm{EILE}}^{(z)}$ & 100 & 3 & 3 & 3 \\
$K_{m,\mathrm{EVAL}}^{(z)}$ & 4 & 4 & 4 & 4 \\
$K_{m,\mathrm{EGLN}}^{(z,28)}$ & 0.001 & 0.001 & 0.001 & 0.001 \\
$K_{m,\mathrm{EGLU}}^{(z,28)}$ & 5 & 5 & 10 & 0.001 \\
$K_{m,\mathrm{EASP}}^{(z,28)}$ & 0.001 & 0.001 & 0.001 & 0.001 \\
$K_{m,\mathrm{EALA}}^{(z,28)}$ & 0.001 & 0.001 & 0.001 & 0.001 \\
$K_{m,\mathrm{ESER}}^{(z,28)}$ & 0.001 & 0.001 & 0.001 & 0.001 \\
$K_{m,\mathrm{ELEU}}^{(z,28)}$ & 0.001 & 0.001 & 0.001 & 0.001 \\
$K_{m,\mathrm{EILE}}^{(z,28)}$ & 0.001 & 0.001 & 0.001 & 0.001 \\
$K_{m,\mathrm{EVAL}}^{(z,28)}$ & 0.001 & 0.001 & 0.001 & 0.001 \\
$K_{m,\mathrm{EGLC}}^{(z,30)}$ & 0.001 & 0.001 & 0.001 & 0.001 \\
$K_{m,\mathrm{EGLN}}^{(z,30)}$ & 0.001 & 0.001 & 0.001 & 0.001 \\
$K_{m,\mathrm{EGLU}}^{(z,30)}$ & 0.001 & 0.001 & 0.001 & 0.001 \\
$K_{m,\mathrm{EASP}}^{(z,30)}$ & 0.001 & 0.001 & 0.001 & 0.001 \\
$K_{m,\mathrm{EALA}}^{(z,30)}$ & 0.001 & 0.001 & 0.001 & 0.001 \\
$K_{m,\mathrm{ESER}}^{(z,30)}$ & 0.001 & 0.001 & 0.001 & 0.001 \\
$K_{m,\mathrm{ELEU}}^{(z,30)}$ & 0.001 & 0.001 & 0.001 & 0.001 \\
$K_{m,\mathrm{EILE}}^{(z,30)}$ & 0.001 & 0.001 & 0.001 & 0.001 \\
$K_{m,\mathrm{EVAL}}^{(z,30)}$ & 0.001 & 0.001 & 0.001 & 0.001 \\
$\kappa_{2}^{(z)}$ & 5 & 10 & 20 & 10 \\
$K_{1,2}^{(z)}$ & $1\times10^{-8}$ & $1\times10^{-7}$ & $1\times10^{-8}$ & $5.623\times10^{-8}$ \\
$K_{2,2}^{(z)}$ & $1\times10^{-8}$ & $1\times10^{-7}$ & $1\times10^{-8}$ & $5.623\times10^{-8}$ \\
$\kappa_{15}^{(z)}$ & 0.5 & 0.8 & -- & 3.5 \\
$K_{1,15}^{(z)}$ & $1\times10^{-7}$ & $1\times10^{-7}$ & -- & $1\times10^{-7}$ \\
$K_{2,15}^{(z)}$ & $1\times10^{-7}$ & $1\times10^{-7}$ & -- & $1\times10^{-7}$ \\
$\kappa_{16}^{(z)}$ & 5 & 3 & 3 & 3 \\
$K_{1,16}^{(z)}$ & $1\times10^{-7}$ & $1\times10^{-7}$ & $1\times10^{-7}$ & $7.943\times10^{-8}$ \\
$K_{2,16}^{(z)}$ & $1\times10^{-7}$ & $1\times10^{-7}$ & $1\times10^{-7}$ & $7.943\times10^{-8}$ \\
$V_{27}^{(z)}$ & 0.03 & 0.02 & 0.03 & 0.02 \\
$V_{17f}^{(z)}$ & 0.1 & 0.04 & 0.15 & 0.01 \\
$V_{17r}^{(z)}$ & 0.03 & 0.05 & 0.015 & $1\times10^{-4}$ \\
$V_{15f}^{(z)}$ & 0.49 & 0.1 & 0.2 & 0.1 \\
$V_{15r}^{(z)}$ & 0.15 & 0.2 & 0.01 & 0.03 \\
$K_{m,\mathrm{GLN}}^{(z)}$ & 1 & 30 & 1 & 5 \\
$K_{m,\mathrm{NH}_3}^{(z,16)}$ & 20 & 100 & 10 & 10 \\
$\kappa_{17}^{(z)}$ & -- & 15 & -- & -- \\
$K_{m,\mathrm{ENH}_3}^{(z)}$ & 20 & 4 & -- & -- \\
$K_{i,\mathrm{NH}_3}^{(z)}$ & 10 & 4 & 8 & -- \\
$K_{i,\mathrm{ENH}_3}^{(z,28)}$ & 100 & 4 & 3 & 100 \\
\end{longtable}

\clearpage
\begin{longtable}{@{}l p{6cm} p{5.5cm}@{}}
\caption{\;Glycosylation-model parameters. Enzyme-indexed vectors are ordered (Man\,I, Man\,II, GnT\,I, GnT\,II, FucT, GalT, SiaT); nucleotide-sugar-indexed vectors are ordered (UDP-GlcNAc, GDP-Fuc, UDP-Gal, CMP-Neu5Ac).}\label{tab:glyco_params}\\
\toprule
Parameter & Description & Value \\
\midrule
\endfirsthead
\multicolumn{3}{c}{\tablename\ \thetable{} -- continued}\\
\toprule
Parameter & Description & Value \\
\midrule
\endhead
\midrule \multicolumn{3}{r}{\textit{continued on next page}}\\
\endfoot
\bottomrule
\endlastfoot
\multicolumn{3}{@{}l}{\textbf{Golgi geometry and pH}}\\
\texttt{a} & Golgi surface area & 99 \\
\texttt{v} & Golgi volume & 25 \\
\texttt{d} & Golgi internal diameter & 7.82 \\
\texttt{l} & Golgi length (not used) & 0.52 \\
\texttt{q} & Volumetric flow rate through Golgi & 1.12 \\
\texttt{phopt} & Optimal Golgi pH & 6.6 \\
\texttt{pka} & Apparent Golgi p$K_a$ & 7.5 \\
\addlinespace
\multicolumn{3}{@{}l}{\textbf{Enzyme concentration profiles (per enzyme)}}\\
\texttt{emax} & Peak enzyme concentration & 0.232, 0.141, 0.114, 0.1022, 0.183, 0.855, 0.626 \\
\texttt{zmaxj} & Location of peak (normalised Golgi position) & 0.255, 0.388, 0.363, 0.495, 0.525, 0.776, 0.782 \\
\texttt{omegaj} & Width of concentration profile ($\times2$ if dynamic) & 0.0785, 0.0575, 0.0783, 0.0781, 0.0753, 0.045, 0.0379 \\
\addlinespace
\multicolumn{3}{@{}l}{\textbf{Enzyme kinetics (per enzyme)}}\\
\texttt{kfmax} & Maximum catalytic turnover rate & 400, 1324, 460, 600, 291, 1200, 2300 \\
\texttt{omegaf} & Enzyme activity (pH-profile) parameter & 1.72, 1.39, 1.08, 0.96, 1.5, 0.5, 1 \\
\addlinespace
\multicolumn{3}{@{}l}{\textbf{Nucleotide-sugar transport (per nucleotide sugar)}}\\
\texttt{tpmax} & Peak transport-protein concentration & $7.6\times10^{-7}$, $3.65\times10^{-7}$, $6.94\times10^{-7}$, $8.23\times10^{-7}$ \\
\texttt{zmaxk} & Location of peak transport-protein conc. & 0.369, 0.496, 0.734, 0.791 \\
\texttt{omegak} & Width of transport-protein profile & 0.0651, 0.0843, 0.0426, 0.0516 \\
\texttt{kcytns} & Cytosolic nucleotide-sugar $K_m$ (antiporter) & 7.13, 7.5, 2.4, 1.3 \\
\texttt{kgolgin} & Golgi nucleotide $K_m$ (antiporter) & 0.135, 0.14, 0.124, 0.133 \\
\texttt{ktk} & Antiporter turnover rate & 1084, 130, 689, 397 \\
\addlinespace
\multicolumn{3}{@{}l}{\textbf{Enzyme dissociation constants}}\\
\texttt{kdman1} & Man I $K_d$ (acceptor Man9/8/7/6) & 60.5, 110, 30.8, 74.1 \\
\texttt{kdman2} & Man II $K_d$ (acceptor Man5/4/5/4) & 200, 100, 200, 100 \\
\texttt{kdgnt1} & GnT I $K_d$ (glycan acceptor, Man5) & 115 \\
\texttt{kdkgnt1} & GnT I $K_d$ (UDP-GlcNAc donor) & 170 \\
\texttt{kdgnt2} & GnT II $K_d$ (glycan acceptor) & 97 \\
\texttt{kdkgnt2} & GnT II $K_d$ (UDP-GlcNAc donor) & 960 \\
\texttt{kdfuct} & FucT $K_d$ (glycan acceptor) & 43.4 \\
\texttt{kdkfuct} & FucT $K_d$ (GDP-Fuc donor) & 46 \\
\texttt{kdgalt} & GalT $K_d$ (agalacto acceptor, FA2G0) & 4160 \\
\texttt{kdgalt\_f} & GalT $K_d$ (mono-galacto acceptor, FA2G1) & 1600 \\
\texttt{kdkgalt} & GalT $K_d$ (UDP-Gal donor) & 70 \\
\texttt{kdsiat} & SiaT $K_d$ (glycan acceptor) & 4810 \\
\texttt{kdksiat} & SiaT $K_d$ (CMP-Neu5Ac donor) & 10 \\
\addlinespace
\multicolumn{3}{@{}l}{\textbf{Manganese dependence}}\\
\texttt{kdmngnt} & Mn$^{2+}$ $K_d$ for GnT & 0.00547 \\
\texttt{kdmngalt} & Mn$^{2+}$ $K_d$ for GalT & 0.0382 \\
\addlinespace
\multicolumn{3}{@{}l}{\textbf{Nucleotide-sugar synthesis and pH coupling}}\\
\texttt{mudpgal} & UDP-Gal maintenance coefficient & 0.0616 \\
\texttt{kudpgal} & Sugar-mediated NS synthesis rate & 0.514 \\
\texttt{na} & Ammonia--Golgi-pH coupling constant & 14 \\
\texttt{kgaludpgal} & Gal/UDP-Gal equilibrium constant & 56.9 \\
\addlinespace
\multicolumn{3}{@{}l}{\textbf{Cell growth and ammonia}}\\
\texttt{mumax} & Max specific growth rate (growth phase) & 1.21 \\
\texttt{mumax\_d} & Max specific growth rate (decline phase) & 0.122 \\
\texttt{k\_amm} & Ammonia inhibition constant (growth) & 2.37 \\
\texttt{k\_amm\_d} & Ammonia inhibition constant (decline) & 8.24 \\
\texttt{q\_amm} & Ammonia specific production rate & 0.0936 \\
\addlinespace
\multicolumn{3}{@{}l}{\textbf{Cytosolic pools and titre}}\\
\texttt{nscyt} & Cytosolic NS conc. (UDP-GlcNAc, GDP-Fuc, UDP-Gal, CMP-Neu5Ac) & 1620, 43, 115.8, 40 \\
\texttt{ncyt} & Cytosolic nucleotide conc. (UMP+UDP, GMP+GDP, CMP+CDP) & 1942, 496, 248 \\
\texttt{mabtiter} & mAb titre (molar), computed & 66.39 \\
\addlinespace
\end{longtable}

\clearpage
\begin{table}[t]\centering
\caption{\;Reaction-index-specific dissociation constants applied in the glycosylation network, expressed as scalings of the base values $\texttt{kdgalt}=4.16\times10^{3}$, $\texttt{kdgalt\_f}=1600$ and $\texttt{kdsiat}=4.81\times10^{3}$~$\mu$M. (A further GalT override at index 21, $1.5\,\texttt{kdsiat}$, is currently disabled.)}\label{tab:kd_overrides}
\begin{tabular}{@{}l l r@{}}
\toprule
Reaction index & $K_d$ expression & Value ($\mu$M) \\
\midrule
\multicolumn{3}{@{}l}{\textbf{GalT ($\beta$-1,4-galactosyltransferase)}}\\
G0, G0F & \texttt{kdgalt\_f} & 1600 \\
G1 & $0.30\,$\texttt{kdgalt} & 1248 \\
G1F & $1.50\,$\texttt{kdgalt} & 6240 \\
G1S1 & $0.05\,$\texttt{kdgalt} & 208 \\
G1FS1 & $0.25\,$\texttt{kdgalt} & 1040 \\
all others & \texttt{kdgalt} (base) & 4160 \\
\addlinespace
\multicolumn{3}{@{}l}{\textbf{SiaT ($\alpha$-2,6-sialyltransferase)}}\\
G2S1 & $1.20\,$\texttt{kdsiat} & 5772 \\
G2FS1 & $0.70\,$\texttt{kdsiat} & 3367 \\
all others (0--32) & \texttt{kdsiat} (base) & 4810 \\
\bottomrule
\end{tabular}\end{table}

\end{document}